\documentclass[fleqn,usenatbib]{mnras}

\usepackage{newtxtext,newtxmath}

\usepackage[T1]{fontenc}
\usepackage{ae,aecompl}

\usepackage{graphicx}	
\usepackage{amsmath}	
\usepackage{amssymb}	

\usepackage{enumitem}
\newlist{legal}{enumerate}{10}
\setlist[legal]{label*=\arabic*.}
\usepackage{epsfig}
\usepackage{color}
\usepackage{natbib}
\usepackage{float}
\usepackage{dblfloatfix}
\usepackage{csvsimple}
\usepackage{natbib,twoopt}
\usepackage{csquotes}

\usepackage{rotating}
\usepackage[table]{xcolor}
\usepackage{setspace}
\usepackage{etoolbox}
\usepackage{threeparttable}

\DeclareMathOperator\erf{erf}
\DeclareMathOperator\median{median}

\title[Optically variable AGNs in GOODS-S]{Robust Identification of Active Galactic Nuclei through \emph{HST} Optical Variability in GOODS-S: Comparison with the X-ray and mid-IR Selected Samples}

\author[E. Pouliasis et al.]{E.~Pouliasis$^{1,2}$\thanks{E-mail: epouliasis@astro.noa.gr},
I.~Georgantopoulos$^{1}$,
A.~Z.~Bonanos$^{1}$,
M.~Yang$^{1}$,
K.~V.~Sokolovsky$^{3,4,1}$,
\newauthor D.~Hatzidimitriou$^{2,1}$, 
G.~Mountrichas$^{1}$,
P.~Gavras$^{5,1}$,
V.~Charmandaris$^{6,7}$,
I.~Bellas-Velidis$^{1}$,
\newauthor Z.~T.~Spetsieri$^{1,2}$
and K.~Tsinganos$^{2,1}$
\\
$^{1}$IAASARS, National Observatory of Athens, 15236 Penteli, Greece\\
$^{2}$Department of Astrophysics, Astronomy \& Mechanics, Faculty of Physics, University of Athens, Zografos, 15783 Athens, Greece\\
$^{3}$Sternberg Astronomical Institute, Moscow State University, Universitetskii pr. 13, 119992 Moscow, Russia\\
$^{4}$Astro Space Center of Lebedev Physical Institute, Profsoyuznaya Str. 84/32, 117997 Moscow, Russia\\
$^{5}$Rhea for ESA/ESAC, Camino bajo del Castillo, s/n, Urbanizacion Villafranca del Castillo, Villanueva de la Ca\~nada, 28692 Madrid, Spain\\
$^{6}$University of Crete, Department of Physics, GR-71003 Heraklion, Greece\\
$^{7}$Institute of Astrophysics, Foundation for Research and Τechnology-Hellas, GR-71110 Heraklion, Greece\\
}

\date{Accepted 2019 May 27. Received 2019 May 24; in original form 2019 February 15}

\pubyear{2019}

\begin{document}
\label{firstpage}
\pagerange{\pageref{firstpage}--\pageref{lastpage}}
\maketitle
\begin{abstract}
Identifying Active Galactic Nuclei (AGNs) through their X-ray emission is efficient, but necessarily biased against X-ray-faint objects.
We aim to characterize this bias by comparing X-ray-selected AGNs to the ones identified through optical variability and mid-IR colours.
We present a catalogue of AGNs selected through optical variability using all publicly available z-band {\em Hubble} Space Telescope images in the GOODS-South field. For all objects in the catalogue, we compute X-ray upper limits or discuss detections in the deepest available $\sim$7\,Ms {\em Chandra} Deep Field South images and present the {\em Spitzer}/IRAC mid-IR colours.
For the variability study, we consider only sources observed over at least five epochs and over a time baseline of up to ten years. We adopt the elevated median absolute deviation as a variability indicator robust against individual outlier measurements 
and identify 113 variability-selected AGN candidates. Among these, 26 have an X-ray counterpart and lie within the conventional AGN area in the F$_X$/F$_{\rm opt}$ diagram. The candidates with X-ray upper limits are on average optically fainter, have higher redshifts compared to the X-ray detected ones and are consistent with low luminosity AGNs. Out of 41 variable optical sources with IR detections, 13 fulfill the IR AGN colour selection criteria.
Our work emphasizes the importance of optical variability surveys for constructing complete samples of AGNs including the ones that remain undetected even by the deepest X-ray and IR surveys.
\end{abstract}




\section{Introduction}\label{intro}

It is widely accepted that massive galaxies and a fraction of lower mass galaxies host a supermassive black hole in their centre \citep{magorrian1998,kormendy2004,filippenko2003,barth2004,greene2004,greene2007,dong2007,greene2008}. Moreover, there is a correlation between the mass of the black holes and the properties of their host galaxies \citep{kormendy2013}, such as the luminosity, the stellar mass, the velocity dispersion or the bulge rotational velocity \citep{dressler1989,kormendy1995,magorrian1998,ferrarese2000,gebhardt2000,tremaine2002,marconi2003,haring2004,ferrarese2005,graham2007,gultekin2009}. 
Hence, either the processes that take place in Active Galactic Nuclei (AGNs) play an important role in star formation and shaping of the galaxy structure, or vice versa the galaxy evolution directly affects the mass and spin of the central black hole. To better understand the relations between the central black hole and its host galaxy, it is necessary to have complete samples of AGNs, not biased against redshift, obscuration, luminosity, etc.

X-rays penetrate deep into material with high column densities and therefore are nearly unaffected by moderate obscuration ($N_H<10^{24}$ $cm^{-2}$) \citep{brandt2015,alexander2016}. Thus wide-field X-ray imaging constitutes the most common and efficient technique to identify AGN. Other methods make use of infrared (IR), ultraviolet and optical single and multiple colour selection criteria to identify AGNs, as the AGN spectral energy distributions differ from those of normal galaxies or stars \citep{richards2002,richards2005,lacy2004,stern2005,alonsoherrero2006,donley2007,schneider2007,schneider2010}. However, at these wavelengths, the AGN samples may be contaminated by foreground stars or biased by the dust emission from the host galaxies compared to X-ray AGN identification where the dilution by the hosts is not that dominant.

 An alternative method to identify AGNs is based on the detection of variability at any wavelengths on timescales from hours to years \citep{ulrich1997,kawa1998,paolillo2004,garcia2014}. There is a correlation between the amplitude of the AGN variability and the timescale of variation \citep{hook1994,trevese1994,cristiani1997,diclemente1996,vandenberk2004,kelly2009,bauer2009,middei2017}, the redshift and the black hole mass (for timescales longer than 100 days; \citealt{cristiani1990,hook1994,trevese1994,vandenberk2004}). On the other hand, the variability amplitude is anticorrelated with the rest-frame wavelength \citep{diclemente1996,cristiani1997,giveon1999,helfand2001,vandenberk2004,zuo2012} and the nuclear luminosity \citep{trevese1994,vagnetti2016}. The latter suggests that low-luminosity AGNs (LLAGNs) will dominate a variability survey compared to more luminous AGNs. 
 
 Several theories and mechanisms have been proposed to explain the variability. For example,  \citet{rees1984} and \citet{kawa1998} proposed that AGN variability may be due to the fact that accretion disks are vulnerable to dynamical instabilities, while \citet{li2008} and \citet{zuo2012} attributed the variability to fluctuations in the accretion rate. In the extreme case of blazars (the most luminous class of AGNs possessing a relativistic jet pointing towards the observer), the variability is modulated by changes in the accretion disk, but also by non-thermal power-law emission from the jets and the (not fully explored) connection between them \citep{gu2013,finke2014,chatterjee2018}. Other interpretations of AGN variability include gravitational microlensing effects \citep{hawkins1993,alexander1995}, tidal disruption events \citep{komossa2015} and multiple explosions of supernovae (SNe) near the nuclei \citep{kawa1998,terlevich1992}. 

 In the last several years, AGN variability has been used in many studies. In the X-ray region, \citet{young2012} and \citet{ding2018} selected LLAGNs in the 4 and 7 Ms {\em Chandra} Deep Field South (CDF-S), respectively. In the optical and near IR bands, \citet{sara2003,sara2011,trevese2008,vill2010,simm2015,falocco2015,graham2014,decicco2015,baldassare2018,kim2018} identified a large sample of AGNs, suggesting that the search for variability at short time scales is efficient in selecting LLAGNs that would have been missed by X-ray surveys. As AGNs exhibit a red-noise behaviour (\citealt{lawrence1993,2017ApJ...834..157P}; i.e. they have more power at low frequencies in the Fourier space), \citet{decicco2015} and \citet{paolillo2017} pointed out 
 that the longer the time baseline (e.g. greater than few years), the larger the variability amplitude and the more complete is the AGN selection. 

However, while blazars tend to show variability at all timescales, the power spectrum and structure function analysis of light curves of many radio-quiet AGNs suggest that their variability amplitude does not rise indefinitely with longer timescales. Their power spectrum flattens below some frequency. Such power spectra were modeled with a damped random walk and continuous auto-regressive moving average models \citep{devries2005,kelly2009,macleod2010,macleod2012,kelly2014,kasliwal2015,kozlowski2016,simm2016}.

In order to study AGN variability over cosmic time one needs a deep field observed multiple times. The Great Observatories Origins Deep Survey Southern field \citep[GOODS-S][]{2004ApJ...600L..93G} centered at $\alpha=$3h32m30s $\delta$ $=-27^{\circ}$48$^{\prime}$20$^{\prime\prime}$ J2000, 
covers an area of $10^\prime \times 16^\prime$. 
It is the most data-rich area of the sky in terms of depth and wavelength coverage and as it has been observed by the {\em Hubble} Space Telescope (HST) multiple times, it perfectly satisfies the requirements of our study.

Variability studies in this field based on  HST multi-epoch data have been performed by \citet{sara2011} and \citet{vill2010,vill2012A}. \citet{sara2011} used the \textit{V}-band (F606W) images over five epochs spanning almost seven months and the standard deviation, $\sigma$, as the statistical variability indicator to identify 42 variable sources. The authors compared their results with the mid-IR and the 2Ms {\em Chandra} X-ray data. \citet{vill2010} identified 88 variable sources (out of $\sim$6,000 sources) using the C-statistic \citep[the ratio of the measured standard deviation, $\sigma$, to the expected one $\sigma_{\rm exp}$, in this case scaled from the estimated photometric errors; ][]{2010AJ....139.1269D} on \textit{z}-band data with the same epochs and time baseline as in \citet{sara2011}. The authors, after removing the false positive detections and the stellar population, validated the AGN nature of 55/88 variable sources through spectral energy distribution fitting, the identification of X-ray counterparts in the CDF-S 4Ms catalogue and auxiliary radio and IR data \citep{vill2012A}.

The field was also targeted in ground-based optical variability studies. 
\citet{trevese2008} studied the variability of sources in AXAF, a larger field that includes GOODS-S. They analysed \textit{V}-band images taken from ground-based telescopes and used magnitude differences between eight epochs over two years of observations to identify 132 variable AGN candidates. Similarly, \citet{falocco2015} applied a multi-epoch variability search spanning six months with the SUDARE-VOICE survey dataset obtained with the VLT Survey Telescope. They selected 175 variable sources over an area of 2\,$deg^2$ around CDF-S using $\sigma$ as the variability index. They compared the optical variable sample with AGNs selected through optical-NIR and IR colour diagnostics and AGNs with X-ray counterparts in the 4Ms CDF-S catalogue. 

We extend the previous HST-based studies of \citet{sara2011} and \citet{vill2010,vill2012A} by using the latest data, variability detection and IR-colour-based AGN selection techniques. We construct a new catalogue of optically variable AGNs based on HST \textit{z}-band observations and comparing it with other selection techniques. We highlight the following novel aspects of this study:
\begin{itemize}
\item We expand the time baseline of the deep HST observations of GOODS-S up to ten years, which should result in a more complete AGN selection.
\item We use the Median Absolute Deviation (MAD) as the variability-detection statistic, which, unlike $\sigma$, is robust against individual outlier measurements \citep{sokol2017b}. We expect MAD to yield a cleaner sample of variable sources compared to the previous studies. 
\item We use the new deepest available 7\,Ms {\em Chandra} image to constrain the X-ray brightness of the variability-selected AGNs.
\item We compare our variable sample with AGN selected in the mid-IR using the \citet{lacy2007} and \citet{donley2012} criteria. 
\end{itemize}

The HST optical observations and the data reduction (astrometry and photometry) along with ancillary data used in this work are presented in Section~\ref{optical}, while in Section~\ref{variability}, we describe the method we used to create the list of variable sources. We also exclude stars and supernovae from the sample of the AGN candidates.
In Section~\ref{properties}, we demonstrate the properties of the AGN candidates (e.g. magnitude and redshift distributions or X-ray luminosities) and construct the mid-IR AGN samples. In Section~\ref{discussiom}, we compare our results with other variability studies and we discuss the differences between optically variable, mid-IR and X-ray selected AGNs, while Section~\ref{summary} presents the summary of the results and conclusions.

\section{Data Reduction}\label{optical}

\subsection{Optical HST data}

\begin{table*}
\caption{The HST Treasury programs included in this study}
\begin{threeparttable}
\begin{tabular}{ c c c c c c c c c c}
\hline \rule{0pt}{1.1\normalbaselineskip} 
Prop. ID & PI Name & Cycle & N\textsubscript{img}\textsuperscript{a} & N\textsubscript{exp}\textsuperscript{b} & Exp. Time (s) & Start\_Obs\textsuperscript{c} & End\_Obs\textsuperscript{c} \\ \hline
    9352   & A. Riess & 11 & 12 & 4 & 1880-4800 & 2002-10 & 2003-02 \\
    9425   & M. Giavalisco & 11 & 78 & 4 & 2040-2120 & 2003-02 & 2003-02\\
    9488   & K. Ratnatunga  & 11 & 2 & 3 & 1800 & 2002-09 & 2003-02\\
    9500   & H-W. Rix & 11 & 58 & 3 & 2160-2286 & 2002-09 & 2003-02\\
    9803   & R. Thompson & 12 & 18 & 6 & 6900 & 2003-08 & 2003-11\\
    9978   & S. Beckwith  & 12 & 68 & 4 & 4660-4860 & 2003-09 & 2004-01\\
    10086   & S. Beckwith & 12 & 3 & 4 & 4660 & 2003-12 & 2003-12\\
    10189   & A. Riess  & 13 & 23 & 3-4 & 1200-2000 & 2004-09 & 2005-01\\
    10258   & C. Kretchmer & 13 & 20 & 4 & 3034 & 2004-10 & 2006-09\\
    10340   & A. Riess  & 13 & 75 & 4 & 1440-1600 & 2004-07 & 2005-02\\
    11144   & R. Bouwens  & 16 & 1 & 4 & 2046 & 2009-10 & 2009-10 \\
    11563   & G. Illingworth & 17 & 31 & 4 & 5102-5332 & 2009-08 & 2010-08\\
    12060   & S. Faber  & 18 & 15 & 4 & 2046-2330 & 2011-03 & 2011-11\\
    12061   & S. Faber  & 18 & 11 & 5 & 1836-2086 & 2010-11 & 2011-06\\
    12062   &  S. Faber & 18 & 15 & 5 & 1886-1986 & 2011-07 & 2011-12\\
    12099   & A. Riess  & 18  & 4 & 4-5 & 1886-2070 & 2010-12 & 2011-08\\
    12461   & A. Riess  & 19 & 2 & 4 & 1943-1992 & 2012-02 & 2012-03\\
    12534   & H. Teplitz  & 19 & 1 & 4 & 5000 & 2012-05 & 2012-05\\
\hline
\end{tabular}
\begin{tablenotes}
\item \textbf{Note.} -- (a): Number of combined (level 2) images. (b): Number of single exposure images. (c): Starting and ending time of the observations. 
\end{tablenotes}
\end{threeparttable}\label{proposal}  
\end{table*}

We analyze all publicly available images of the GOODS-S region obtained with the Wide Field Channel of the HST Advanced Camera for Surveys \citep[ACS,][]{ford1998} in the F850LP filter (\textit{z}-band). 
The images are collected from the \textit{{\em Hubble} Legacy Archive (HLA) Data Release~10}\footnote{\url{http://hla.stsci.edu/}}. 
Each image corresponds to an individual HST visit and results from a combination of three or more individual exposures with the purpose of rejecting the cosmic rays. The observations were collected in the framework of the observing programs listed in Table~\ref{proposal}. We analyze totally 437 individual images spanning up to 10 years in some regions.

We used the code developed by M.~Tewes\footnote{\url{http://obswww.unige.ch/~tewes/cosmics_dot_py/}} which is based on P.~G.~van~Dokkum's \texttt{L.A.Cosmic} algorithm \citep{vandokkum2001} to further reduce the cosmic-ray contamination of the visit-combined images, especially on the edges of the combined frames. This algorithm is based on variations of the Laplacian edge detection and is capable of rejecting any cosmic ray, regardless of its shape and size, keeping at the same time the faint point-like sources untouched.

Source detection and photometry was performed using SExtractor \citep{1996A&AS..117..393B,bertin2010}. We applied the \textit{mexican hat} spatial filter for detection and set the minimum contrast parameter for deblending (\texttt{deblend\_mincont}) to 0.0075 in order to avoid multiple detections for individual extended sources. This did not affect the unresolved sources, since the GOODS-S field is not crowded. For the photometry and the variability analysis, we used a circular aperture with a radius of 0.36$^{\prime\prime}$. This was the radius used by \citet{vill2010} as for smaller radii the photometry is affected by changes in the point-spread function and aperture centring issues. In addition, we measured the magnitudes for two more radii (0.05$^{\prime\prime}$ and 0.15$^{\prime\prime}$), which correspond to the ones used for the {\em Hubble} Source Catalogue \citep[HSC,][]{whitmore2016} apertures (\texttt{MagAper1} and \texttt{MagAper2}). The latter were used to compute the concentration index (\texttt{CI}) and to validate our photometry against the HSC. \texttt{CI} is an indicator of the extension of a source and is described below.

After visually inspecting images associated with outlier points appearing in many light curves, we noticed that many outliers were situated near the frame edge or the gap between CCD chips in these images. This is related to the background estimation, which is essential for aperture photometry. Because the images have been resampled to a north-up east-left orientation, blank areas appear around the actual image (e.g. CCD gaps and image edges). SExtractor uses these blank areas and gets incorrect background estimates. To avoid this effect and, consequently, outliers and false-positive variable sources, we used the weight images provided by HLA and excluded all detections located within 10 pixels, or ${\sim}$0.5$^{\prime\prime}$, from the edges ($\sim$1\% of all the detections).

\begin{figure}
\begin{center}
\includegraphics[width=0.49\textwidth]{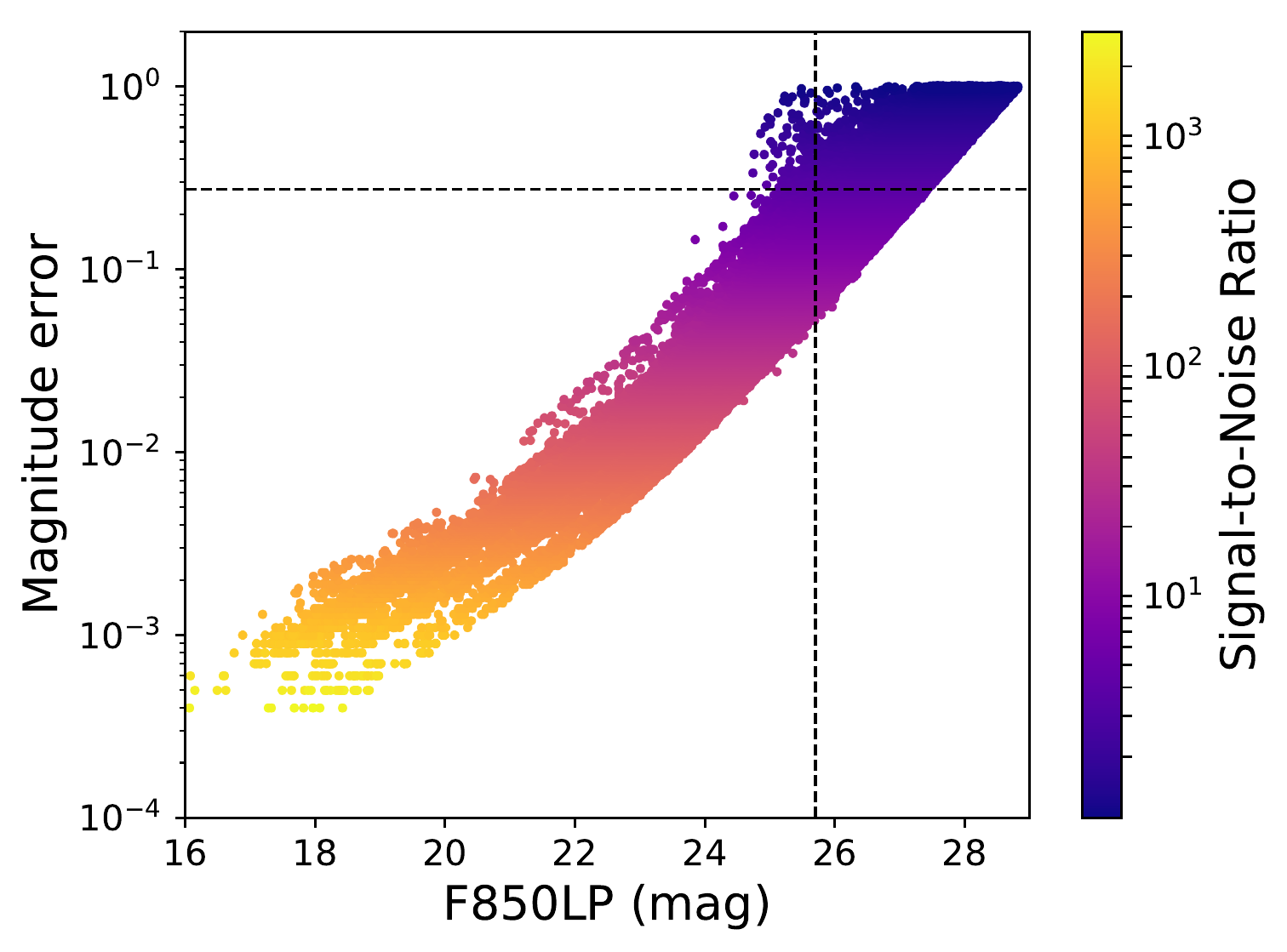}
\end{center}
\caption{The magnitude errors as a function of the magnitude for each detection. The measurements are colour coded by the SNR. The vertical and horizontal lines represent the limits in magnitude and magnitude error, respectively, after the SNR filtering.}\label{SNR}
\end{figure}

 \begin{figure}
\centering
\includegraphics[width=0.50\textwidth]{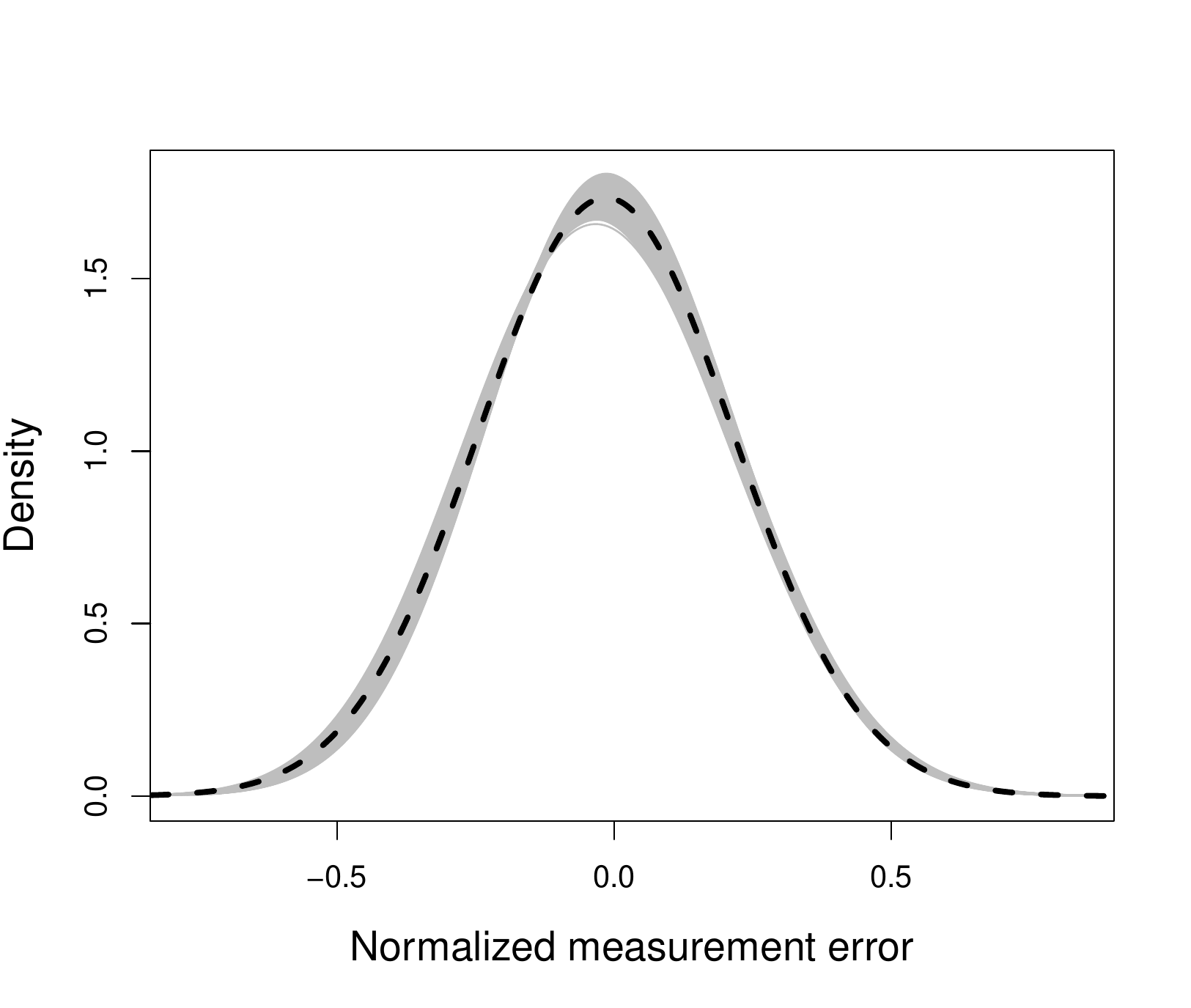} \\
\caption{The distributions of the normalized measurement errors for each magnitude bin (gray). The black dashed line represent the Gaussian fit to the data. The units are given in magnitudes.}
\label{densityerror}
\end{figure}

To ensure the quality of the data but also enable the search for small amplitude variations, we selected measurements with a signal-to-noise ratio (SNR) greater than five. The SNR was calculated for each of the detections for all images using the fluxes and the corresponding flux errors derived with SExtractor. In Figure~\ref{SNR}, we plot the magnitude error as a function of the magnitude, colour-coded with SNR. The relation between magnitude and magnitude error is approximately linear as expected if the errors are mainly statistical at the faint end. After applying the SNR cut-off, the faintest sources have an average error of about 0.25 mag.

We also divided all the detections into magnitude bins and over-plotted the distributions of the normalized magnitude errors (after subtracting the median) for each of the magnitude bins in order to check if all detections of the same magnitude have similar errors. 
Figure~\ref{densityerror} shows that these distributions are approximately Gaussian. The distribution would have been skewed, if there was a group of sources measured with systematically larger uncertainties (e.g. due to bright local background).

Since the astrometric accuracy of the HST is limited by the positional accuracy of individual Guide Star Catalogue stars \citep{lasker2008}, we applied a triangle matching technique based on the \citet{valdes1995} algorithm to find the astrometric solution. We used the \texttt{match\_v1} program by M.~W.~Richmond\footnote{\url{http://spiff.rit.edu/match/match-1.0/}} to automatically determine the coordinate system corrections using the 50 brightest sources in each source list. We used the second version of HSC (HSCv2) as the reference catalogue for the astrometry and the resulting positional errors are less than 0.1". 

\begin{figure}
\centering
\includegraphics[scale=0.50]{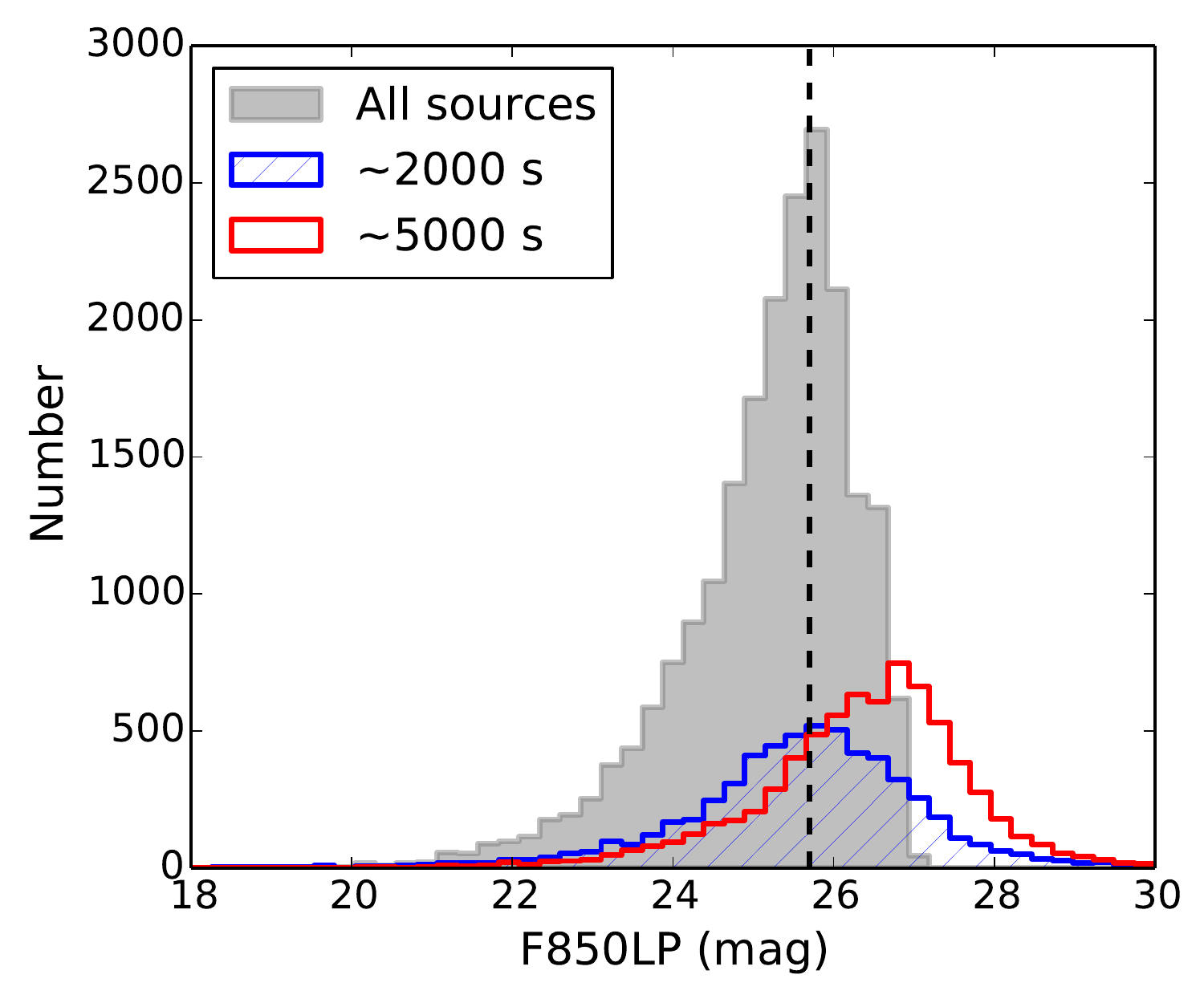} \\
\caption{The distribution of the median magnitude, <F850LP>, of all the sources (gray shaded histogram) along with the completeness curves for images with exposure times of ~2000 s (blue hatched) and ~5000 s (red), respectively. The dashed line indicates the completeness limit of our final sample.}
\label{maghist}
\end{figure}

\begin{figure}
\begin{center}
\includegraphics[width=0.47\textwidth]{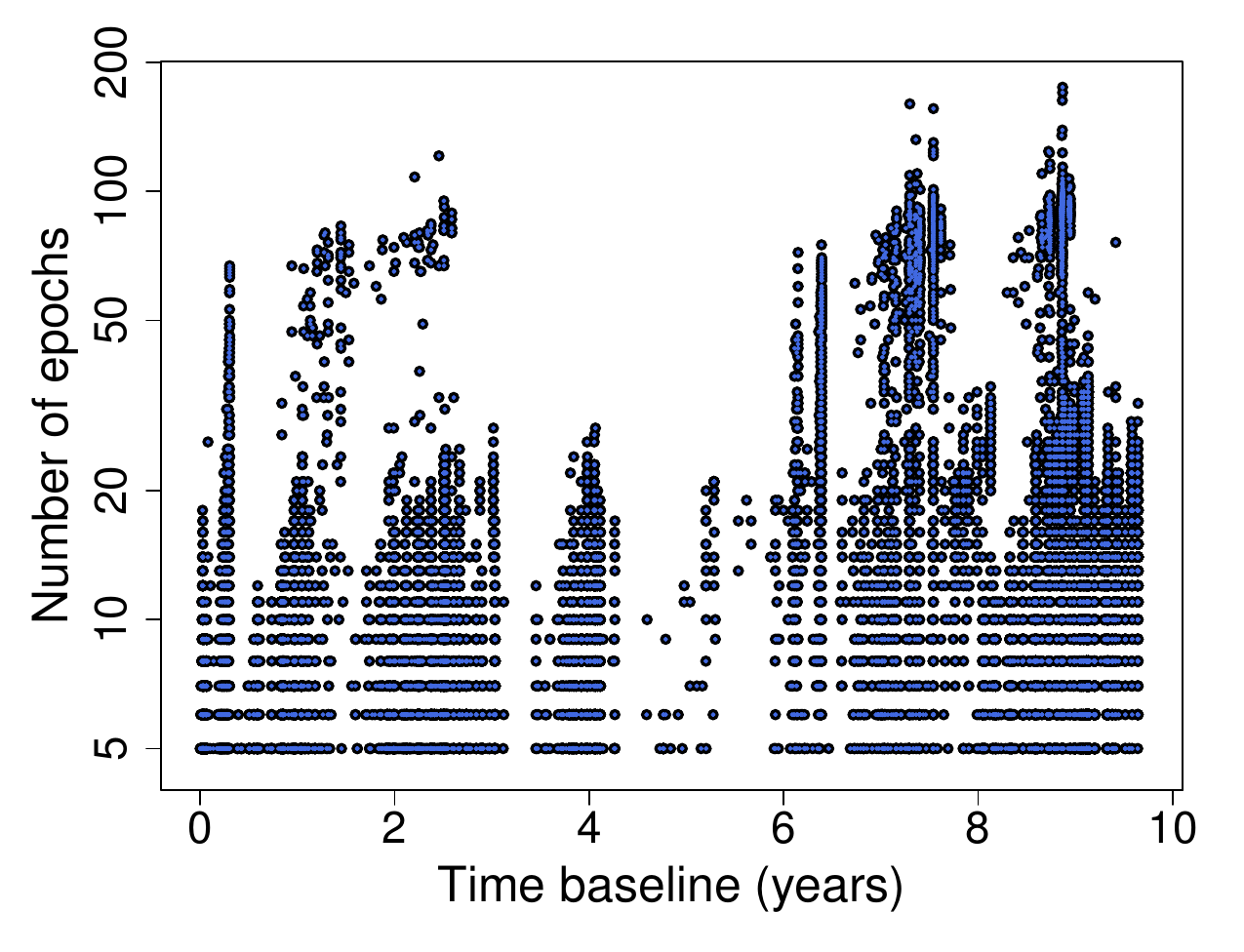}
\end{center}
\caption{The number of data points as a function of the maximum time baseline covered by the light curve.}\label{Ndistr}
\end{figure}

We then cross-matched the coordinates-corrected source lists with each other to construct a light curve for each source. We kept only sources with at least five measurements. Figure~\ref{maghist} presents the histogram of the median magnitude, <F850LP>, of all the sources after the SNR filtering. Due to the drop-off of detected sources beyond 25.7 mag, our sample is photometrically complete down to this magnitude. Since the images used to derive the source lists have different depth (Table~\ref{proposal}), we over-plot in Figure~\ref{maghist} the completeness curves of different images with extreme exposure times. We summed the detections of images with exposure times of $\sim$2000 s (Prop. ID: 12062) and $\sim$5000 s (Prop. ID: 11563), respectively. For both data sets, the number counts of detections decline at magnitudes fainter than the magnitude completeness limit of this work. Thus, the variable depth of the images did not affect our results.
The resulting catalogue consists of 21,647 sources. Figure~\ref{Ndistr} presents the distribution of the number of data points in the light curve, N\textsubscript{p}, as a function of the time baseline, T\textsubscript{bas}, which is defined as the time difference between the first and the last observation of a source. A large fraction of sources ($\sim80\%$) has been observed for more than two and up to ten years with a median time baseline of 8.5 years. The average and median number of data points in the light curves are 15 and 12, respectively.

\begin{figure}
\begin{center}
\includegraphics[width=0.47\textwidth]{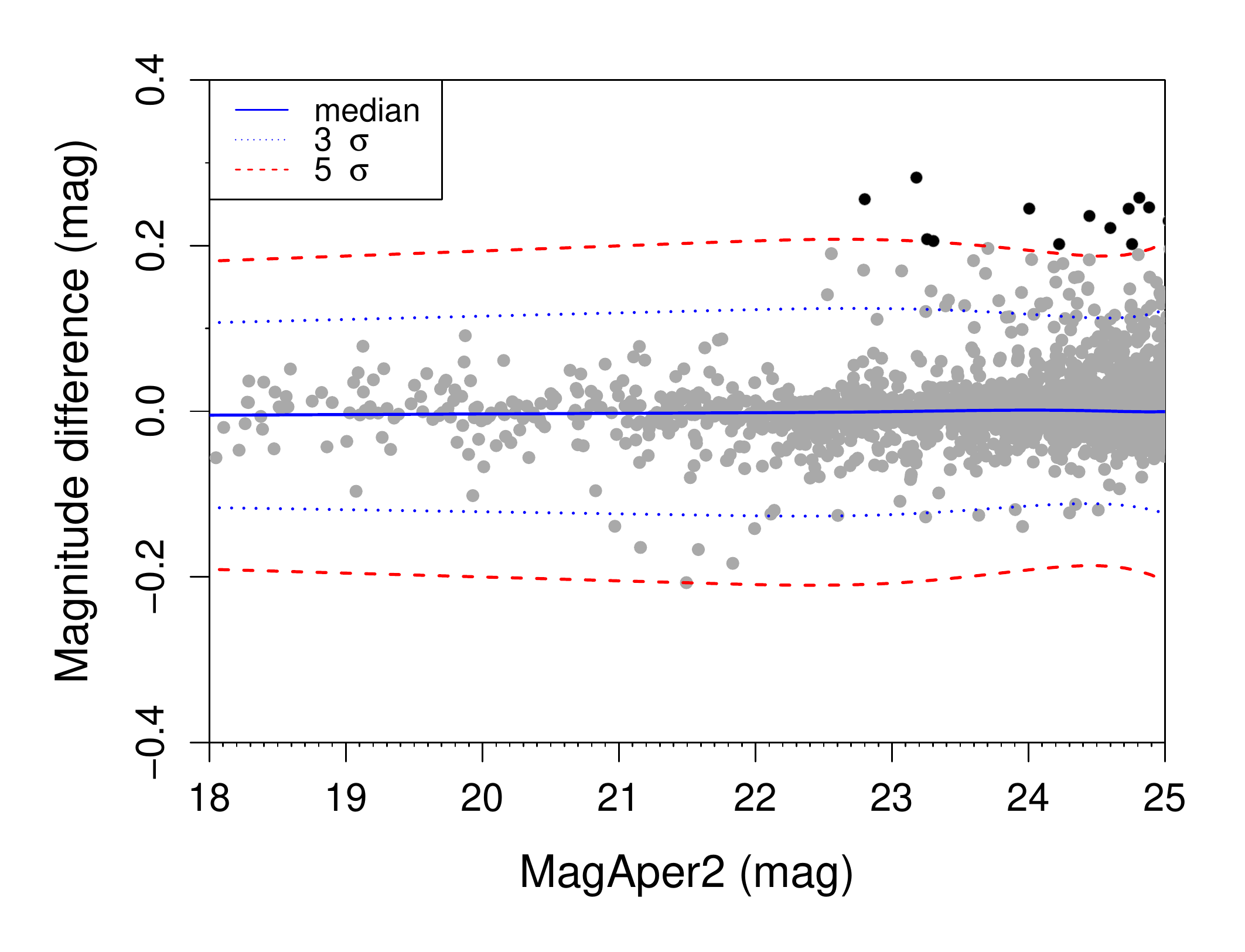}
\end{center}
\caption{A comparison between the <F850LP> magnitude of our photometry and HSCv2. The grey points represent all the sources in common. The blue solid line shows the median, while the blue dotted and red dashed lines represent the 3$\sigma$ and 5$\sigma$ values of the magnitude difference, respectively. Sources exceeding the 5$\sigma$ value are highlighted in black.}\label{diffMag}
\end{figure}

The photometric accuracy was tested by comparing our \texttt{MagAper2} magnitudes to those in HSCv2. First, we cross-matched all the sources with HSCv2 using a radius of 1$^{\prime\prime}$, resulting in 7245 matches out of 21,647 final sources (our source list is much deeper than HSCv2 in this region). In Figure~\ref{diffMag}, we plotted the difference in magnitude, <F850LP>\textsubscript{this work}--<F850LP>\textsubscript{HSCv2} as a function of magnitude between this study and HSCv2 and find the values to be comparable. The relation is linear through the full magnitude range as expected. We visually inspected the images and the light curves of the outliers (marked with black filled circles on Figure~\ref{diffMag}) and attributed the magnitude discrepancies to multiple detections of the same extended source in the HSCv2.

\begin{figure}
\centering
\includegraphics[scale=0.45]{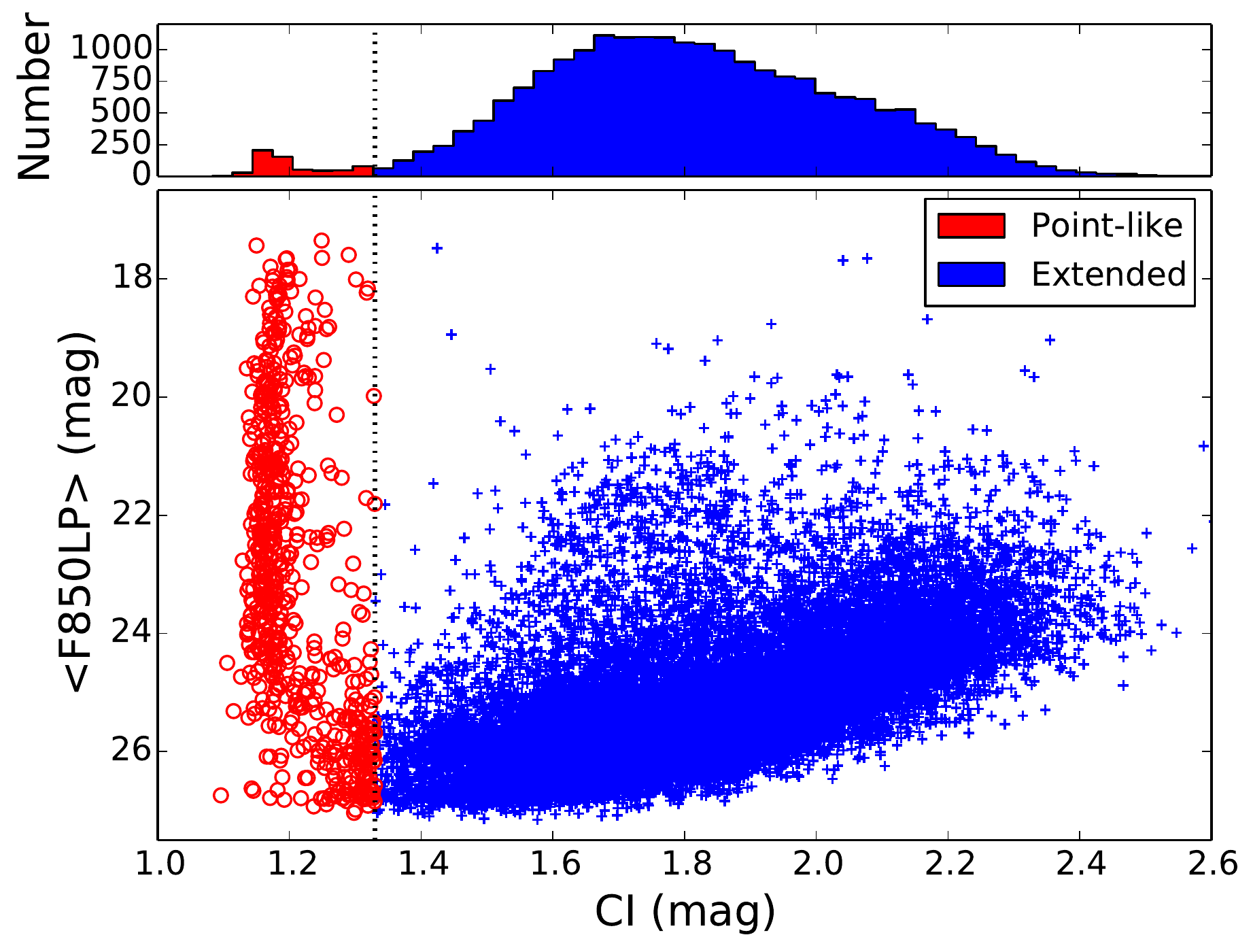} \\
\caption{The median magnitude, <F850LP>, as a function of the $CI$ for point-like (red circles) and extended sources (blue crosses). The dashed line represents the chosen threshold at $CI=1.33$ mag that separates the two populations. The upper panel shows the histogram of the $CI$.}
\label{CIall}
\end{figure}

The separation of the extended and point-like sources was performed using \texttt{CI} as defined in the HSC, \texttt{CI=MagAper1$-$MagAper2} \citep{whitmore2016}. The \texttt{CI} histogram reveals two well-defined areas (Figure~\ref{CIall}, top panel). We fitted two Gaussians to the two populations (point-like and extended) and the point where these two fits come across is at \texttt{CI}=1.33 mag. Adopting it as the separation threshold results in 21,022 extended and 625 point-like sources. Figure~\ref{CIall} (bottom panel) shows the \texttt{CI} as a function of magnitude, where the two populations are plotted using different colours.

\subsection{X-ray and IR data sets}\label{auxdata}

\begin{figure}
\includegraphics[width=0.47\textwidth]{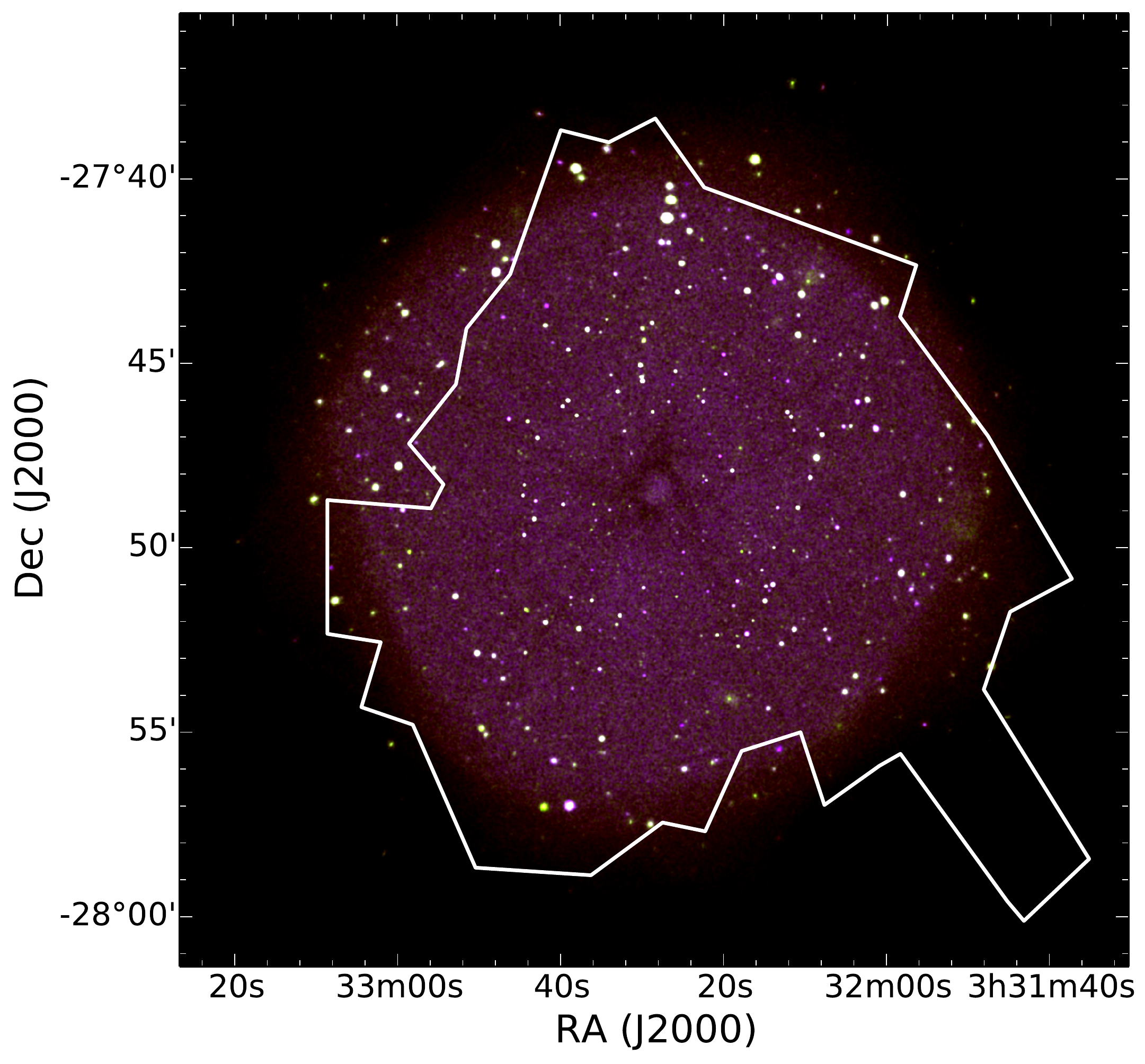}
\caption{Colour composite image of the exposure corrected and smoothed 7\,Ms CDF-S. The white polygon represents the GOODS-S footprint.}\label{xrayimage}
\end{figure}

GOODS-S is among the deepest and best studied fields in the sky and over the last decades there is a variety of imaging and spectroscopic data available from radio to X-ray wavelengths. In this study, apart from the HST optical observations, we utilize X-ray and IR catalogues and, also, photometric or spectroscopic data when available to validate the nature of the variable sources.

\begin{table*}
\caption{Summary of the basic information of the CDF-S X-ray catalogues.}
\centering
\scalebox{0.90}{
\begin{threeparttable}
\label{tableXrays}  
  \begin{tabular}{ccccccccc}
\hline
   X-ray & Number of & Observation  & \# sources   & F\textsubscript{0.5-8 keV} & F\textsubscript{0.5-2 keV} & F\textsubscript{2-8 keV} & \# matches & \# matches \\ 
catalogue & observations & dates & CDF-S  & (ergs cm\textsuperscript{-2} s\textsuperscript{-1}) & (ergs cm\textsuperscript{-2} s\textsuperscript{-1}) & (ergs cm\textsuperscript{-2} s\textsuperscript{-1}) & optical & variable \\\hline \hline
    250\,ks ECDF-S & 9 & 2004 &  430  & $>3.5\times10^{-16}$ & $>1.1\times10^{-16}$ & $>6.7\times10^{-16}$ & 144 & 14\\ 
    2\,Ms CDF-S & 23 & 1999 - 2000 &   578 & $>7.1\times10^{-17}$ & $>1.9\times10^{-17}$ & $>1.3\times10^{-16}$ & 298 & 16\\ 
    4\,Ms CDF-S & 54 &  1999 - 2010 & 776 & $>3.2\times10^{-17}$ & $>9.1\times10^{-18}$ & $>5.5\times10^{-17}$ & 464 & 21\\
    7\,Ms CDF-S & 102 &  1999 - 2016 &  1055 & $>1.9\times10^{-17}$ & $>6.4\times10^{-18}$ & $>2.7\times10^{-17}$ & 621 & 24\\
\hline
\end{tabular}
\begin{tablenotes}
\item \textbf{Note.} -- The flux limits of the 7\,Ms CDF-S catalogue in the broad and the hard band are derived up to 7 keV.
\end{tablenotes}
\end{threeparttable}}
\end{table*}

We use four X-ray catalogues from CDF-S and the Extended Chandra Deep field South (ECDF-S) with different depths: 250\,ks \citep{xue2016}, 2\,Ms \citep{luo2008}, 4\,Ms \citep{xue2011} \& 7\,Ms \citep{luo2017}. The area studied in this work partly overlaps with CDF-S, which is centered at $\alpha=$3h32m28.06s and $\delta=-27^{\circ}$48$^{\prime}$26.4$^{\prime\prime}$ (J2000) and covers an area of $\sim$464.5 arcmin$^2$. ECDF-S covers an area of $0.54$ deg$^2$ in the sky centered at $\alpha=$3h32m24.0s  $\delta=-27^{\circ}$48$^{\prime}$47.0$^{\prime\prime}$ (J2000). Figure~\ref{xrayimage} shows CDF-S and GOODS-S projected on the sky. The catalogues are produced from the X-ray images taken by the Advanced CCD Imaging Spectrometer camera \citep[ACIS,][]{garmire2003} aboard the {\em Chandra} \textit{X-Ray Observatory}. We also used these four catalogues to test the dependence of the number of optical counterparts on the depth of the X-ray image.

Table~\ref{tableXrays} presents the number of sources detected in at least one band (the supplementary sources -- lower significance X-ray sources with bright IR counterparts --  of each catalogue are also included), the number of observations and the observation dates with the on-axis sensitivity limits in the soft (0.5$-$2 keV), hard (2$-$8 keV) and broad (0.5$-$8 keV) bands of the catalogues. The 250\,ks catalogue is from ECDF-S, which covers a much larger region, but here we only present the sources that lie within CDF-S. The 7\,Ms CDF-S catalogue extends up to 7 keV to avoid the background noise present at higher energies. The full list of the observations of each catalogue and the detailed description of the data reduction can be found in the aforementioned papers and the references therein. All published catalogues include additional information on the sources, such as X-ray properties, multi-wavelength counterparts and redshifts.

For the variable optical sources identified in Sec.~\ref{variability} that have no X-ray counterparts, we independently reduced the {\em Chandra} images using the \texttt{CIAO} software v2.0.1\footnote{\url{http://cxc.harvard.edu/ciao}} to construct the 7\,Ms image and calculated the X-ray flux upper-limits. First, we created the 7\,Ms co-added images in the three bands -- broad, soft and hard, starting with the level 2 event files. We used 99 observations taken from October 1999 to March 2016. We kept only the central CCD chips (\texttt{ccd\_id}=0,1,2,3) and filtered out flares affecting the background in the light curve of each observation by masking the sources and using the \texttt{deflare} tool with the \texttt{clean\_lc} option. For each observation, we ran the \texttt{wavedetect} tool to create source catalogues, so we could reproject the images at the same reference point in the sky to achieve a good absolute astrometric solution. The final step was to combine the event files with the \texttt{dmmerge} tool and create images and exposure maps for all the bands. 

The smoothed image including all three bands is shown in Figure~\ref{xrayimage}. For the non-detected sources in the X-rays, we measured the counts and the exposure effective areas in a circular region centered on the position of the optical counterparts. The radius, $r_i$, used for each optical source was calculated in a way that to enclose a specified fraction of the point spread function. The fraction adopted here decreases from the on-axis (0.95) to off-axis sources (0.5). The background was extracted from 500 circular regions (with aperture half of $r_i$) at random positions around the optical source (within a distance from 1.5 to 5 times of $r_i$) that do not overlap with other X-ray or the optical variable sources. We derived the count rate for each background region and finally normalized the mean value of all of them to the area of the source. Then we derived the upper limits with a confidence interval of 99.7\%. The count rates were converted to fluxes, using an energy conversion factor equal to $2.8\times10^{-9}$, $1.5\times10^{-9}$ and $6.2\times10^{-9}$ ergs photon\textsuperscript{-1} for the broad, soft and hard band, respectively by assuming a power-law model with photon index of $\Gamma=1.7$.

GOODS-S overlaps, also, with the Spitzer IRAC/MUSYC Public Legacy Survey in the Extended Chandra Deep Field South (PI: Pieter van Dokkum, SIMPLE). SIMPLE covers an area of $\sim$ 1,600 arcmin\textsuperscript{2} surrounding GOODS-S and contains photometry for $\sim$45,000 sources from deep Spitzer/IRAC \citep{fazio2004} observations combined with other UV to mid-IR data from the Multiwavelength Survey by Yale-Chile (MUSYC). In this work, we use the four IRAC bands (3.6 $\mu m$, 4.5 $\mu m$, 5.8 $\mu m$ and 8.0 $\mu m$) to construct the IR selected AGN sample (Section~\ref{IR}), while the r, J, K and 3.6 $\mu m$ bands from the same catalogue were used to separate the stellar from the extra-galactic objects in Section~\ref{stars}. The full description of these data can be found in \citet{damen2011}.

\section{AGN selection based on optical variability}\label{variability}

\begin{figure*}
\begin{center}
\begin{tabular}{  c  c }
\includegraphics[trim=1cm 0.04cm 0.cm 0cm, width=0.47\textwidth]{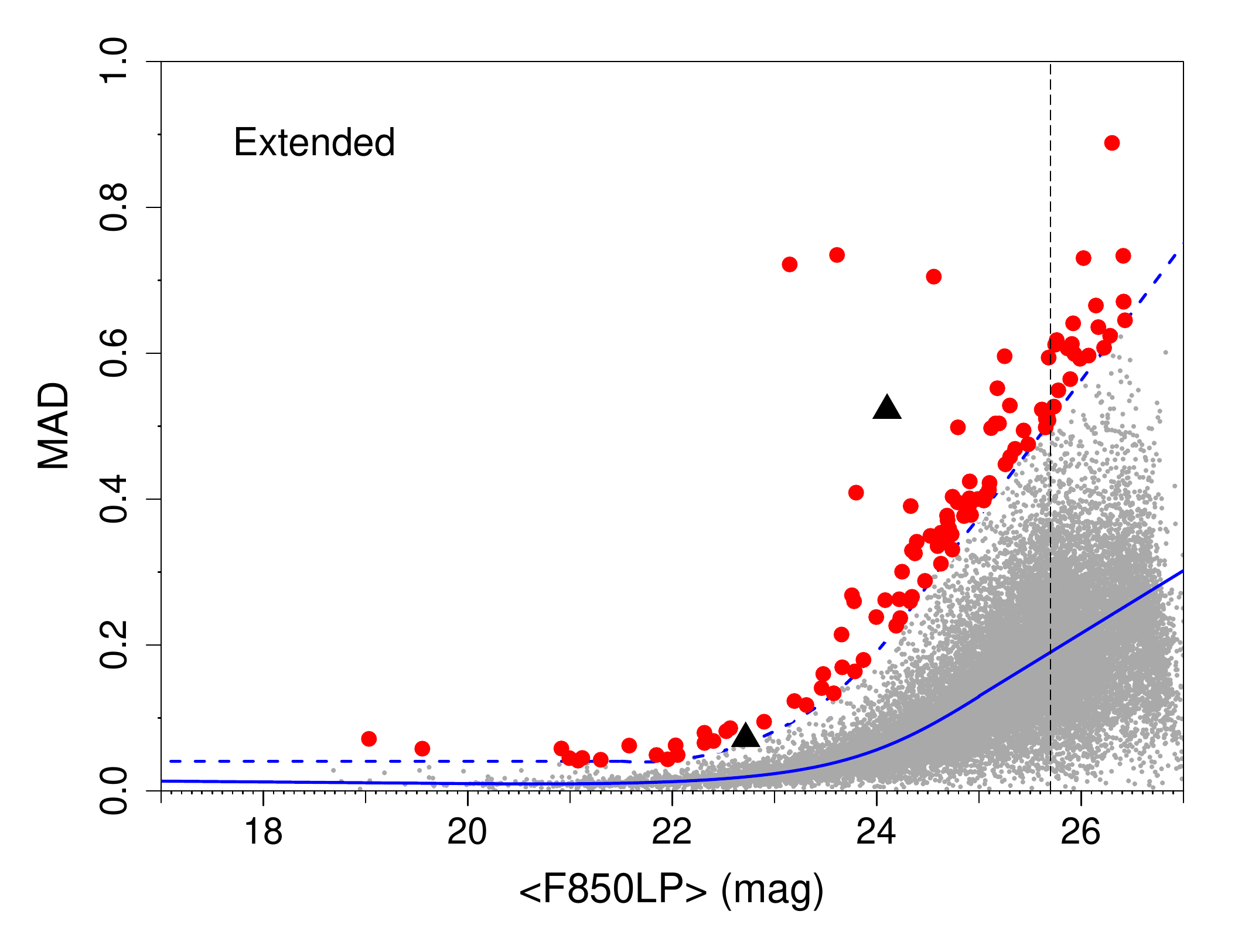} &
\includegraphics[trim=1cm 0.04cm 0.cm 0cm, width=0.47\textwidth]{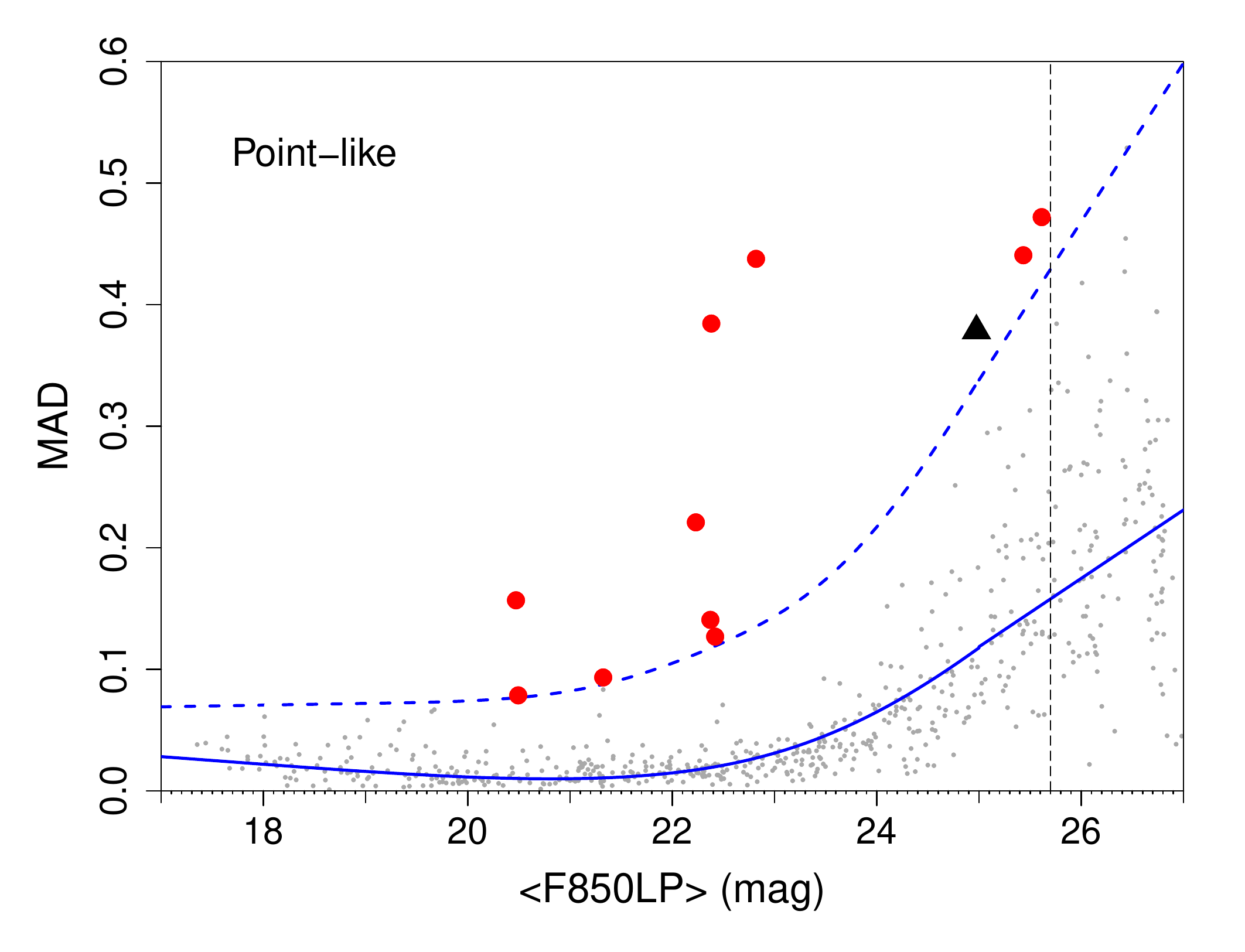} \\
\includegraphics[trim=1cm 0cm 0cm 0.9cm, width=0.47\textwidth]{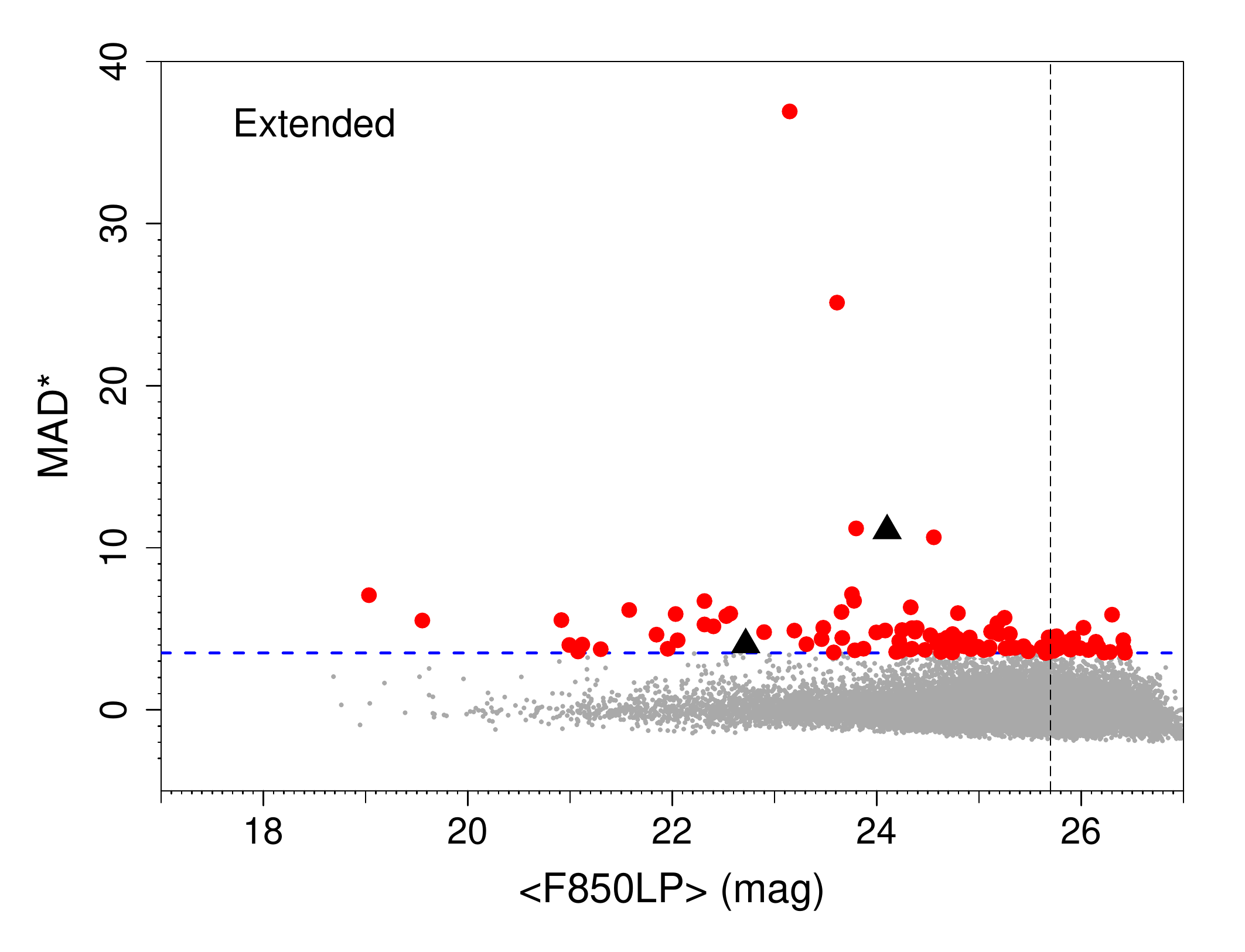} &
\includegraphics[trim=1cm 0cm 0cm 0.9cm, width=0.47\textwidth]{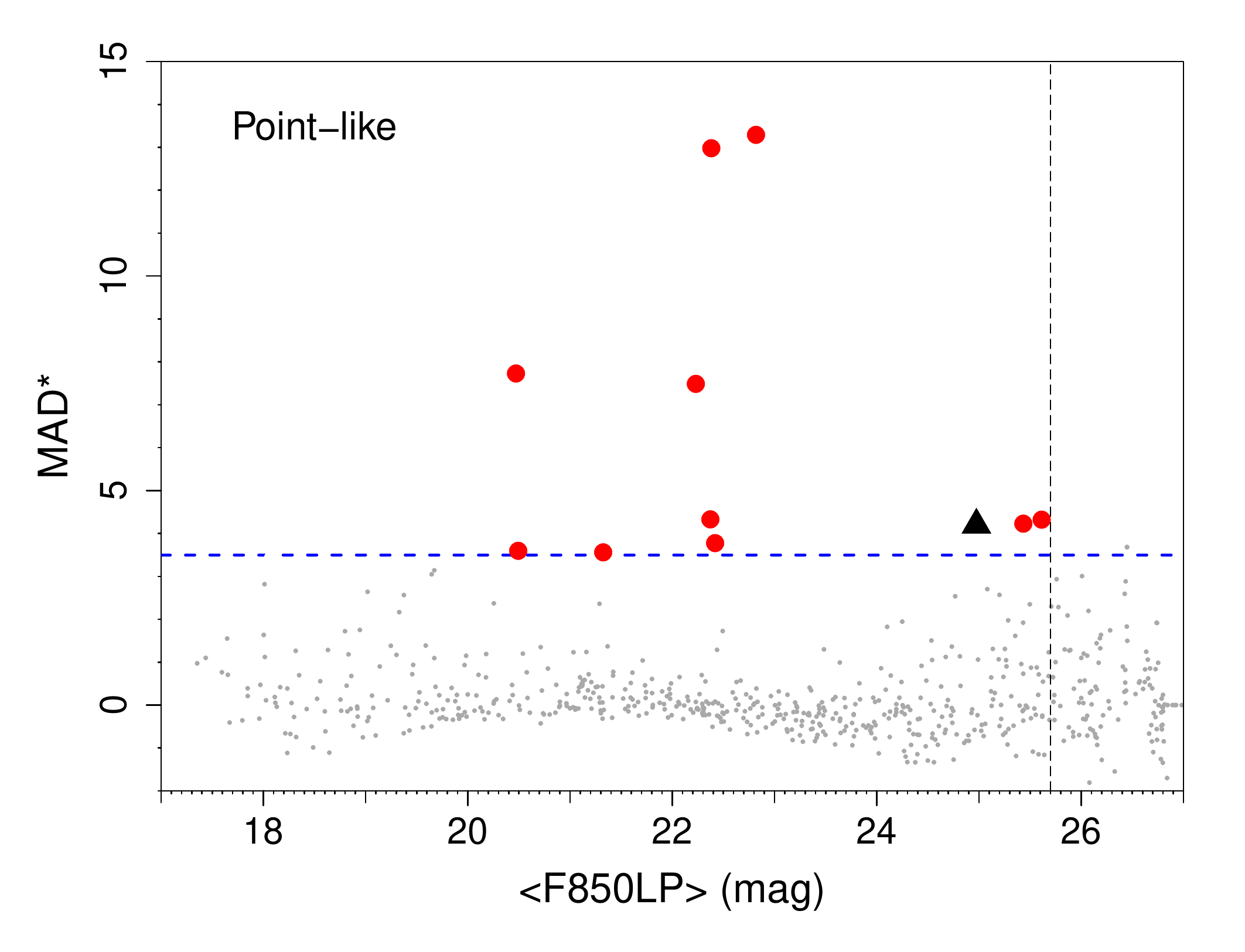}
\end{tabular}
\end{center}
\caption{The ${\rm MAD}$ (upper panels) and normalized significance ${\rm MAD}$* (lower panels) as a function of the median magnitude (<F850LP>) for the extended (left) and point-like sources (right). All the sources in our survey are shown with grey points, while the AGN candidates are shown with red circles, and the confirmed  SNe with black triangles. The blue solid and dashed lines represent the median and the threshold, while the vertical black dashed line the completeness limit of our sample.}\label{VIsign}
\end{figure*}

\subsection{Variability method}\label{vi}

\citet{sokol2017b} discussed two classes of statistical methods that quantify  variability of a source. The first class quantifies the scatter of the magnitudes within a light curve, while the methods of the second class quantify the smoothness of a light curve by taking into account the order and time at which the magnitude measurements were obtained. Regular variability can be detected that way too, if the observing cadence is shorter than the variability timescale \citep{ferreira2016}, or, if the scatter is higher than what is expected from noise \citep{ferreira2017}.

Light curve simulations by \citet{sokol2017b} suggest that the scatter-based methods are more suitable for detection of variability in light curves having a small number of points compared to the methods that characterize the light curve smoothness. Median Absolute Deviation\footnote{\url{https://en.wikipedia.org/wiki/Median_absolute_deviation}} \citep[${\rm MAD}$]{rousseeuw1993} belongs to the first class of methods. It is the most robust to outliers among the variability indices discussed by \citet{sokol2017b}. In Appendix~\ref{simulationssection}, we compare the performance of various variability-detection statistics in the presence of photometric outliers and find that MAD is a reliable method, resistant to individual outlier measurements.

${\rm MAD}$ is defined as the median value of the absolute deviations of the measurements, $m_{j}$, from the median:
\begin{equation}\label{MADformula}
{\rm MAD} = b\times \median(|m_{j}-\median(m_{j})|),
\end{equation}
where $b=1/(\sqrt{2}\erf^{-1}(1/2))\simeq1.4826$ is the factor scaling the median absolute deviation to the standard deviation (assuming the normal distribution of $m_{j}$); $\erf^{-1}$ is the inverse error function.

Specifically, our variability detection algorithm works as follows: we divided the sources into magnitude bins and by assuming a Gaussian distribution, we calculated for each bin the median magnitude, the median MAD and the standard deviation ($\sigma$) of MAD. The bin size is adjusted so as to have at least 50 sources in each bin. To get a smooth magnitude dependence, we fitted a cubic spline to the median and the threshold values. We, also, extrapolated toward fainter magnitudes to account for the completeness limit (Section~\ref{optical}). Taking into account that the majority of the sources are normal galaxies and no variations are expected, the variable sources are those that exceed a cut-off above the median. We note that \cite{sara2011} rely on the same critical assumptions as we do here: that the majority of sources are non-variable and that sources of similar brightness have similar photometric errors. \cite{vill2010} also rely on this assumption indirectly when they derive the scaling factors for the estimated photometric errors that they use to compute the $C$-statistic.

Following \citet{bershady1998}, or more recently \citet{sara2011}, we determine the threshold separately for the point-like and the extended sources (except we determine the threshold in MAD scaled to $\sigma$ rather than in $\sigma$ as \citealt{sara2011}). Figure~\ref{VIsign} (upper panels) shows the variability index, ${\rm MAD}$, as a function of the median magnitude, <F850LP> for the extended (left) and point-like (right) sources. We also calculated the normalized significance, ${\rm MAD}^{*}$, for each source through the following formula:
\begin{equation}\label{significanceformula}
{\rm MAD}^{*}_{i}=\frac{{\rm MAD}_{i}-\median({\rm MAD})_{b}}{\sigma({\rm MAD})_{b}},
\end{equation}
\noindent where ${\rm MAD}_{i}$ is the MAD for the $i^{th}$ source and b the corresponding magnitude bin. The significance has units of $\sigma$. The plots of ${\rm MAD}$* as a function of magnitude are shown in the lower panels of Figure~\ref{VIsign}.

 We set the threshold of $3.5\sigma$ in MAD* above which we consider the sources to be variable. Assuming the normal distribution of MAD* we estimate the fraction of sources that are expected to have the value of MAD* above the threshold, i.e. the false positive rate, as:  
\begin{equation}
{\rm FP}_{\rm rate} = 1 - \frac{1}{2}(1+\erf(\frac{3.5}{\sqrt{2}})) \simeq 2.3 \times 10^{-4},
\end{equation}
so out of 21,022 extended and 625 point-like sources of the initial sample we expect 5 and $<1$ false positives among the extended and point-like candidate variable sources, respectively. This corresponds to $\sim$3.2\% of the total number of 187 variable candidates. The false positives are expected to have magnitudes near the completeness limit where the majority of sources are found (Fig.~\ref{maghist}). Furthermore, as our sample is not homogeneous concerning the number of data points in their light curves, we simulated under-sampled data to check the dependence of the adopted variability threshold on the number of data points (Appendix~\ref{simulationsN}). We found that the thresholds derived from the simulated data were either very close to, or below the adopted variability threshold, indicating the goodness of our threshold.

\begin{figure}
\includegraphics[width=0.50\textwidth]{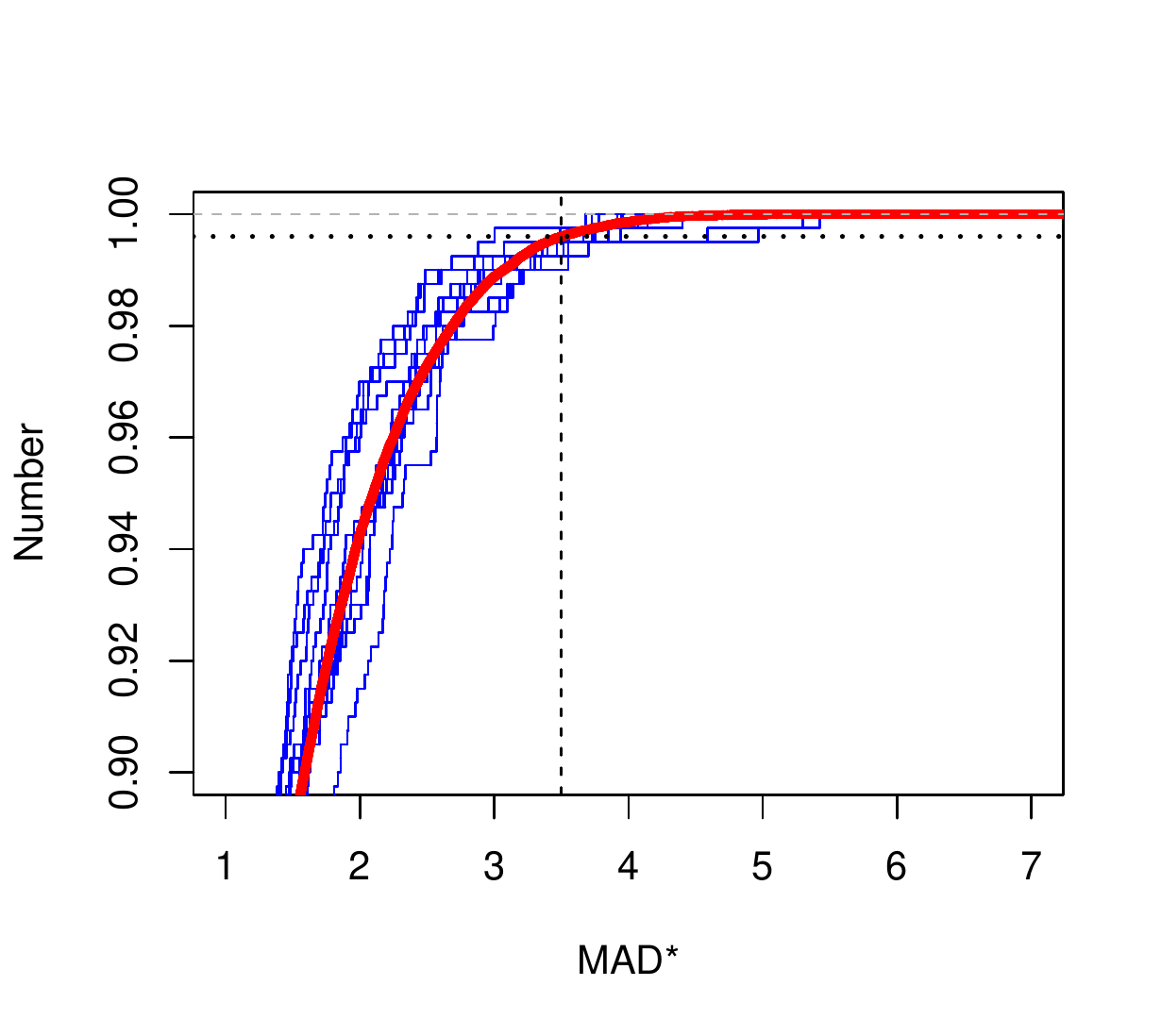}
\caption{The median cumulative distributions of the normalized significance for under-sampled data with different number of data points in the light-curve (blue thin lines). The red thick line represents the distribution from our final sample, while the horizontal dashed and dotted line represents the statistical significance of 100\% and 99.65\%, respectively. The vertical line is our 3.5 $\sigma$ threshold.}\label{cumul1}
\end{figure}

Moreover, to further explore the genuine statistical significance of our threshold, we calculated the cumulative distributions of the normalized MAD for several magnitude bins with different number of data points (N=5, 6, 7, 8, 9, 10, 20, 30 \& 40). In Figure~\ref{cumul1}, we show the median cumulative distributions for different number of data points as long as with the real distribution of our sample. The average statistical significance of the 3.5$\sigma$ thresholds is 99.65\%. However, the true significance should be greater than this value, since the distribution also includes the variable sources.
We visually inspected the light curves of all the candidate variable sources and the associated images. We checked for diffraction spikes from nearby foreground stars, close neighbors, poorly removed cosmic rays, proximity to a frame edge and saturation or misalignment of individual exposures. All these factors may introduce false variability. Following this procedure, we classified all the variable sources into three categories. Sources with clear variability in their light curve, far from other sources with no artifacts or potential problems recorded and with accumulated significance higher than 99.9\%, were assigned grade A (86 sources). Sources with minor problems that may affect the reliability were assigned a grade B (32 sources). This category includes sources that might have centring issues, caused by the extension of the object or that are too faint and dispersed and sources between 3.5 sigma and accumulated significance of 99.9\%. Finally, all sources that were found to be affected by saturation, diffraction spikes, blending or other significant problems were  assigned a grade C (69 sources).

\subsection{Stars and supernovae}\label{stars}

\begin{figure*}
\begin{center}
\begin{tabular}{ c  c }
\includegraphics[trim=0cm 0cm 0.cm 0cm, width=0.51\textwidth]{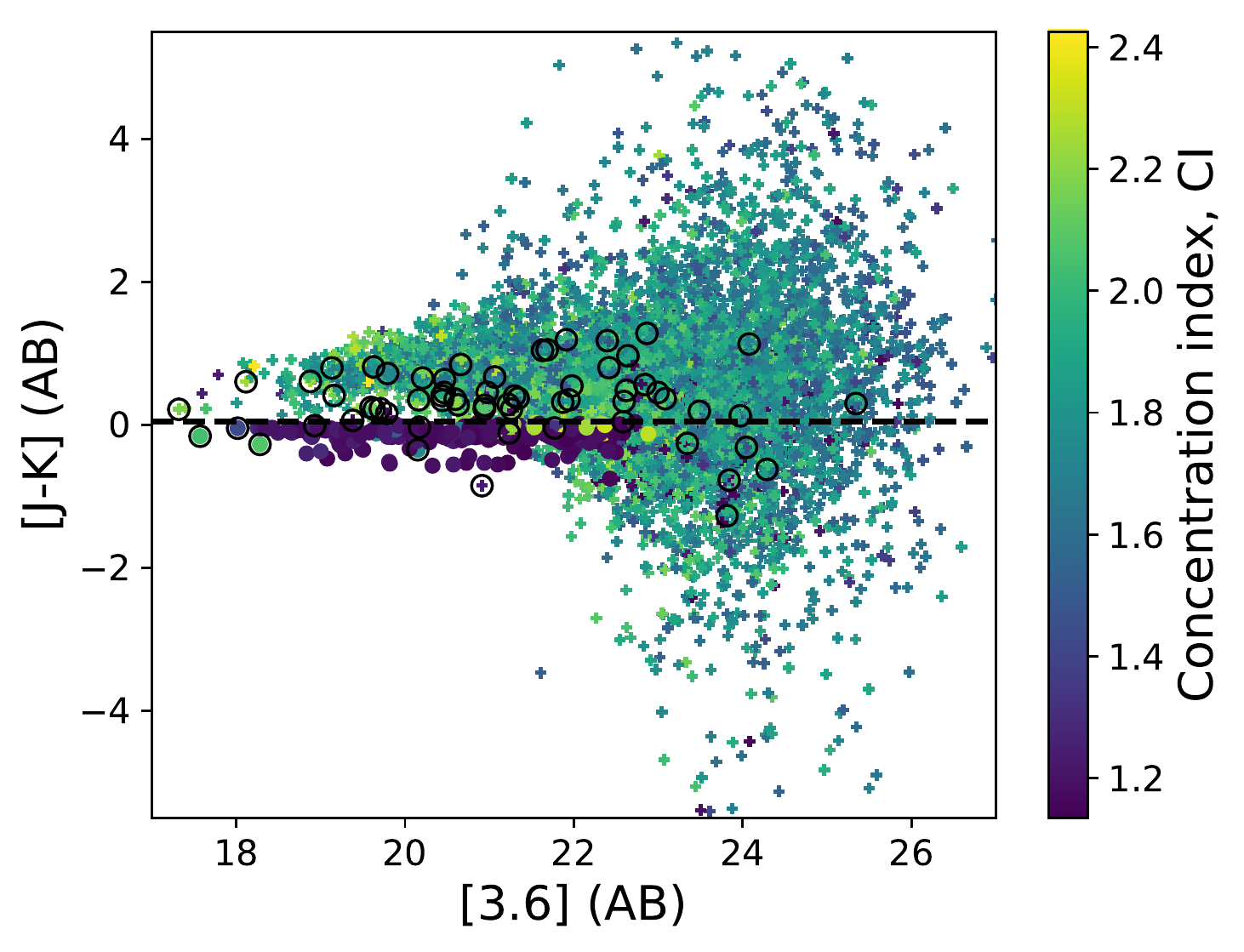}
\includegraphics[trim=0cm 0cm 0.cm 0cm, width=0.52\textwidth]{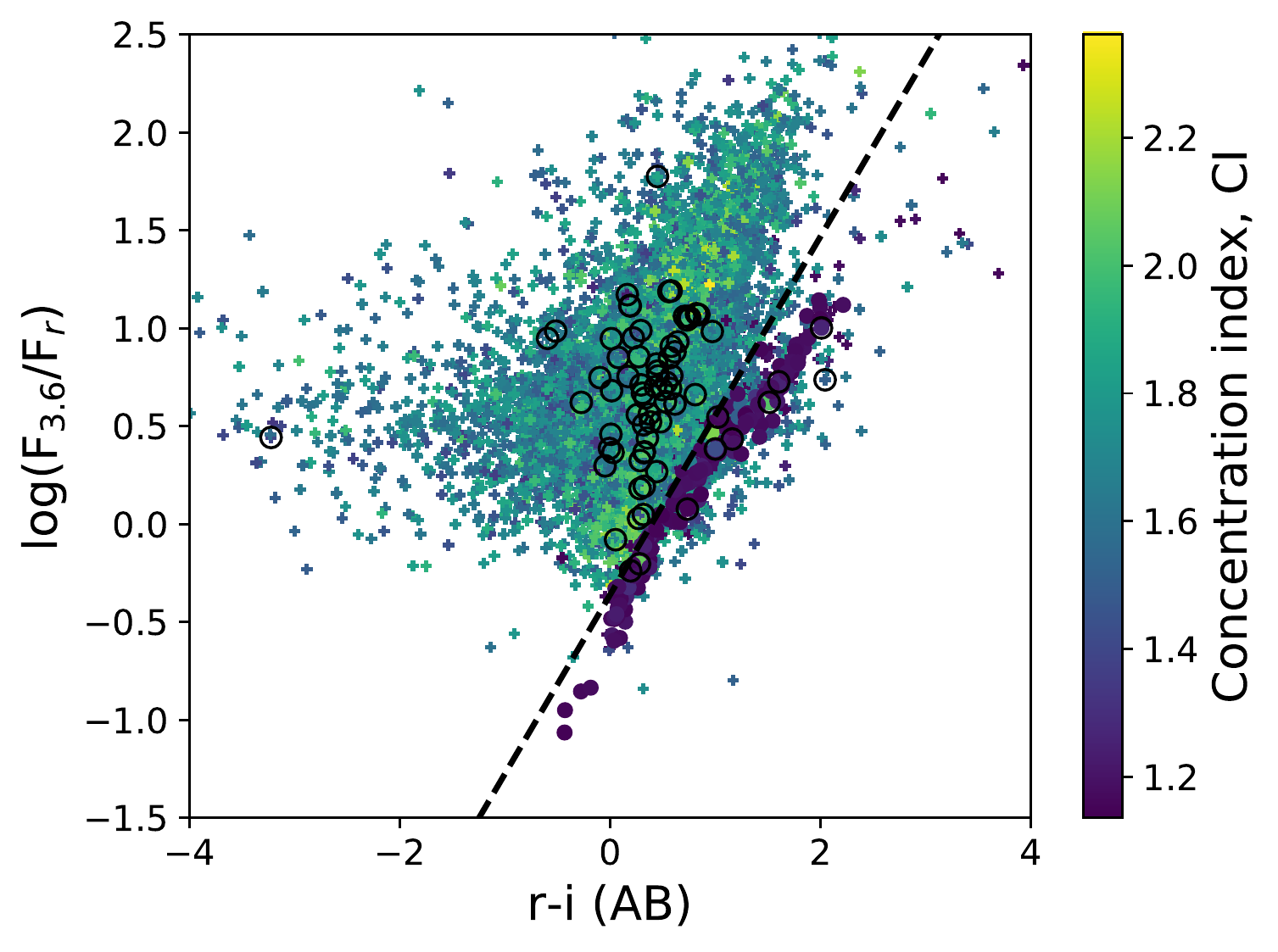} 
\end{tabular}
\end{center}
\caption{The $[J-K]~versus~3.6\mu m$ diagram (left) and the $3.6\mu m/r$ flux ratio versus $r-i$ diagram (right). The crosses represent the common sources between our sample and the SIMPLE data. The filled circles are the sources classified as stars. All sources are colour coded by the $CI$. The open black circles indicate the variable sources.}\label{colorcolor}
\end{figure*}

In order to separate AGN candidates from stars, we followed \citet{rowan2005} and \citet{damen2011}. According to \citet{rowan2005}, stars can be distinguished from the extra-galactic objects, such as QSOs and very distant AGNs by their brightness in the r band (r<23 mag) and their position in the $3.6\mu m/r$ flux ratio versus the $r-i$ diagram (Figure~\ref{colorcolor}, right). The two populations occupy different regions in the diagram and can be easily separated. This method was also used by \citet{falocco2015} and \citet{rowan2013}. On the other hand, \citet{damen2011} excluded the stellar population using a colour cut-off ($[J-K](AB)<0.04 mag$) and applied certain quality criteria to their initial sample: signal-to-noise ratio in K band $(S/N)_K>5$ and their relative weight in the K band versus the z- band, $wK>0.5$.

We cross matched our initial catalogue of 21,647 sources with the SIMPLE data (Sec.~\ref{auxdata}) to obtain the colours for our sources and applied the above diagnostics. Figure~\ref{colorcolor} shows the $[J-K]$ vs. $3.6\mu m$ and the $3.6\mu m/r$ vs. $[r-i]$ diagrams with the sources colour coded by the \texttt{CI}. The open black circles presented in the plot are used to show the variable sources in each diagram. Both diagnostics indicated eight classified variable sources to be stars. All of them are grade C variable sources, as six were saturated and two were blended sources. We therefore identify no high-confidence candidate variables among the foreground stars.

For the identification of SNe, we relied on visual inspection of the light curves and the corresponding images. We found three sources with light curves of Grade A resembling SNe (ID~7343, 9581 \& 14446), which have all been previously reported in the literature by \citet{strolger2004} and \citet{riess2007}. Their light curves and observational properties can be found in Figure~\ref{lc2} and Table~\ref{SN} in the Appendix~\ref{AppendixSNe}. Out of the SNe catalogues, there are 13 more SNe that have a counterpart in our initial sample, but they are all below the variability threshold. Possible explanations could be the differences in the observation dates (the peak was not observed), or that our variability algorithm could not detect variability if the number of the data points in the light curve that correspond to the peak was small and they were considered as outliers by MAD. For these sources, we also calculated the standard deviation, but they were still below the threshold, thus the first case is more likely to happen.

\begin{figure}
\includegraphics[width=0.50\textwidth]{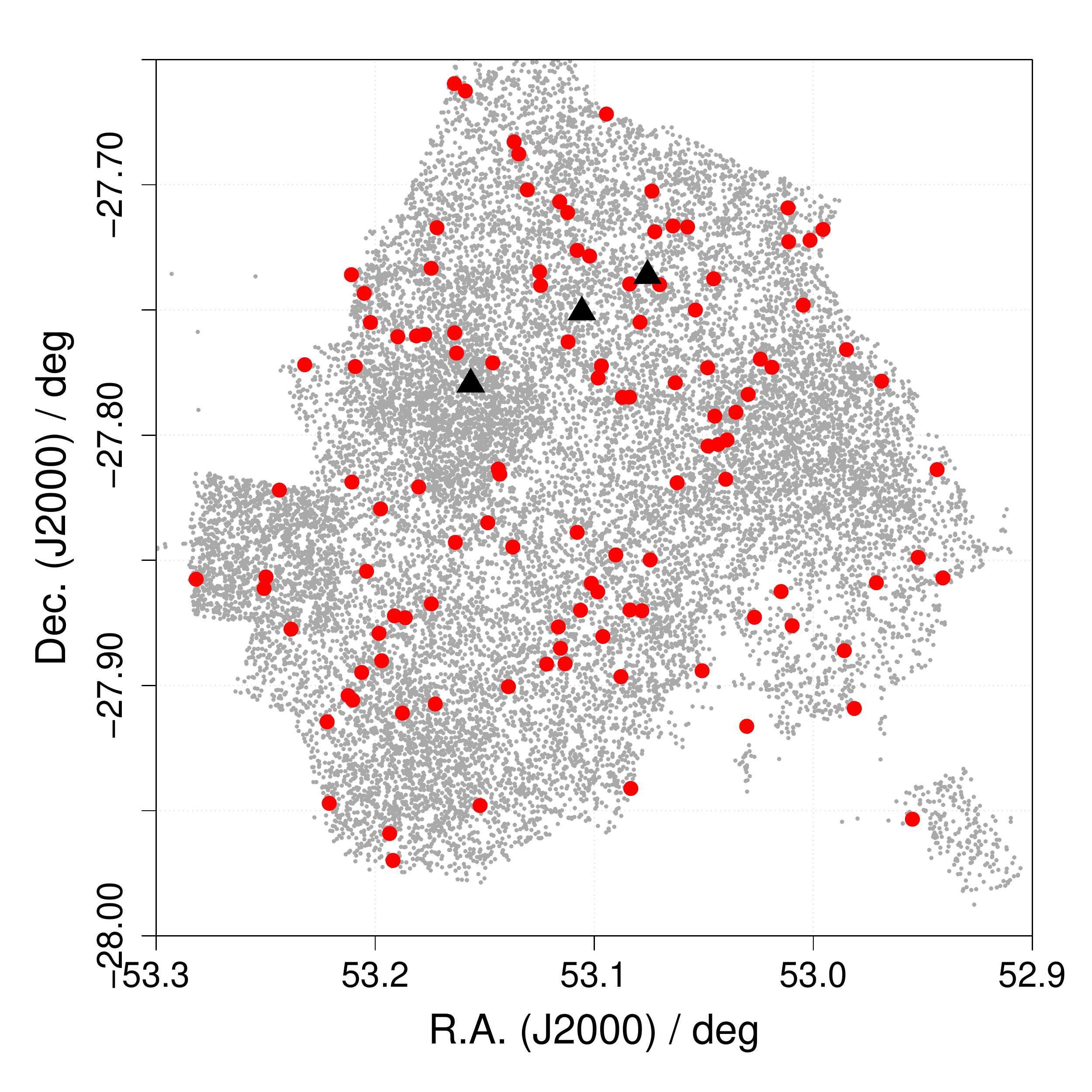}
\caption{Spatial distribution of the AGN candidates (red circles). The confirmed SNe are shown with black triangles. The background grey points represent the whole sample consisting of 21,647 sources.}\label{finderchart}
\end{figure}

\begin{figure*}
\centering
\begin{tabular}{  l l }
\includegraphics[trim=0cm 0cm 0.cm 0cm,width=0.50\textwidth]{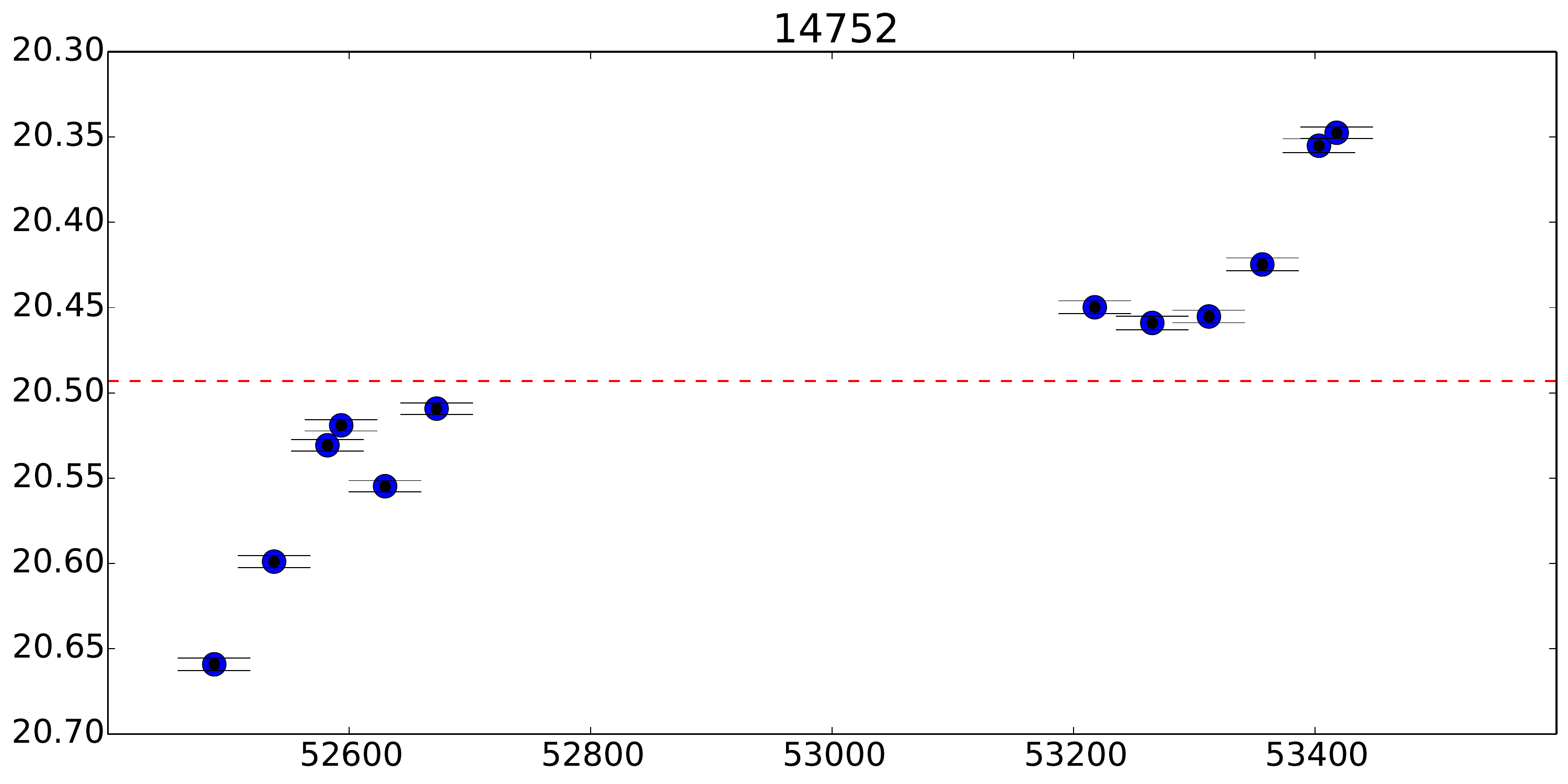} &
\includegraphics[trim=0cm 0cm 0.cm 0cm,width=0.5\textwidth]{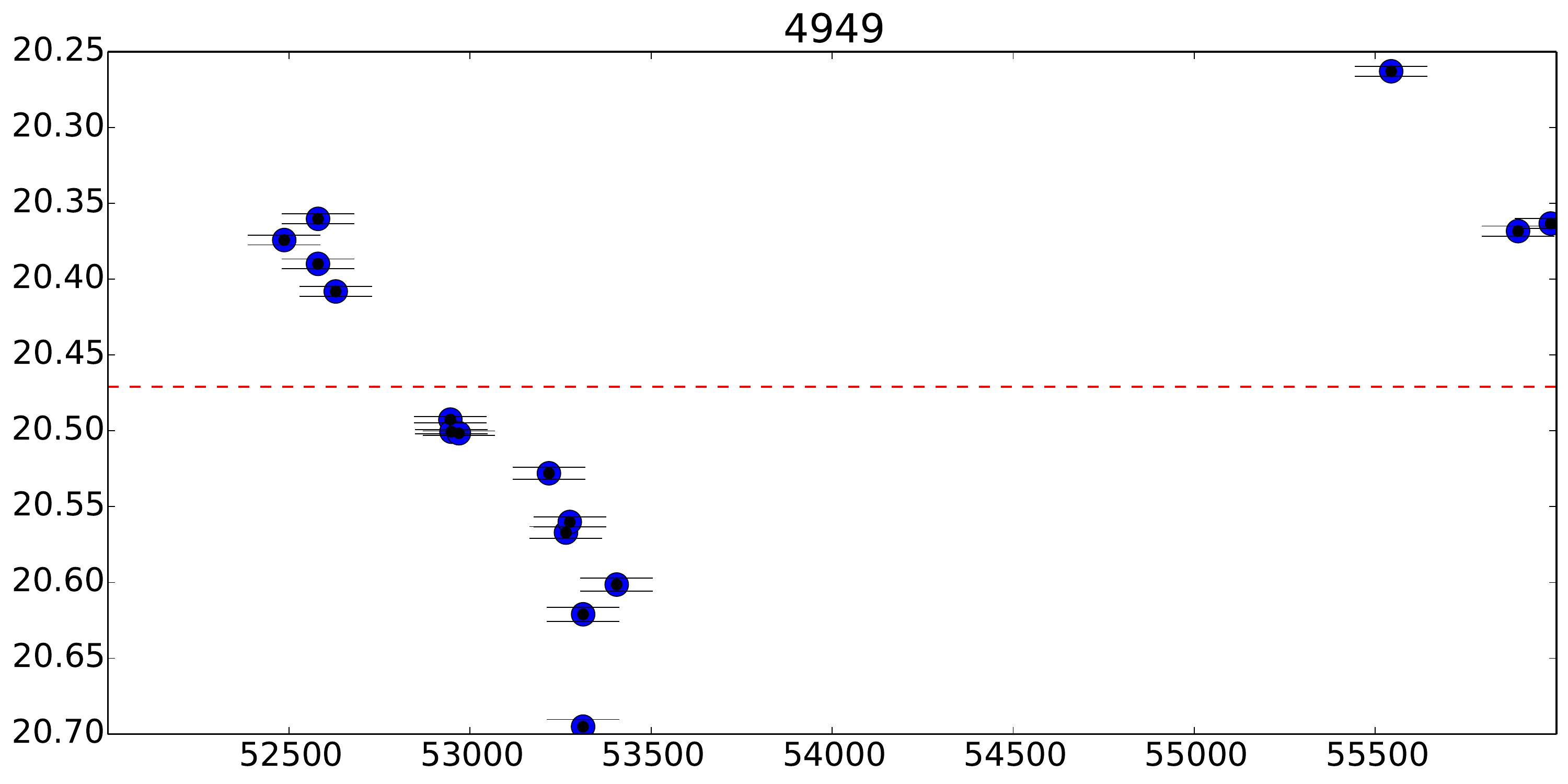} \\
\includegraphics[trim=1.25cm 0cm 0.cm 0cm,width=0.5\textwidth]{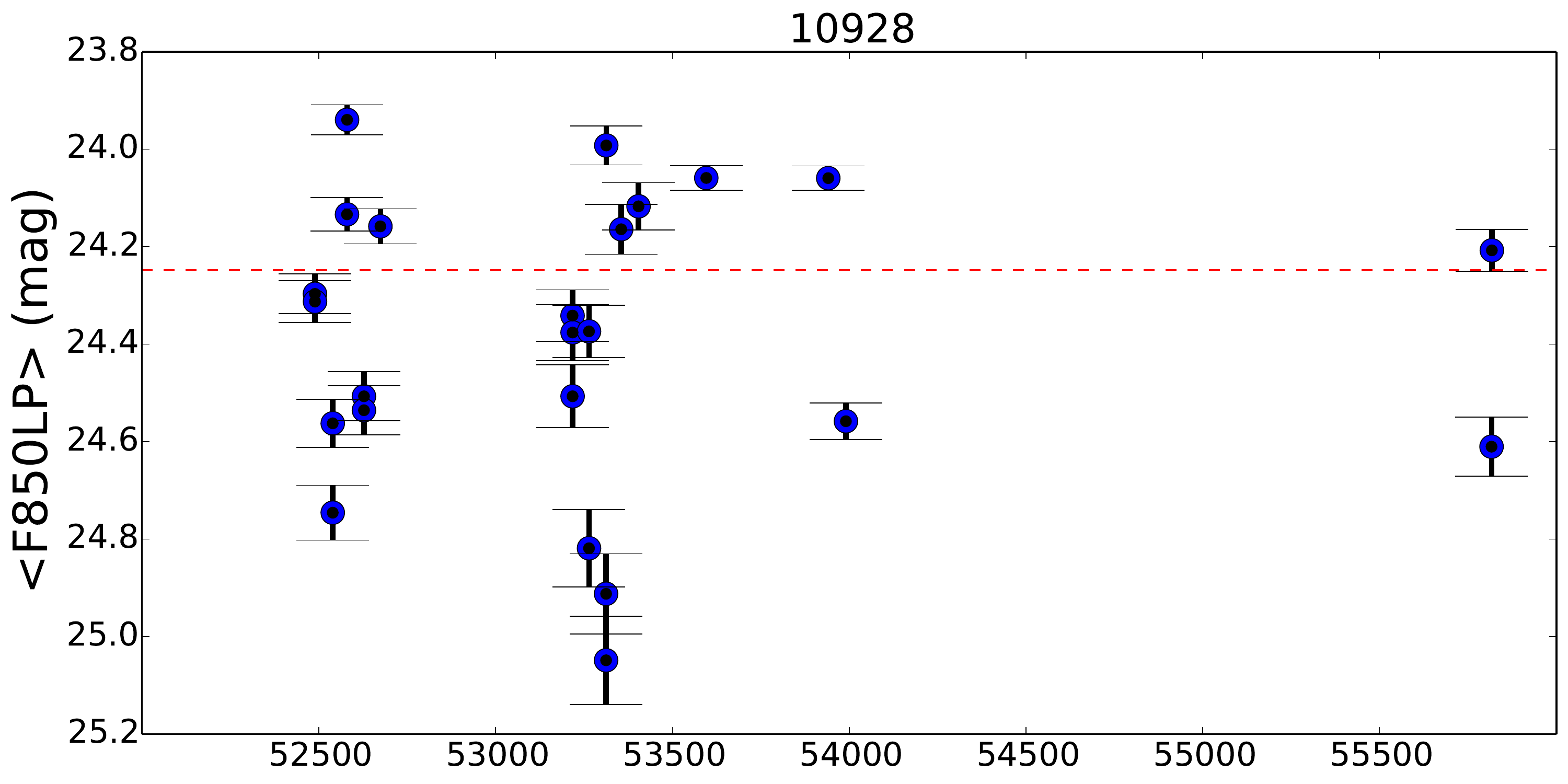} &
\includegraphics[trim=1.4cm 0cm 0.cm 0cm,width=0.5\textwidth]{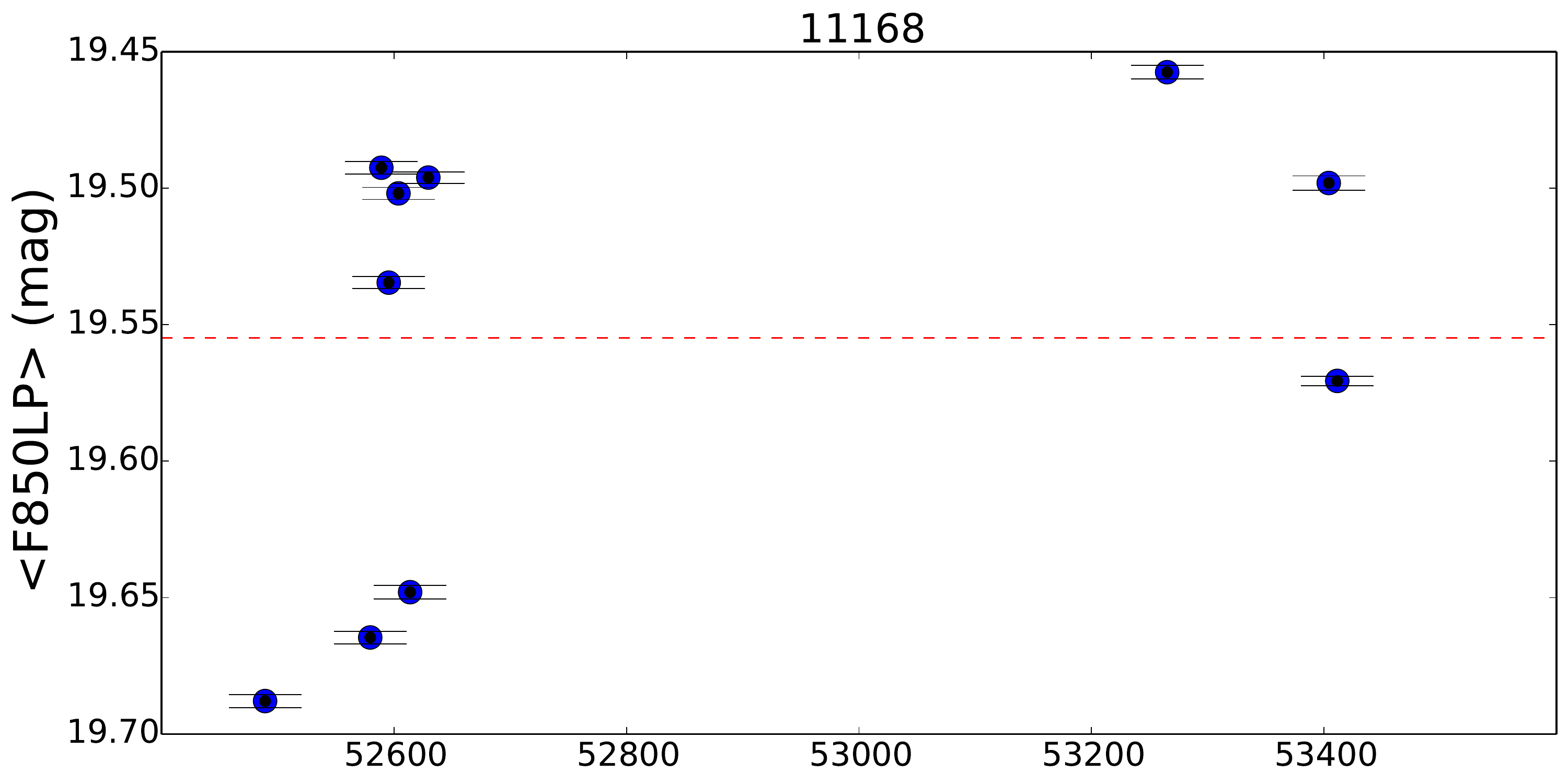} \\
\includegraphics[trim=0.cm 0cm 0.cm 0cm,width=0.5\textwidth,height=0.28\textwidth]{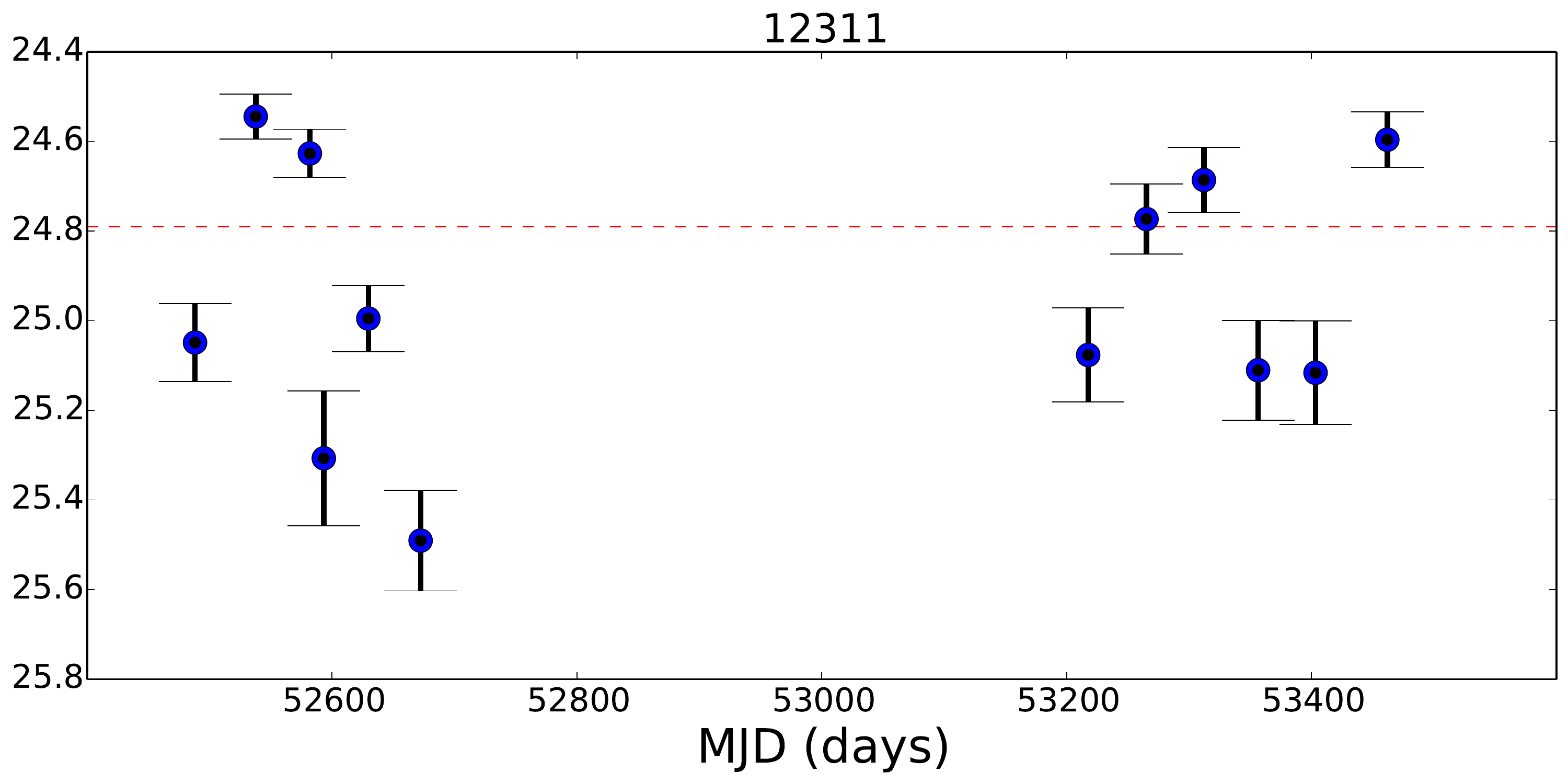} &
\includegraphics[trim=0.4cm 0cm 0.cm 0cm,width=0.5\textwidth,height=0.28\textwidth]{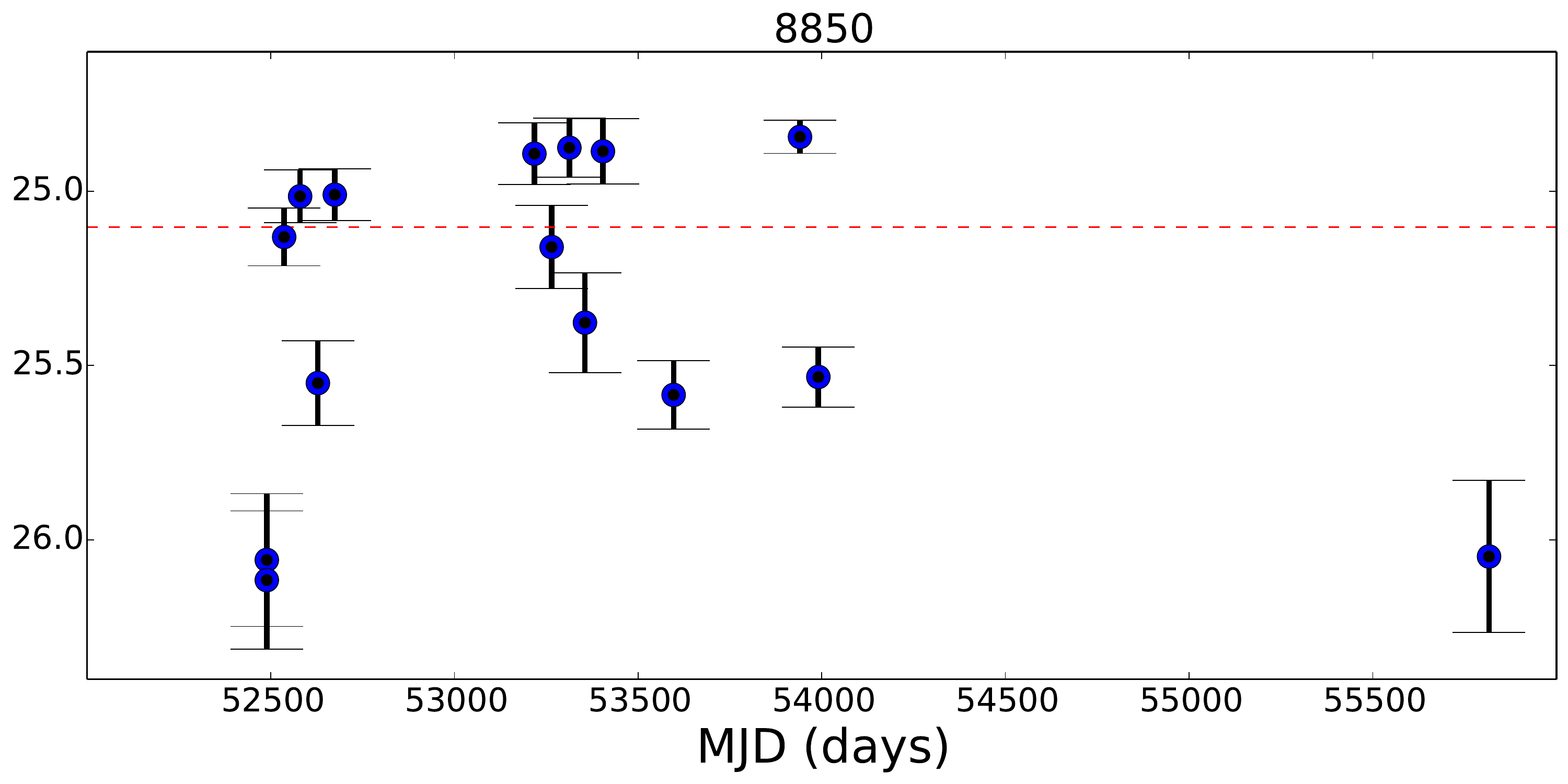} 
\end{tabular}
\caption{Example light curves of the AGN candidates. The dashed line indicates the median magnitude and the number on the top of each plot indicates the identifier of the source.}\label{AGNlightcurves}
\end{figure*}

In the next sections, we proceed with the analysis of the remaining 113 variable sources (10 point-like and 103 extended), which are presumed to be AGN candidates (Table~\ref{table175}). 
Figure~\ref{finderchart} illustrates the positions of the variable sources on the sky, while some examples of the AGN light curves can be found in Figure~\ref{AGNlightcurves}.

\begin{table*}
\caption{Catalogue of the variable AGN candidates.}
\scalebox{0.90}{
\begin{threeparttable}
\begin{tabular}{c c c c c c c c c c c c} 
\hline
ID & Grade & RA & Dec & N\textsubscript{p} & T\textsubscript{bas} & CI & <F850LP> & ${\rm MAD}$* & z & z Ref. & F\textsubscript{x} [0.5-8 keV]\\
 & & (J2000) & (J2000) &  & (years) & (mag) & (mag) & ($\sigma$) & & & (ergs cm\textsuperscript{-2} s\textsuperscript{-1})  \\ 
(1) & (2) & (3) & (4) & (5) & (6) & (7) & (8) & (9) & (10) & (11) & (12)\\ \hline\hline
347& B &52.94086&-27.85703&5&9.1&2.22&22.56&5.93&0.58& 3,c & <7.40E-16 \\
398& A &52.94348&-27.81377&6&9.12&2.19&23.31&4.04&0.89& 3,c & <4.86E-16 \\
561& A &52.95213&-27.84881&8&9.11&1.7&24.55&10.6&0.2& 3,g  & <4.44E-16 \\
615& A &52.95474&-27.95340&6&1.92&2.14&22.4&5.15&1.22& 1,e  &2.10E-015\\
1081& A &52.96888&-27.77844&5&8.18&2.01&22.03&5.91&0.67& 1,a  &3.17E-015\\
1181& B &52.97121&-27.85896&5&9.11&2.03&24.32&3.71&1.3& 3,c & <2.40E-16 \\
1560& B &52.98128&-27.90919&5&9.13&1.64&25.67&3.64&0.95& 3,c & <4.65E-16 \\
1749& A &52.98486&-27.76583&6&1.95&1.57&23.14&36.9&4.83& 3,c & <2.72E-16 \\
1799& A &52.98585&-27.88607&6&9.11&2.26&22.31&6.71&0.24& 1,c & <1.31E-16 \\
2300& B &52.99559&-27.71783&6&8.3&2.28&23.77&6.72&1.03& 3,d & <3.05E-16 \\
2644& A &53.00151&-27.72218&9&2.51&1.28&22.23&7.5&1.04& 1,a   &1.46E-014\\
2816& A &53.00463&-27.74807&9&8.46&1.29&25.43&4.23&0.66& 3,a   & <2.01E-16  \\
3129& A &53.00967&-27.87611&5&9.14&1.47&23.78&3.67&1.43& 1,a  &1.06E-015\\
3240& A &53.01125&-27.72273&9&8.58&1.69&25.76&4.54&2.17& 3,b & <1.92E-16 \\
3268& A &53.01156&-27.70921&7&8.58&1.64&25.65&3.63&2.9& 3,b & <4.20E-16 \\
3485& B &53.01471&-27.86247&5&9.14&2.17&23.65&6.03&1.33& 3,d & <3.52E-16 \\
3753& A &53.01887&-27.77287&6&8.46&1.57&26.16&3.88&1.16& 3,b & <2.98E-16 \\
4080& A &53.02418&-27.76962&7&1.06&1.56&25.86&4.19&1.3& 3,b & <1.44E-16 \\
4210& A &53.02684&-27.87276&5&9.14&2.01&24.59&4.08&2.33& 3,c & <3.02E-16 \\
4388& B &53.02974&-27.78371&5&8.03&1.66&25.89&3.71&3.67& 3,b & <1.67E-16 \\
4422& A &53.03039&-27.91629&5&8.99&2.16&24.08&4.89&1.14& 3,c & <2.66E-16 \\
4757& A &53.03535&-27.79084&5&9.07&1.68&25.92&4.41&3.08& 3,d & <1.38E-15 \\
4949& A &53.03939&-27.80194&16&9.58&1.18&20.47&7.73&2.81& 1,a  &6.51E-015\\
4981& A &53.03997&-27.81762&5&1.11&1.73&26.3&5.87&0.3& 3,b & <1.65E-16 \\
5182& A &53.04340&-27.80370&5&0.9&1.57&26.02&5.06&1.47& 3,b & <1.59E-16 \\
5281& A &53.04498&-27.79240&19&9.58&1.99&24.9&3.99&0.81& 3,b & <9.88E-17 \\
5322& A &53.04550&-27.73754&11&8.59&1.26&22.42&3.78&1.62& 1,a    &1.95E-014\\
5497& A &53.04805&-27.80435&14&9.57&1.86&21.84&4.63&0.54& 1,a  &3.78E-017\\
5510& B &53.04824&-27.77309&11&9.44&2.21&23.47&5.06&1.21& 2,b & <7.72E-16 \\
5664& A &53.05076&-27.89412&5&9.19&2.08&23.75&7.13&1.1& 3,c & <1.82E-16 \\
5856& B &53.05383&-27.75002&22&2.52&2.13&24.34&3.77&1.2& 2,b & <2.87E-16 \\
6061& B &53.05739&-27.71684&5&2.5&1.69&25.73&3.64&1.55& 3,b & <2.74E-16 \\
6381& A &53.06216&-27.81900&12&9.39&1.79&25.06&3.73&1.05& 3,b & <9.72E-17 \\
6446& A &53.06299&-27.77906&34&9.33&2.06&24.39&5.04&0.96& 2,b & <5.29E-17 \\
6515& A &53.06407&-27.71645&12&8.59&1.82&24.62&4.27&2.55& 3,d & <1.45E-16 \\
6932& B &53.07015&-27.73988&5&2.44&1.62&25.48&3.6&0.76& 3,b & <2.17E-16 \\
7112& B &53.07239&-27.71873&14&8.59&2.33&22.89&4.79&0.64& 1,b & <6.74E-15 \\
7224& A &53.07367&-27.70253&10&8.59&2.05&23.66&4.43&1.18& 1,d & <2.06E-16 \\
7270& A &53.07447&-27.84984&17&9.51&2.39&20.91&5.53&0.12& 1,a  &1.58E-016\\
7531& B &53.07825&-27.87013&14&9.64&2.17&23.99&4.77&0.66& 1,d & <2.67-E16 \\
7588& A &53.07909&-27.75488&12&9.33&1.86&25.43&3.92&1.18& 3,b & <1.89E-16 \\
7906& B &53.08332&-27.94113&5&3.82&1.85&23.19&4.88&0.65& 3,b & <1.46E-16 \\
7924& A &53.08366&-27.86983&6&9.51&1.78&25.16&4.77&0.73& 1,c & <1.24E-16 \\
7938& A &53.08386&-27.73963&20&2.51&2.26&23.86&3.77&1.21& 1,a  &1.46E-016\\
7947& B &53.08388&-27.78481&5&9.08&1.44&25.77&3.79&2.3& 3,b & <1.47E-16 \\
8184& A &53.08716&-27.78490&18&9.33&2.14&22.31&5.26&0.66& 1,a  &4.15E-017\\
8240& A &53.08781&-27.89646&7&3.98&1.58&25.98&3.8&4.33& 3,b & <1.32E-16 \\
8413& A &53.09017&-27.84784&20&9.64&2.19&24.18&3.56&1.03& 3,c &3.75E-017\\
8728& A &53.09443&-27.67175&8&2.6&1.9&24.9&4.46&1.35& 3,c & <3.54E-16 \\
8850& A &53.09608&-27.88041&15&9.11&1.51&25.1&3.86&0.02& 3,d & <3.08E-16 \\
8917& A &53.09674&-27.77237&26&9.33&1.93&24.73&3.51&0.6& 2,b & <1.76E-16 \\
9014& B &53.09825&-27.77718&5&2.21&2.14&24.22&3.63&0.42& 1,f  & <1.74E-16 \\
9021& A &53.09840&-27.86258&7&9.08&1.62&25.68&4.47&1.12& 3,b & <1.17E-16 \\
9239& A &53.10137&-27.85917&6&9.08&1.56&25.74&4.51&0.9& 3,b & <1.45E-16 \\
9288& B &53.10213&-27.72846&5&2.26&1.74&25.25&3.79&1.08& 1,f  & <1.96E-16 \\
9630& B &53.10628&-27.86994&5&2.25&1.92&25.11&4.82&1.33& 3,d & <2.41E-16 \\
9757& A &53.10780&-27.83883&32&9.64&2.25&23.57&3.54&1.09& 1,a &3.75E-017\\
9764& A &53.10784&-27.72622&15&8.59&1.8&24.47&3.69&1.21& 1,a  &9.52E-017\\
10122& B &53.11194&-27.76271&8&8.86&1.98&25.17&5.34&1.3& 1,d & <1.49E-16 \\
10144& A &53.11215&-27.71111&12&8.59&2.27&23.79&11.2&1.6& 2,b & <3.22E-16 \\
10240& A &53.11330&-27.89131&26&9.11&1.89&24.62&3.56&2.19& 2,b & <1.28E-16 \\
10426& A &53.11540&-27.88511&5&3.72&1.52&25.9&4.16&1.29& 3,b & <1.88E-16 \\
10463& A &53.11578&-27.70676&11&8.59&1.43&25.19&4.67&0.12& 3,d & <1.78E-16 \\
\hline
\end{tabular}
\begin{tablenotes}
\item \textbf{Note.} -- (1): Identifier. (2): Quality grade. (3): Right ascension. (4): Declination. (5): Number of data points in the light curve. (6): Time baseline. (7): Concentration index. (8): Median magnitude in the F850LP filter. (9): Normalized ${\rm MAD}$. (10): Redshift. (11): Method used to compute z (1: spectroscopy, 2: grism and 3: photometry), while the letter refers to the paper where the redshift obtained (a: \citet{luo2017}, b: \citet{mom2016}, c: \citet{cardamone2011}, d: \citet{straatman2017}, e: \citet{xue2016}, f: \citet{taylor2009} and g: \citet{wolf2008}). (12): Flux in the X-ray [0.5-8 keV] band. The '<' symbol represent the flux upper limit.
\end{tablenotes}
\end{threeparttable}}
\label{table175}
\end{table*}

\begin{table*}
\contcaption{}
\scalebox{0.90}{
\begin{threeparttable}
\begin{tabular}{c c c c c c c c c c c c} 
\hline
ID & Grade & RA & Dec & N\textsubscript{p} & T\textsubscript{bas}  & CI & <F850LP> & ${\rm MAD}$* & z & z Ref. & F\textsubscript{x} [0.5-8 keV] \\
 & & (J2000) & (J2000) &  & (years) & (mag) & (mag) & ($\sigma$) & & & (ergs cm\textsuperscript{-2} s\textsuperscript{-1})  \\ 
(1) & (2) & (3) & (4) & (5) & (6) & (7) & (8) & (9) & (10) & (11) & (12)  \\ \hline\hline
10511& A &53.11638&-27.87660&14&9.32&2.21&22.52&5.78&0.38& 1,a & <8.85E-17 \\
10928& B &53.12174&-27.89144&24&9.11&2.23&24.24&4.92&1.03& 3,d & <2.70E-16 \\
11168& A &53.12448&-27.74021&10&2.52&2.31&19.55&5.51&0.07& 1,a &6.95E-016\\
11213& A &53.12494&-27.73476&5&2.27&2.35&19.03&7.07&0.07& 1,a &5.12E-016\\
12113& A &53.13448&-27.68765&11&2.66&2.1&24.69&4.32&0.96& 2,b &3.61E-016\\
12311& B &53.13659&-27.68284&12&2.66&1.9&24.78&4.39&0.67& 3,d & <4.09E-16 \\
12361& A &53.13720&-27.84469&22&9.33&2.04&23.46&4.36&2.03& 1,a &8.05E-017\\
12540& A &53.13916&-27.90047&26&9.12&1.93&24.34&5.03&1.33& 2,b & <5.83E-17 \\
12911& A &53.14316&-27.81556&28&8.73&1.91&24.85&3.91&4.14& 1,b & <1.35E-16 \\
12977& A &53.14385&-27.81352&52&8.73&2.28&21.11&4.02&0.2& 3,b & <2.06E-16 \\
13240& B &53.14633&-27.77111&8&2.29&2.21&24.73&3.88&1.31& 1,d & <1.32E-16 \\
13505& A &53.14860&-27.83498&7&8.47&1.96&25.09&3.74&0.06& 3,b & <1.67E-16 \\
13941& B &53.15214&-27.94796&12&3.91&2.24&21.29&3.74&0.12& 1,c & <2.92E-16 \\
14752& A &53.15884&-27.66252&12&2.55&1.18&20.49&3.59&0.84& 1,a  &4.63E-014\\
15280& A &53.16286&-27.76722&82&8.87&1.23&21.32&3.3&1.22& 1,a   &7.93E-015\\
15342& A &53.16340&-27.84283&13&8.85&1.57&24.98&3.89&0.13& 3,b & <2.63E-16 \\
15390& B &53.16373&-27.75912&94&8.86&1.98&24.37&4.82&1.03& 3,d & <2.51E-16 \\
15405& A &53.16386&-27.65959&8&2.54&1.71&25.25&5.68&1.1& 3,b & <1.44E-15 \\
16363& B &53.17179&-27.71713&5&2.28&1.75&25.3&4.68&1.28& 3,b & <2.54E-16 \\
16466& A &53.17258&-27.90738&11&2.5&1.51&26.41&4.3&0.08& 3,b & <2.99E-16 \\
16705& A &53.17438&-27.86740&13&9.12&1.23&22.37&4.33&3.61& 1,a  &5.74E-015\\
16721& A &53.17445&-27.73336&22&8.85&1.34&24.9&4.13&2.57& 1,a  &2.68E-015\\
17058& A &53.17742&-27.75976&10&7.29&1.62&26.14&4.19&2.51& 3,b & <1.42E-16 \\
17370& A &53.18014&-27.82066&15&8.86&1.3&22.82&13.33&1.92& 1,a &9.28E-015\\
17472& B &53.18116&-27.76034&5&6.9&1.71&26.28&3.57&2.62& 3,b & <1.20E-16 \\
17999& A &53.18626&-27.87292&28&4.06&1.92&24.52&4.6&2.05& 1,b & <1.17E-16 \\
18125& A &53.18755&-27.91102&23&9.12&2.39&21.07&3.6&0.45& 1,a  &2.07E-016\\
18467& A &53.19124&-27.87216&14&9.12&2&24.21&4.23&0.98& 2,b & <3.50E-16 \\
18541& A &53.19188&-27.96988&6&2.73&1.61&24.79&5.96&0.21& 3,c & <6.89E-16 \\
18687& B &53.19341&-27.95913&6&0.85&1.96&22.05&4.29&0.65& 3,c & <5.09E-16 \\
19006& B &53.19706&-27.89010&16&9.12&1.99&24.67&3.92&1.29& 2,b & <2.42E-16 \\
19047& A &53.19754&-27.82945&6&2.69&1.55&25.67&3.55&0.24& 3,b & <1.16E-16 \\
19102& A &53.19828&-27.87916&12&9.13&1.25&25.61&4.32&1.04& 3,b   & <3.02E-16 \\
19429& B &53.20216&-27.75500&5&8.6&1.9&25.3&3.8&2.19& 3,d & <1.78E-16 \\
19579& A &53.20396&-27.85437&21&9.12&2.16&24.33&6.32&1.37& 2,b & <1.76E-16 \\
19673& A &53.20503&-27.74339&12&8.57&2.24&21.95&3.76&0.21& 1,a  &1.71E-016\\
19762& B &53.20616&-27.89478&5&0.61&1.68&25.93&3.97&1.45& 3,b & <5.02E-16 \\
19976& A &53.20908&-27.77260&7&7.29&1.48&26.42&3.51&5& 3,b & <1.43E-16 \\
20057& A &53.21029&-27.90583&15&3.98&2.28&20.99&4&0.12& 1,b & <3.90E-16 \\
20076& A &53.21064&-27.81873&7&8.35&1.52&26.22&3.52&2.84& 3,b & <4.87E-17 \\
20085& A &53.21084&-27.73590&6&8.57&1.72&24.92&3.75&1.08& 3,b & <2.36E-16 \\
20190& A &53.21235&-27.90401&24&3.98&2.04&24.68&4.45&1.68& 2,b & <3.47E-16 \\
20698& A &53.22099&-27.94706&6&0.85&1.51&23.61&25.1&0.35& 3,c & <4.47E-16 \\
20760& B &53.22195&-27.91449&5&2.5&1.77&25.65&3.5&0.85& 3,d & <2.57E-15 \\
21288& A &53.23217&-27.77188&6&8.71&1.75&24.74&4.67&0.58& 3,c & <2.16E-16 \\
21541& A &53.23844&-27.87746&10&9.08&2.07&24.7&4.09&0.77& 2,b & <3.49E-16 \\
21733& A &53.24374&-27.82197&14&1.04&1.53&26.07&3.69&0.95& 3,g  & <2.07E-16 \\
21944& B &53.24986&-27.85666&24&8.98&2.03&25.04&3.68&1.25& 3,d & <2.74E-16 \\
21983& B &53.25070&-27.86125&27&8.98&2.27&21.57&6.16&0.01& 2,b & <5.51E-16 \\
22884& A &53.28173&-27.85756&9&8.81&1.24&22.38&13.02&1.61& 1,a   &2.07E-014\\
\hline
\end{tabular}
\begin{tablenotes}
\item \textbf{Note.} -- (1): Identifier. (2): Quality grade. (3): Right ascension. (4): Declination. (5): Number of data points in the light curve. (6): Time baseline. (7): Concentration index. (8): Median magnitude in the F850LP filter. (9): Normalized ${\rm MAD}$. (10): Redshift. (11): Method used to compute z (1: spectroscopy, 2: grism and 3: photometry), while the letter refers to the paper where the redshift obtained (a: \citet{luo2017}, b: \citet{mom2016}, c: \citet{cardamone2011}, d: \citet{straatman2017}, e: \citet{xue2016}, f: \citet{taylor2009} and g: \citet{wolf2008}). (12): Flux in the X-ray [0.5-8 keV] band. The '<' symbol represent the flux upper limit.
\end{tablenotes}
\end{threeparttable}}
\end{table*}

\section{Properties of the AGN candidates}\label{properties}

\subsection{X-ray detections \& upper limits}\label{xraysection}

We cross-matched the final sample of the 113 AGN candidates with the four X-ray catalogues described in Section~\ref{auxdata}. We used a search radius of 2" as the maximal positional error of the X-ray sources reaches values of $\sim$1.8" that corresponds to high off-axis angles. The positional errors of our optically variable sources are less than 0.1". In order to check if the X-ray counterparts are the correct ones, we visually checked both optical and X-ray images. We excluded two X-ray counterparts that corresponded to neighbouring sources, resulting in a total of 26 AGN candidates with X-ray emission. The counterparts in the 7\,Ms catalogue include all the sources from the lower depth catalogues, except for two sources detected in the 250 ks and 4\,Ms catalogues. According to \citet{luo2017}, they do not exist in the 7\,Ms catalogue, because they are variable sources or they are spurious detections resulting from background fluctuations or they did not pass the binomial no-source probability threshold used. The number of the X-ray counterparts is listed in Table~\ref{tableXrays} and increases with depth, as expected. All the variable point-like sources, except for two (ID: 2816 and 19102 with optical magnitude greater than 25 mag), have X-ray counterparts and also have been identified as QSOs or AGNs in other studies, including \citet{vill2010}, \citet{trevese2008} and \citet{sara2011}. It is worth noting that four optically variable AGN candidates (ID 7270, 11168, 11213 \& 19673) have been selected as X-ray variable sources in \citet{young2012}, while two out of those (ID 11168 \& 11213) are present also in the variability-selected LLAGN catalogue of \citet{ding2018}.

\begin{figure*}
\includegraphics[width=1\textwidth]{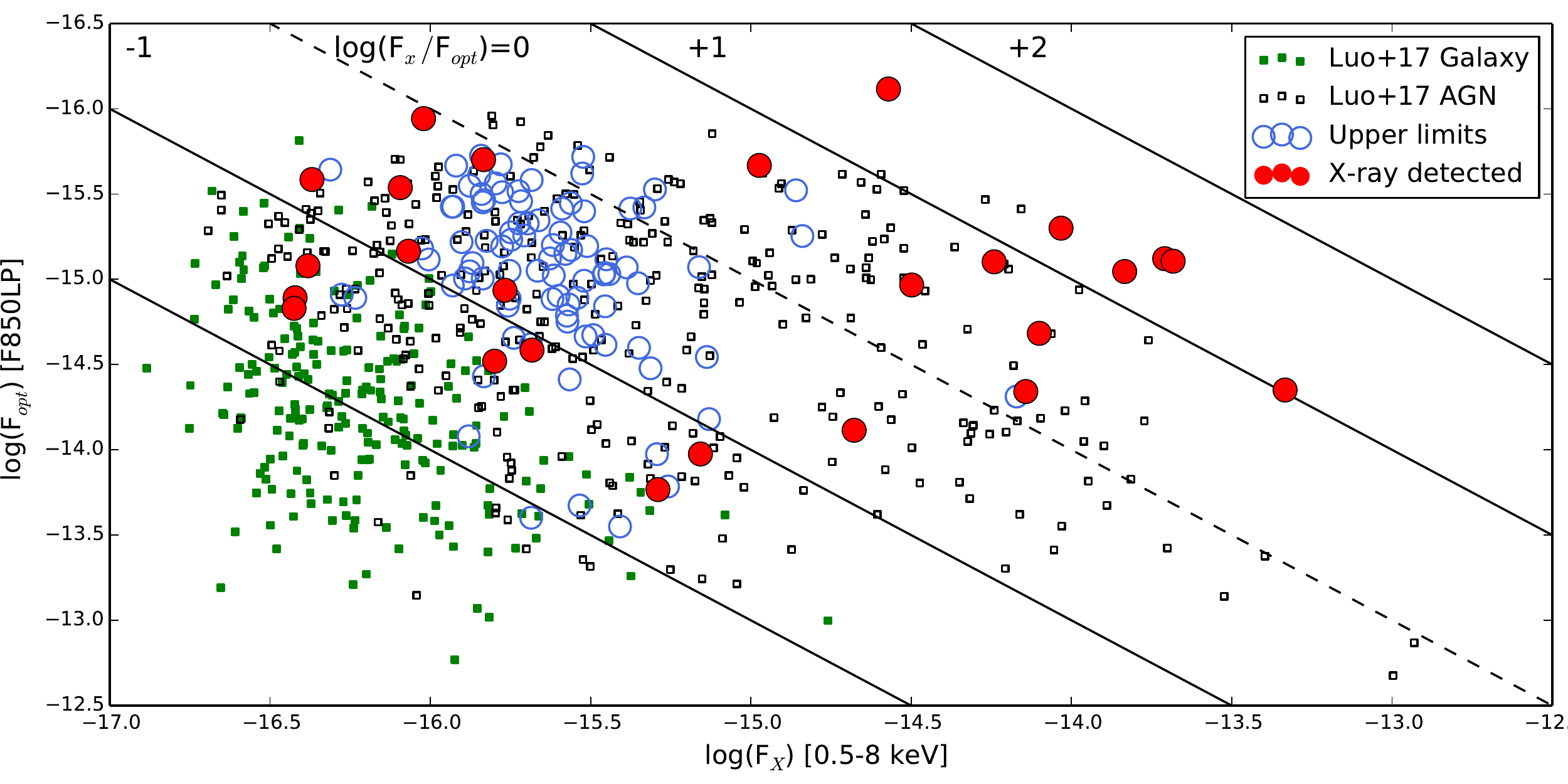}
\caption{Broad (0.5-8 keV) X-ray vs. optical (<F850LP>) flux for the 26 AGN candidates with X-ray counterparts (red circles). Open blue circles represent those sources for which only upper limits were derived. The open black and filled green squares in the background represent the sources classified as AGN or normal galaxies in \citet{luo2017}, respectively. The dashed line indicates the $\log(F_{\rm X}/F_{\rm opt})=0$ and the solid lines from left to right correspond to $\log(F_X/F_{\rm opt})=-2, -1, +1, +2$, respectively. The fluxes are given in units of ergs cm\textsuperscript{-2} s\textsuperscript{-1}.}\label{fxfopt}
\end{figure*}

A well-known diagnostic to identify AGNs is the $F_{\rm X}/F_{\rm opt}$ diagram \citep{maccacaro1988,barger2003,hornschemeier2003}, where $F_{\rm X}$ and $F_{\rm opt}$ are the X-ray and optical flux, respectively. The conventional AGN population lies in the area between $\log(F_{\rm X}/F_{\rm opt})=\pm1$, while spectroscopically confirmed AGN have been reported up to $\log(F_{\rm X}/F_{\rm opt})=\pm2$. The normal galaxies are expected to have $\log(F_{\rm X}/F_{\rm opt})\leqslant-2$. In Figure~\ref{fxfopt}, we plot the optical flux (F850LP) as a function of the X-ray (0.5-8 keV) flux of our optically variable AGNs compared with the normal galaxy (filled green squares) and AGN (open black squares) populations from \citet{luo2017}. The optically variable AGNs with X-ray detections (filled red circles) lie over the whole area within $F_{\rm X}/F_{\rm opt}=\pm2$. We could be quite confident that sources with $F_{\rm X}/F_{\rm opt}>-1$ are AGNs, while between -2 and -1 their flux ratio is still consistent with AGNs, but their nature is confirmed by the combination of variability with X-ray emission.

The flux upper limits (open blue circles) are consistent with AGNs according to their $F_{\rm X}/F_{\rm opt}$ ratio, though deeper X-ray images are needed to detect them. In particular, the majority of the flux upper limits lie below the average $F_{\rm X}/F_{\rm opt}=0$ line, indicating that these AGNs are mostly X-ray weak. Concerning the sources that overlap with the normal galaxies, \citet{luo2017} mentioned that a fraction of the normal galaxy population may also include LLAGNs. For the sources detected only in the soft or hard band, we transformed the fluxes into the [0.5-8 keV] band using the \texttt{WebPIMMS}\footnote{\url{https://heasarc.gsfc.nasa.gov/docs/software/tools/pimms.html}}~\texttt{v4.8d} software, assuming the photon index $\Gamma=1.4$ and the Galactic HI column density $n_{H}=10^{20}$cm\textsuperscript{-2}. The same transformations were applied to the broad band ($0.5-7$ keV) of \citet{luo2017}.

\begin{figure}
\includegraphics[width=0.5\textwidth]{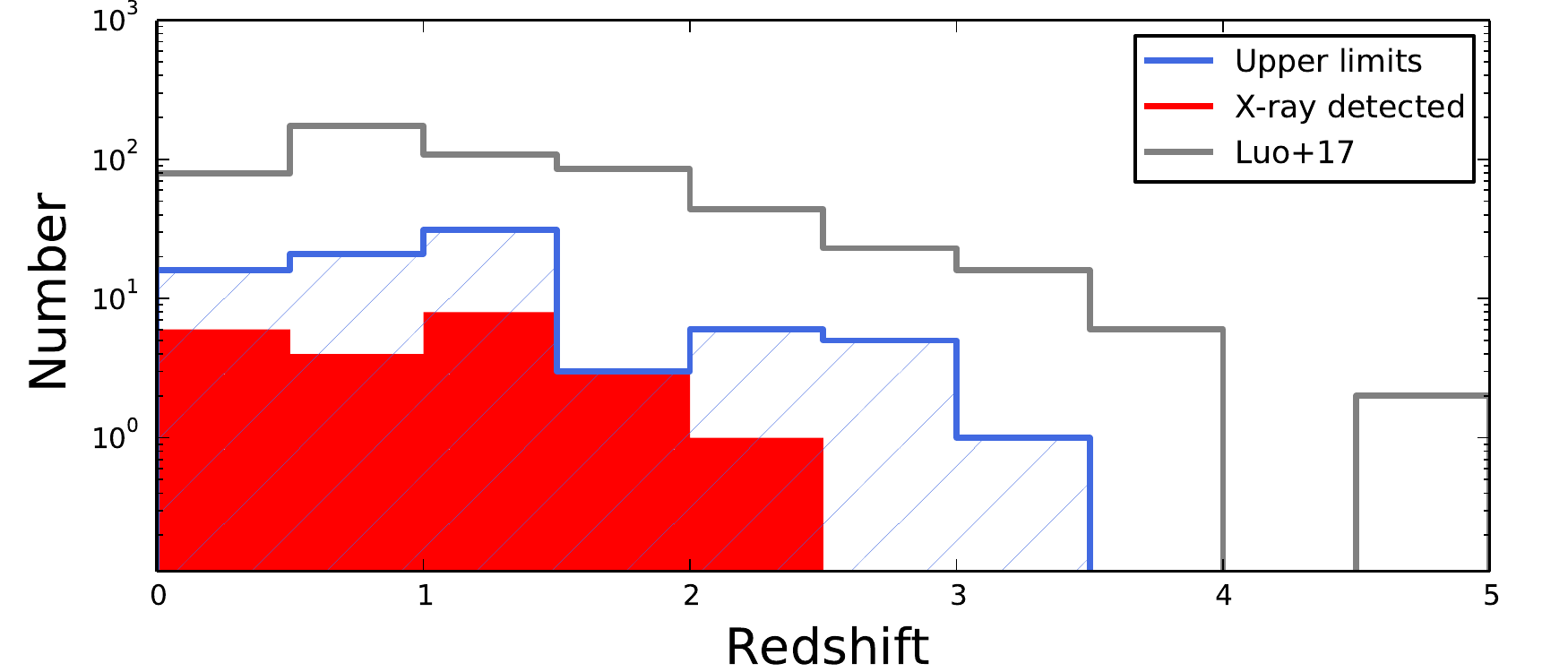}
\caption{Redshift distribution for the candidate AGNs with (filled red) and without (hatch-filled blue) X-ray counterparts. The grey histogram indicates all the X-ray sources with optical counterparts in GOODS-S.}\label{redshift}
\end{figure}

\begin{figure}
\includegraphics[width=0.5\textwidth]{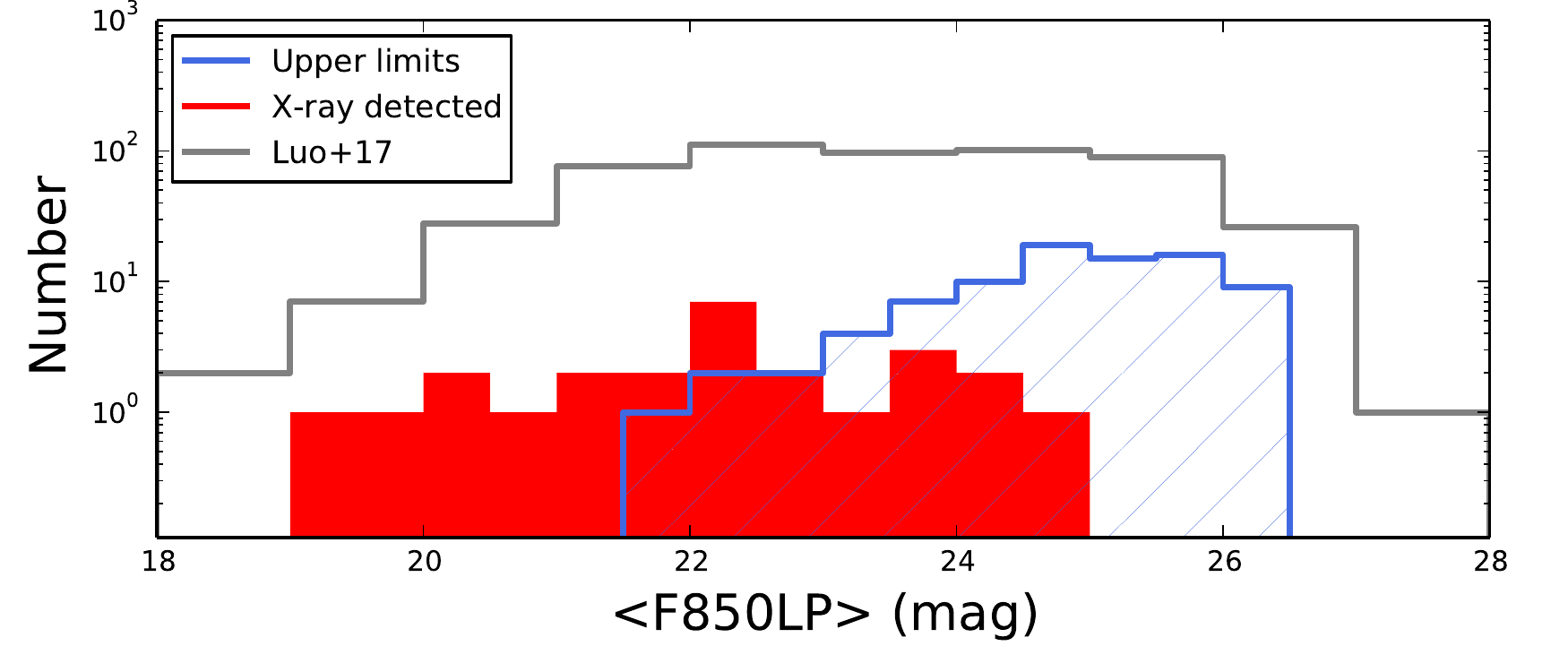}
\caption{Median F850LP magnitude distribution for the candidate AGNs with (red filled) and without (blue hatch-filled) X-ray counterparts. The grey histogram indicates all the X-ray sources with optical counterparts in GOODS-S.}\label{magdistr}
\end{figure}

\begin{figure}
\includegraphics[width=0.48\textwidth]{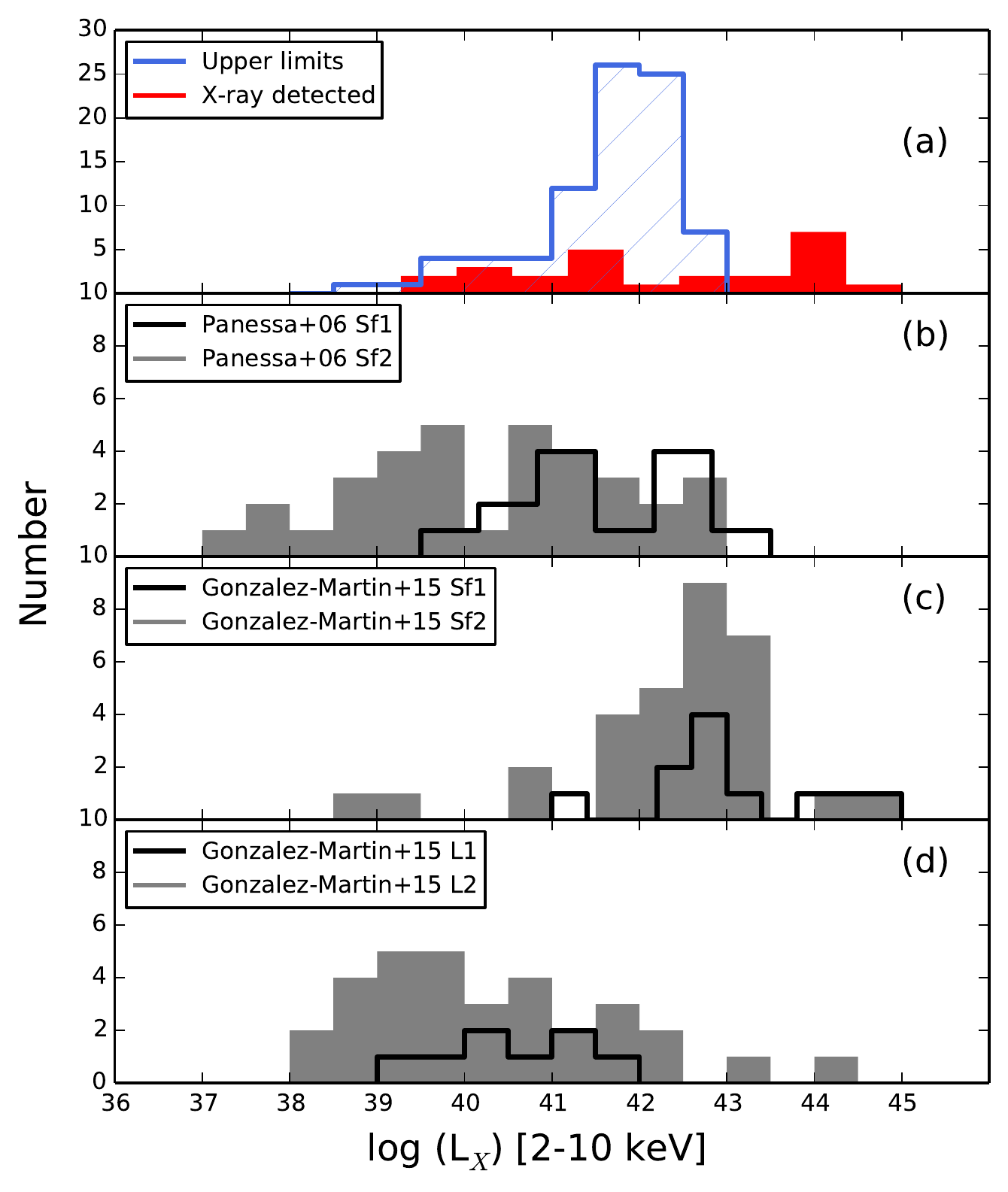}
\caption{X-ray luminosities in the [2-10 keV] band (panel a) for the AGN candidates, with (red filled) and 3$\sigma$ upper limits (blue hatch-filled) X-ray counterparts. In panel (b) the luminosity distributions of Seyfert galaxies are shown from \citet{panessa2006}. Panel (c) and (d) represent the galaxy population of Seyferts and LINERs derived from \citet{martin2015}. The black and shaded grey histograms indicate the type \uppercase\expandafter{\romannumeral1} and \uppercase\expandafter{\romannumeral2}, respectively. }\label{lumin}
\end{figure}

To further understand the nature of these sources, we cross-matched our AGN candidates with the catalogue of \citet{mom2016} to associate each source with the corresponding redshift of the host galaxy. Their catalogue provides the best redshift among grism, ground-based spectroscopic or photometric redshifts \citep{skelton2014}. For the candidates with X-ray counterparts, we used the spectroscopic redshifts provided by \citet{luo2017}. The redshifts of some sources that did not have a match in the previous catalogues, were recovered from \citet{straatman2017}, \citet{cardamone2011}, \citet{taylor2009} and \citet{wolf2008}. We found published redshifts for all the variability-selected AGN candidates: 63 sources have photometric redshifts, while for the remaining 60 sources the redshifts were derived from spectra.

Even though the available spectra from \citet{mom2016} and the other catalogues are capable to provide secure redshifts, the width of the lines is too noisy for the classification of our optically faint AGN sample (broad or narrow lines). The redshift distribution for the AGN candidates is presented in Figure~\ref{redshift} with the distribution of all the X-ray sample of the 7\,Ms CDF-S catalogue that have an optical counterpart in GOODS-S for comparison. The AGNs with not yet detected X-ray emission extend to higher redshifts.

Figure~\ref{magdistr} shows the magnitude distribution of the candidate AGNs and also the magnitude distribution of the X-ray sample reported in \citet{luo2017} with an optical counterpart in GOODS-S (both AGNs and normal galaxies). It is very clear that the sample with X-ray upper limits is optically fainter than the AGN candidates with X-ray detections. The redshift and magnitude distributions suggest that the optical surveys, such as HST, may be able to identify faint high-redshift AGNs through variability. These AGNs would have been missed by current X-ray studies. Since faint nuclear emission can be observed in the optical, there are no obscuration at all or very weak obscuration effect by dust, thus these AGNs are likely LLAGNs. Their position in the $F_{\rm X}/F_{\rm opt}$ diagram suggests that these high-redshifted intrinsically X-ray weak AGNs lie below the conventional AGN population (around $F_{\rm X}/F_{\rm opt}=0$), and thus, the dependence on redshift and X-ray flux should be considered when working with $F_{\rm X}/F_{\rm opt}$ diagrams.

We next estimated the X-ray luminosity [2-10 keV] of all the AGN candidates. We found the sample of variable sources with X-ray detections and upper limits to be distributed over $\sim$5.5 and $\sim$4.5 orders of luminosity, from $10^{39.5}$ to $10^{45}$ ergs s$^{-1}$ and $10^{38.5}$ to $10^{43}$ ergs s$^{-1}$, respectively, with mean values of $1.77\times10^{42}$ and $4.57\times10^{41}$ ergs\textsuperscript{-1} s\textsuperscript{-1}. In Figure~\ref{lumin}, we compare our luminosity distribution with the distribution of different populations of AGNs located in the nearby universe (z$\sim$0): Seyfert galaxies \citep{panessa2006} and LLAGNs \citep{martin2015} that include type \uppercase\expandafter{\romannumeral1} and \uppercase\expandafter{\romannumeral2} of LINERs and Seyferts. The luminosities of the AGN candidates detected here on the basis of optical variability lie within the range of the latter sub-classes of AGNs, denoting further evidence of AGN activity even at these low X-ray luminosities.

\subsection{Mid-infrared selected AGN}\label{IR}

\begin{figure}
\includegraphics[width=0.49\textwidth]{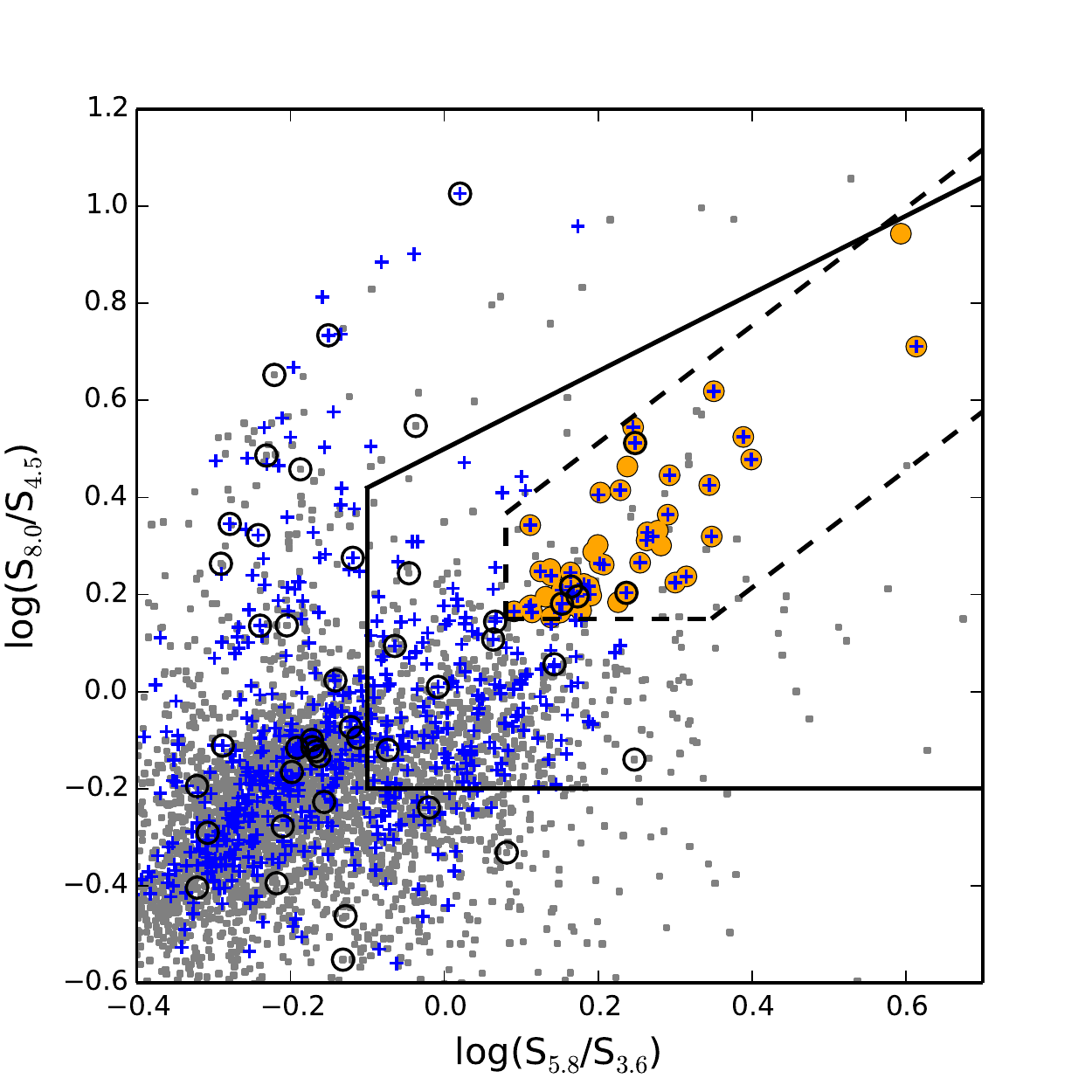}
\caption{IRAC colour-colour diagram of the IR sample (gray points). The Lacy IR AGNs defined by the solid line. The Donley IR AGNs are those inside the dashed line and follow an IR power-law (filled orange circles). The optically variable and the X-ray selected AGN samples are represented by open circles and blue crosses, respectively.} \label{IRplot}
\end{figure}

In addition to studying their X-ray properties, we explore whether our optically variable AGN candidates show evidence of accretion onto a supermassive black hole via their infrared emission. The mid-IR emission from AGNs, in particular after the advent of sensitive  spectrographs in space telescopes such as {\em ISO} and {\em Spitzer}, has proven extremely useful in revealing the presence of an AGN and characterising whether it is of type \uppercase\expandafter{\romannumeral1} or type \uppercase\expandafter{\romannumeral2} \citep{clavel2000,verma2005,weedman2005,wu2009,alonsoherrero2016}. It is now widely accepted that the AGN continuum emission appears as a power law from the 3 to 10 $\mu$m range, since the strong UV and X-ray radiation destroys the molecules responsible for the Polycyclic Aromatic Hydrocarbon (PAH) emission, while heating the surrounding dust particles in thermal equilibrium to near dust sublimation temperatures. The mid-IR AGN spectrum may also display absorption features with variable strength (due to astronomical silicates at  9.7 and 18$\mu$m) depending on the geometry of the obscuring dust as well as the luminosity of the active nucleus compared with the host galaxy \citep{nenkova2008a,nenkova2008b}. 

Even when mid-IR spectra are not available, one may use mid-IR broad-band colours to trace this slope. A number of such diagnostics have been proposed using the IRAC instrument \citep{fazio2004} on board the {\em Spitzer} Space Telescope \citep{werner2004} which provided imaging at  3.6, 4.5, 5.8 and 8.0 $\mu$m for a large sample of galaxies. These include the "Lacy wedge" \citep{lacy2004,lacy2007,sajina2005}, the "Stern wedge" \citep{stern2005} and more recently the "Donley wedge" \citep{donley2007,donley2012}. Similar methods have also been proposed for sources observed with WISE \citep{stern2012,assef2013,mateos2012}. We examine our candidates using the diagnostic of \citet{donley2012}, which has proven to be the most robust for a rather wide redshift range. The criteria by \citet{lacy2007} are also used, for comparison. 

We used the SIMPLE data mentioned in Section~\ref{auxdata}. This sample is photometrically complete at 5 $\mu Jy$, where there is a turn-over in the number density plot of the fluxes in the [5.8] band. Furthermore, we restricted our analysis to 3,904 mid-IR sources (IR sample) that have detections at all four IRAC channels as well as an optical HST counterpart with five or more data points in the light curve identified by our survey. Out of the IR sample, there are 41 optically variable sources. Following the AGN selection criteria by \citet{donley2012}:

\begin{legal}
\item $x > 0.08$  and  $y>0.15$
\item $y >(1.21 * x)-0.27$ and $y<(1.21 * x)+0.27$
\item $f[4.5] > f[3.6]$ and $f[5.8] > f[4.5]$ and $f[8.0] > f[5.8]$,
\end{legal}

\noindent where x=log(f[5.8]/f[3.6]), y=log(f[8.0]/f[4.5]) and f[band] is the flux of the corresponding band, we found 53 sources (hereafter, Donley IR AGNs). Out of those, 37 have X-ray counterparts, while five sources are optical variables. The latter five sources have also been detected in X-rays (in the 7Ms image) and are classified as QSOs in the literature. The optical variability can be explained by their QSO nature; i.e. direct view to the central source. At the same time the powerful AGN heats the dusty torus (seen face on) and its reprocessed emission dominates the infrared emission from the host galaxy. 

\citet{lacy2007} used a similar mid-IR colour-colour diagram with somewhat relaxed limits and without the power law condition:
\begin{legal}
\item $x > -0.1$  and  $y>-0.2 $
\item $y >(0.8 * x)-0.5$.
\end{legal}

\noindent Among the 770 sources that fulfill these criteria (hereafter, Lacy IR AGNs), there are 188 X-ray detections and 13 optically variable AGNs according to our analysis. The Lacy IR AGNs contains all 53 Donley IR AGNs. 
Figure~\ref{IRplot} shows the IRAC 4-band colour-colour plot for the IR sample. The lines represent the wedges as defined in \citet{lacy2007} and \citet{donley2012}. We also over-plotted the optically variable, Donley IR and the X-ray selected AGNs.

\section{Discussion}\label{discussiom}

\subsection{Comparison with previous variability studies}\label{studies}

We compared our variable sources with other variability studies of GOODS-S, including \citet{vill2010} and \citet{sara2011}, who also searched for optical variability in this field. Out of the 88 variable sources reported by \citet{vill2010}, 86 sources were included in our initial sample of 21,647 sources. Out of these, $\sim8\%$ were identified as variable with our method. Similarly, out of the 42 variable sources of \citet{sara2011} in common with our initial sample, we recovered $\sim17\%$. In a larger field of view, \citet{trevese2008} found 132 variable sources, 23 of which lie in the area studied in this work and are included in our survey; we find eight sources to be variable in our catalogue. Regarding the sample of \citet{falocco2015}, there is only one common source with our survey, which is classified as non-variable by our variability detection algorithm.

\begin{figure}
 \includegraphics[width=0.5\textwidth]{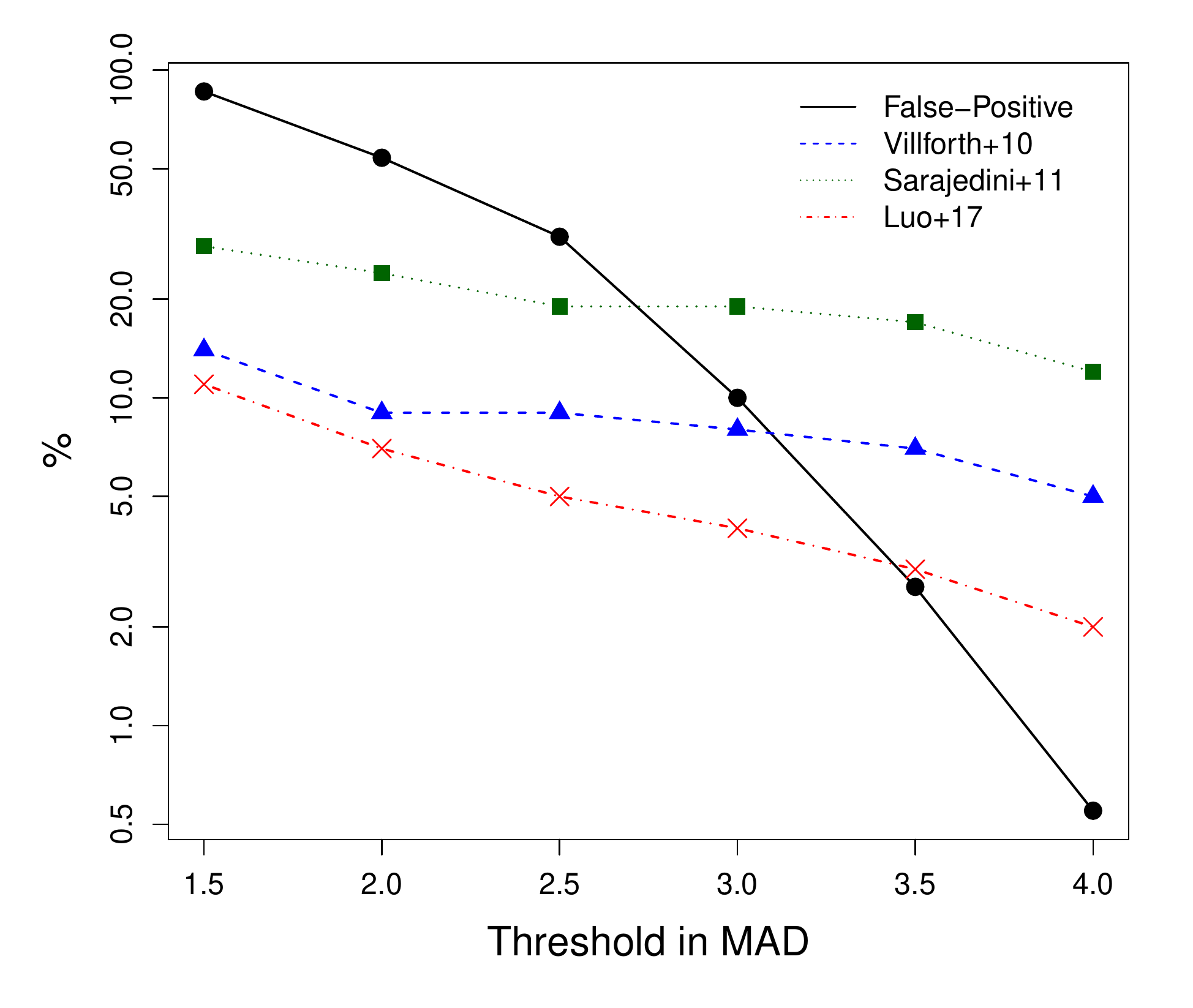}
\caption{Percentage of false positive rate (black circles), recovery rate of variable sources identified by \citet{vill2010} (blue triangles) and \citet{sara2011} (green squares) and percentage of X-ray sources by \citet{luo2017} (red crosses) identified as variables in this study as a function of different values of thresholds. The y-axis is in logarithmic scale.} \label{percentage}
\end{figure}

The main differences between our study and the studies of \citet{vill2010} and \citet{sara2011} lie in the source detection algorithm and the larger amount of data. Our approach to identifying variable objects among a set of light curves is similar to the one used by \citet{sara2011} with two important modifications: 1) we used MAD as the measure of the light-curve scatter to filter out individual outliers and 2) we used the median instead of mean to determine the expected value of scatter in a given magnitude bin. \citet{vill2010} used C statistics that rely mostly on the estimated photometric uncertainties to select variables, while \citet{sara2011} used the classic standard deviation on \textit{V}-band imaging data. Both used a 3$\sigma$ vs. a more secure threshold of 3.5$\sigma$ employed in this work. 
In order to check the dependence of the recovery rate and the false-positive contamination by the adopted variability threshold, we calculated the recovery rate of both studies and also the percentage of the false positive rate out of the variable sources (as described in Section~\ref{vi}) for different values of the threshold. In Figure~\ref{percentage}, we plot the results. By relaxing the threshold to lower values (MAD$\simeq$2.5), the recovery rate remains almost the same for both studies (1-2\% difference), though the false positive rate rapidly increases. The threshold of 3.5 that we have adopted ensured that the false-positive contamination was kept below 5\%.

The larger amount of data are due to the larger area included in this work, which doubled the number of sources, and the longer time baseline of the light curves. In particular, we used observations spanning up to ten years, instead of six months, increasing the number of points in the light curves, so we expect to detect new sources with higher sensitivity at longer timescales. Furthermore, the data reduction -- source detection and photometry -- differs from that employed in the previous studies. We used SExtractor, since it is better suited for detecting extended sources compared to IRAF. Moreover, we used images from the latest HLA data release, which is the first one to take into account the misalignments between both single exposures and filters\footnote{\url{http://hla.stsci.edu/hla_faq.html}}. Thus, the quality of the images used in this work and, consequently, the reliability of our photometry are supposed to be much higher than previous studies.

\subsection{X-ray, mid-IR and optical variability  selected AGNs}\label{studies}

\begin{figure}
 \includegraphics[trim=0.0cm 1cm 3cm 1cm,width=0.40\textwidth]{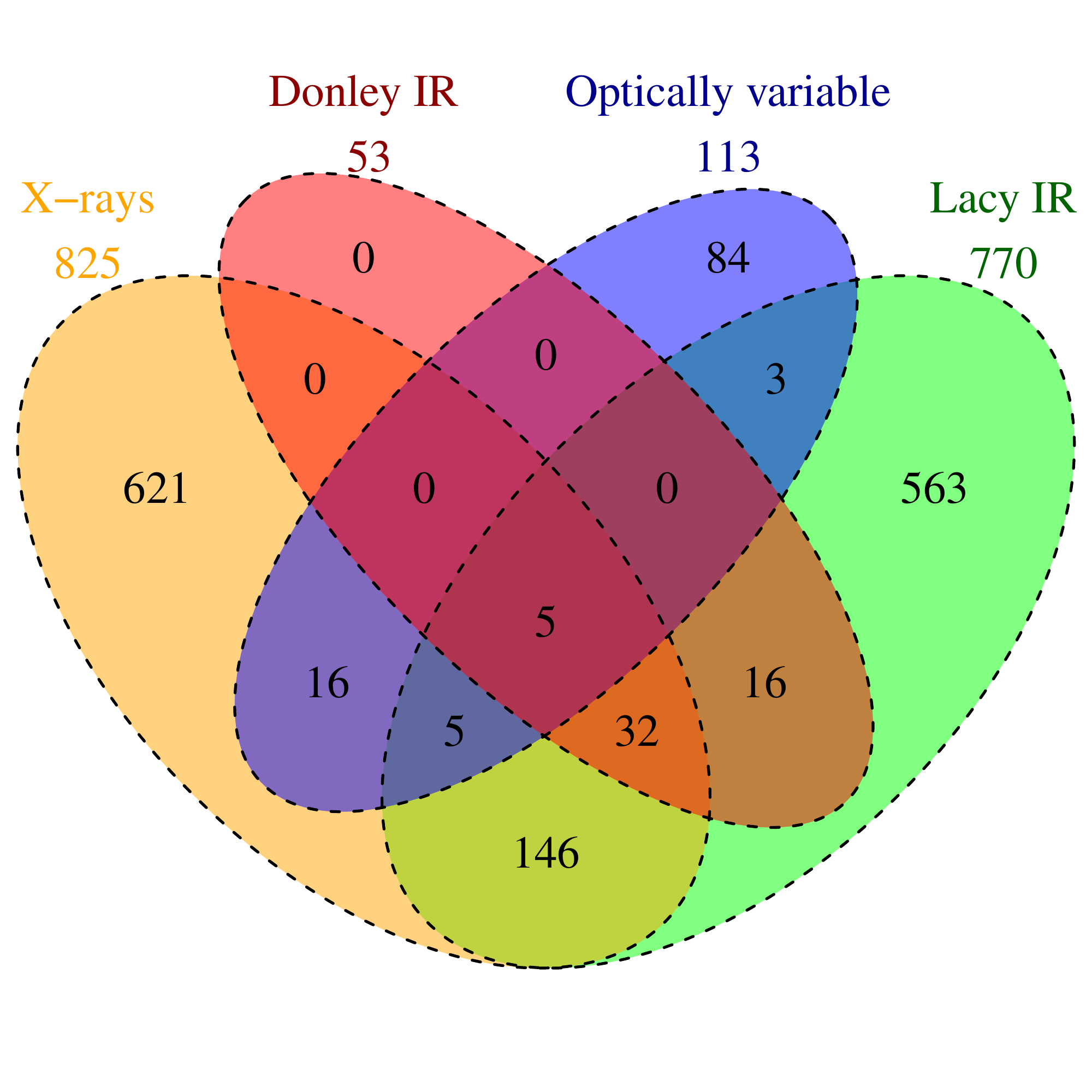}
\caption{Venn diagram of the AGN samples selected through optical variability (blue), X-rays (orange), \citet{donley2012} (red) and \citet{lacy2007} (green) IR criteria.} \label{Venn}
\end{figure}

To facilitate a direct comparison and present the various selection methods in a uniform manner, we selected the Donley IR, Lacy IR and X-ray detected AGNs that lie inside the area of GOODS-S along with the optical variable AGN candidates. In Figure~\ref{Venn}, we demonstrate the overlapping of the optical variability (113), Donley IR (53), Lacy IR (770) and X-ray selected (825) AGN samples with a Venn diagram.

\begin{figure}
 \includegraphics[trim=1cm 2cm 0cm 1cm,width=0.50\textwidth]{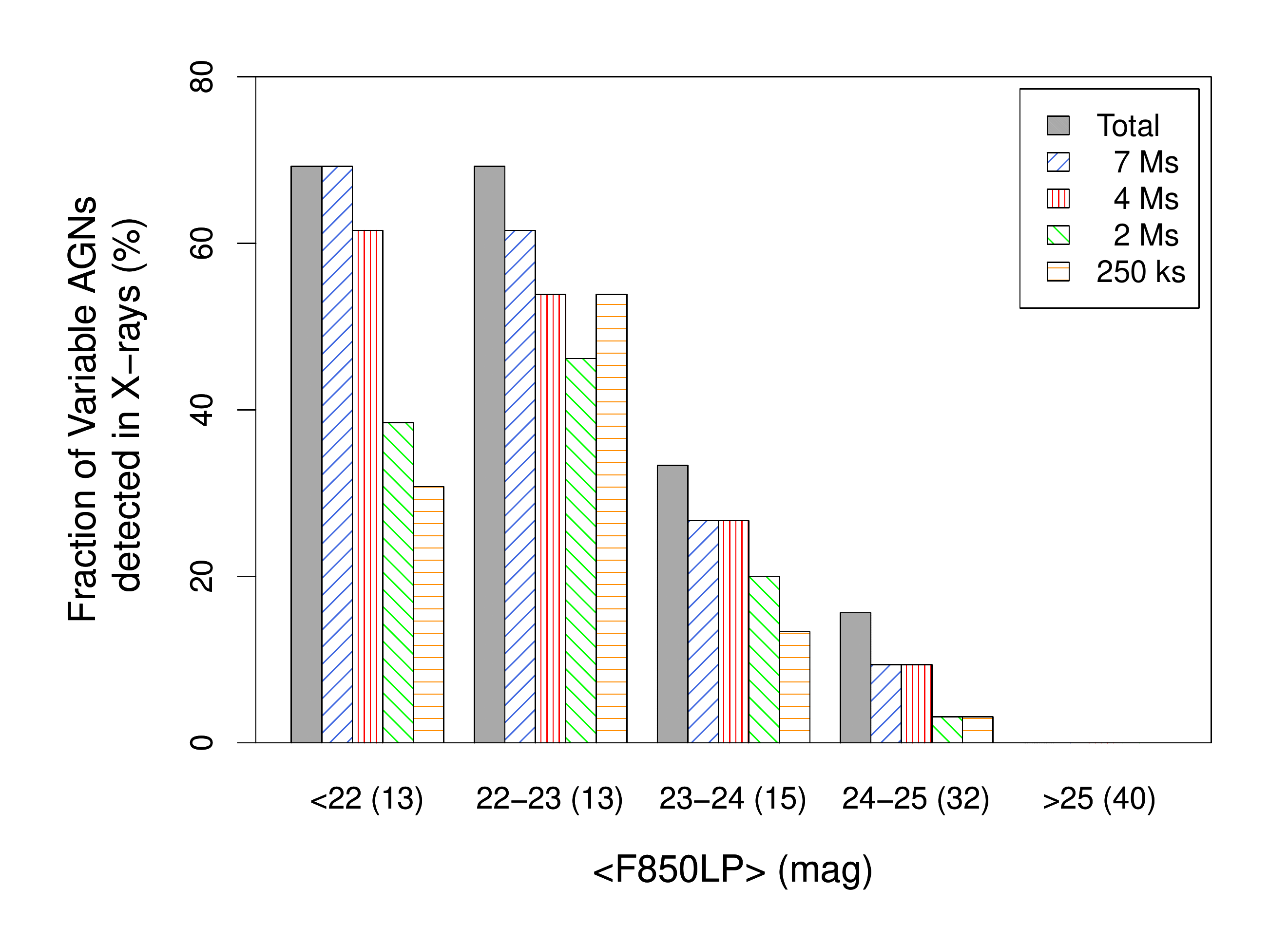}
 \caption{The fraction of AGNs selected through optical variability divided into five magnitude bins that are X-ray detected in catalogues of different depths. The number of optically variable AGNs in each bin are shown in the parentheses.} \label{Xdepths}
\end{figure}

The deepest available X-ray catalogue contains 825 sources in the area of GOODS-S. 621 sources have optical counterparts with five or more data points in their light curve, while 587 have both optical and IR detections. We found $\sim$3.5\% of the X-ray sources having significant optical variability (Figure~\ref{percentage} shows also this percentage as a function of the variability threshold). On the other hand, among the 113 optical variable AGN candidates, 26 have X-ray counterparts ($\sim$23\%). Optical variability could identify AGNs through a wide range of magnitudes and, especially, the AGN population that is missed by X-rays ($\sim$75\%). This result comes from current X-ray survey depths. To visually demonstrate this dependence, in Figure~\ref{Xdepths} we show the fraction of variable sources that are X-ray detected in the four catalogues with various depths, divided into different magnitude bins. For the magnitude bins in the bright end, the fraction reaches values of about $\sim$70\%, while at fainter magnitude bins there are no X-ray detections. \citet{vill2012A} and \citet{trevese2008} found similar results, while \citet{sara2011} had a higher rate of variable sources with X-ray detections. Out of the variable sample of \citet{falocco2015}, less than 10\% had X-ray counterparts, and in the COSMOS field, \citet{decicco2015} reached a high percentage, up to $\sim$75\%. However, their sample had a magnitude limit in the r band at $r\leq23$ mag. Given the same limits in the magnitude, we derive almost the same completeness with respect to X-rays (Figure~\ref{Xdepths}).

Furthermore, in this work we set a 3.5-$\sigma$ cut-off in MAD to identify variables. Given a less conservative value of the cut-off, the percentage of X-ray detected sources in the variable sample increases. In particular, a 3-$\sigma$ cut-off increases the X-ray detected sources to 30\%. However, at the same time the false positive variability rate also increases significantly. To avoid a high incidence of false variables, we necessarily miss a population of X-ray detected AGNs which display lower optical variability. Future surveys with higher sensitivities and better sampled data with longer time baselines will allow us to identify individually variable objects and fully characterize their variability properties without taking into account the whole population statistics. In that case, variability will be able to recover the X-ray detected AGN population with low significance (either low redshifted or high luminous AGNs).

Regarding the IR selected AGNs, the \citet{lacy2007} method selected a large number of AGN candidates through colour-colour criteria (770), comparable to that of X-ray AGNs. 25\% of these are X-ray detected, while the majority of the X-ray sources fall outside the Lacy wedge (Fig.~\ref{IRplot}). Despite the large number of AGN candidates, the contamination of star-forming galaxies is expected to be as high as 80\% \citep{donley2012}, as the sensitivity limit of our IR sample is at 5 $\mu Jy$. \citet{donley2012} studied the star-forming contamination of the IR selected AGNs defined by \citet{lacy2007} and \citet{stern2005} for samples with different depths and revised these criteria by adding an additional power law criterion. Thus, the Donley IR sample of 53 sources is expected to only have $\sim$10\% contamination as stated by \citet{donley2012}. Out of those, 37 have X-ray emission ($\sim$70\%) and all of them are included in the Lacy IR sample. It is noteworthy to say that the Donley wedge contains sources that are selected by the \citet{lacy2007} criteria, but fail the power law criterion of \citet{donley2012}. One third of the variable sample had detections at all four IRAC bands (41/113). The percentage of the variable sources common between the Donley and Lacy IR AGNs is $\sim$12\% and $\sim$32\%, respectively, while the majority of the variable sources fall outside both wedges (Figure~\ref{IRplot}) compared to previous ground-based variability studies \citep{falocco2016}. This implies that the optical variability based on HST observations is capable to identify AGNs deep into the IR region where other selection methods fail.

It is obvious that different methods are at some level sensitive to different types of objects (different luminosities, different redshifts, different dust content along the line of sight, etc.). X-ray selection is by far the most robust technique resulting in a high number of AGNs. However, there is a large fraction of optically variables and IR selected AGNs that are not detected in X-rays. These sources are likely highly obscured or low-luminosity AGNs. The \citet{lacy2007} diagnostic also returns a large sample of AGN candidates, but the contamination from star-forming galaxies is very high considering the depth of this study. At shallower depths, the contamination is minimized \citep{donley2012}. On the other hand, the AGN sample selected by \citet{donley2012} is highly complete and reliable having no contamination. However, it picks only a small sample of luminous, unobscured and obscured AGNs and misses the LLAGNs. Finally, optical variability returns a large sample of AGNs less obscured, moderate luminosity and redshift and is capable of identifying LLAGNs, which are missed by the other methods due to the currently survey depths.

\section{Summary and Conclusions}\label{summary}

Variability is a basic property of AGNs and has been proven to be a reliable method to reveal non-obscured LLAGNs. Many studies assembled multi-epoch data and used variability to identify AGNs. The need, though, for highest photometric accuracy and long-term observational monitoring impose limits on the completeness of such surveys. Previously in the GOODS South field, \citet{vill2010,vill2012A} and \citet{sara2003,sara2011} used a five epoch dataset (\textit{i}- and \textit{z}-band, respectively) spanning six months. In a larger area and from ground-based telescopes, \citet{falocco2015} used, also, a six month baseline, while \citet{trevese2008} used a longer time baseline of two years.

In this work, we substantially increased the time baseline by up to ten years using deep HST observations (\textit{z}-band).We created a new catalogue of optically variable AGN candidates in the GOODS-S. We used SExtractor to construct the light curves of $\sim$22,000 sources. The Median Absolute Deviation was utilized to search for variability as it has the highest performance among the other variability indices in the presence of outliers. A 3.5$\sigma$ cut-off was applied to identify variable sources. Our results can be summarized as follows:

\begin{itemize}
\item We identified 116 high confidence variable sources. After removing three known Supernovae, we ended up with 113 AGN candidates (103 extended and ten point-like). 

\item We explored the mid-IR properties of our AGN candidates. 41 sources have been detected in all four {\em Spitzer}/IRAC bands and, out of those, 13 and five sources are classified as AGNs through the colour selection adopted by \citet{lacy2007} and \citet{donley2012}, respectively. Also, the space observations compared to ground-based studies identify AGNs deep into the IR region when other methods fail.

\item We cross-matched our AGN sample with the published X-ray catalogues (CDF-S 2, 4 \& 7\,Ms and ECDFS 250 ks) and found 26 variable sources with X-ray counterparts. This corresponds to $\sim$23\% (26/113). 

\item For all the sources without X-ray detections, we used the 7\,Ms image in CDF-S and estimated the flux upper limits using a confidence level of 99.7\%. These sources are optically fainter with higher redshifts up to z=4. 

\item The X-ray to optical flux ratios revealed that the variable sources are consistent with the AGN population, as they lie within the area of $-2<\log(F_X/F_{\rm opt})<+2$, but their average $\log(F_{\rm X}/F_{\rm opt})$ ratio suggests that high-redshifted intrinsically X-ray weak AGNs lie below the conventional $\log(F_{\rm X}/F_{\rm opt})=0$ area.

\item The X-ray luminosities of our variable AGN candidates are comparable to those of LLAGNs in the Local Universe \citep{panessa2006,martin2015}. Hence, the variability in deep optical photometric data is a  promising method of finding optically low luminosity AGNs, which the X-ray observations may miss.
\end{itemize}

We conclude that the different methods (optical variability, IR, X-rays) used to identify AGNs are complementary to each other and equally important to constrain the full picture of the AGN demographics. In particular, optical variability is able to identify a large number of LLAGNs at high redshifts. These are critical for studying the faint end of the AGN luminosity function and it might be the key between normal galaxies and AGNs.

This work is part of the European Space Agency (ESA) project \enquote{{\em Hubble} Catalogue of Variables} \citep[HCV,][]{sokol2017a,gavras2017}, which aims to identify variable sources (stars, transients, Supernovae, AGN, etc.) from the Hubble Source Catalogue \citep[HSC,][]{whitmore2016}, through different filters and instruments. The HCV targeted fields are more than 250 and the number of sources included exceeds 3.5 million. Specifically for AGNs, there are more than 30 bona-fide deep fields covered by multi-wavelength data with observing time baselines more than two years. Extrapolating our results to these fields, we expect to identify more than 2,000 new AGNs with a high fraction of them being LLAGNs. The variability detection technique used in this work may be applied not only to HST observations but also to other surveys such as the Large Synoptic Survey Telescope \citep[LSST]{izevic2008}.

\section*{Acknowledgments}\label{ackn}

The authors are grateful to the anonymous referee for valuable suggestions that significantly improved the manuscript. E. Pouliasis acknowledges financial support by ESA under the HCV programme, contract no. 4000112940. This  research  has  made  use  of  the  VizieR  catalogue access  tool,  CDS, Strasbourg, France. The original description of the VizieR service is presented by \citet{ochsenbein2000}. This research has made use of the SIMBAD database \citep{wenger2000}, operated at CDS, Strasbourg, France and, also, of NASA's Astrophysics Data System. This   research   made   use   of   Astropy,   a   community-developed core Python package for Astronomy (Astropy Collaboration et al.   2013,   \url{http://www.astropy.org}). This  publication  made  use  of  TOPCAT \citep{taylor2005} for all table manipulation. This work is based (in part) on observations made with the Spitzer Space Telescope, which is operated by the Jet Propulsion Laboratory, California Institute of Technology under a contract with NASA.
The  plots  in  this publication  were  produced  using  Matplotlib, a Python library for publication quality graphics \citep{hunter2007} and R\footnote{R Core Team (2016). R: A language and environment for statistical computing. R Foundation for Statistical Computing, Vienna, Austria. URL \url{https://www.R-project.org/}.}. This work was supported in part by Michigan State University through computational resources provided by the Institute for Cyber-Enabled Research.






\appendix

\section{Variability detection simulations}
\subsection{Variability detection in the presence of outlier measurements}
\label{simulationssection}

For all but the faintest optical sources, the accuracy of their brightness measurements is limited by the poorly constrained systematic effects rather than the number of collected photons (``shot noise'') and uncertainties in the background level estimations. This means we typically do not have a reliable error bar attached to a photometric measurement. Of a particular concern in the context of HST photometry are the residual cosmic rays that were not cleaned-out perfectly in the process of image stacking (``drizzling''; \citealt{2002PASP..114..144F}) that overlap with the measured image of the object.
To circumvent the above issues, we may {\em 1)}\,assume that in a non-crowded field like GOODS-S (Sec.~\ref{intro}) objects of similar brightness will have similar measurement errors and the majority of objects are non-variable; {\em 2)}\,employ a variability detection statistic that is robust against individual outlier measurements (similar to those caused by cosmic ray hits).

\begin{figure}
\centering
\includegraphics[trim=0cm 0cm 0.cm 0cm,width=0.48\textwidth]{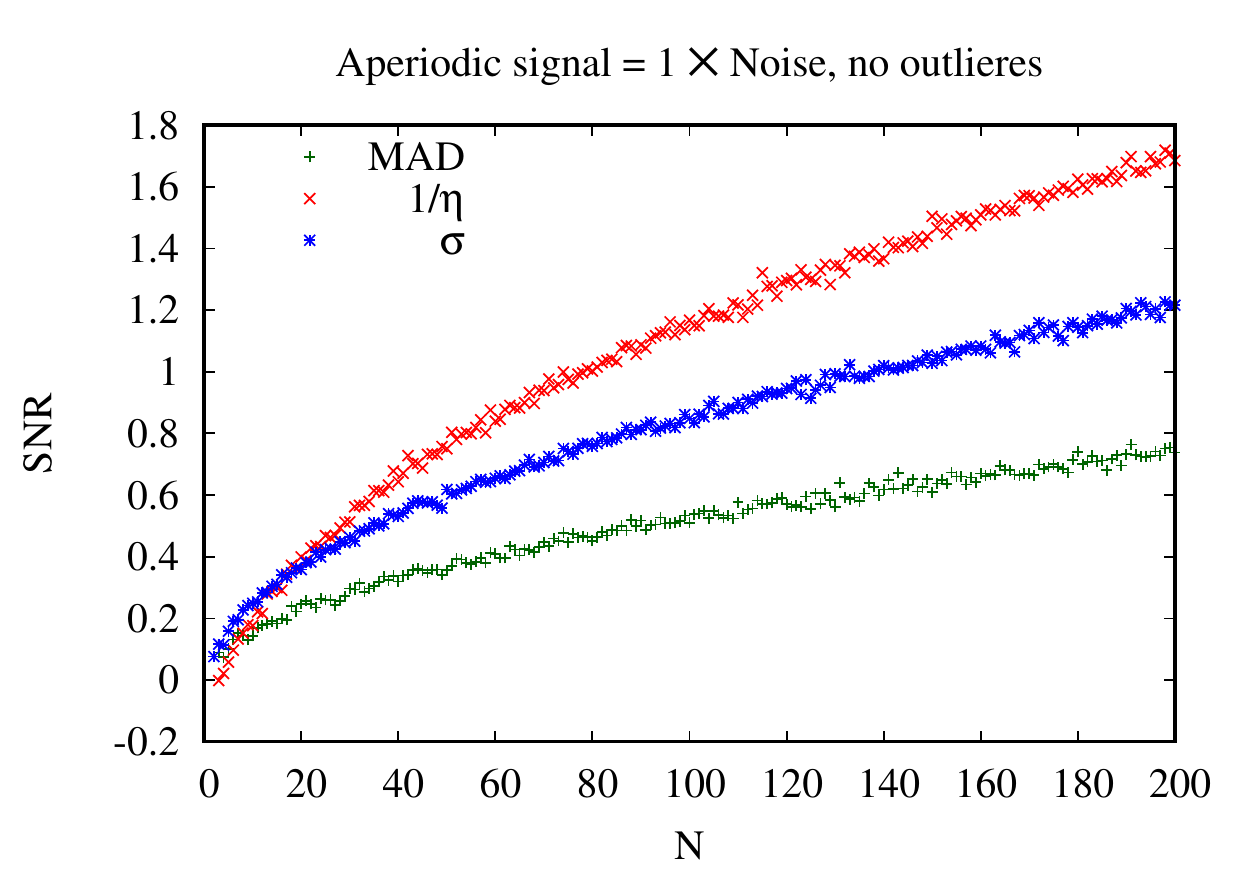}
\includegraphics[trim=0cm 0cm 0.cm 0cm,width=0.48\textwidth]{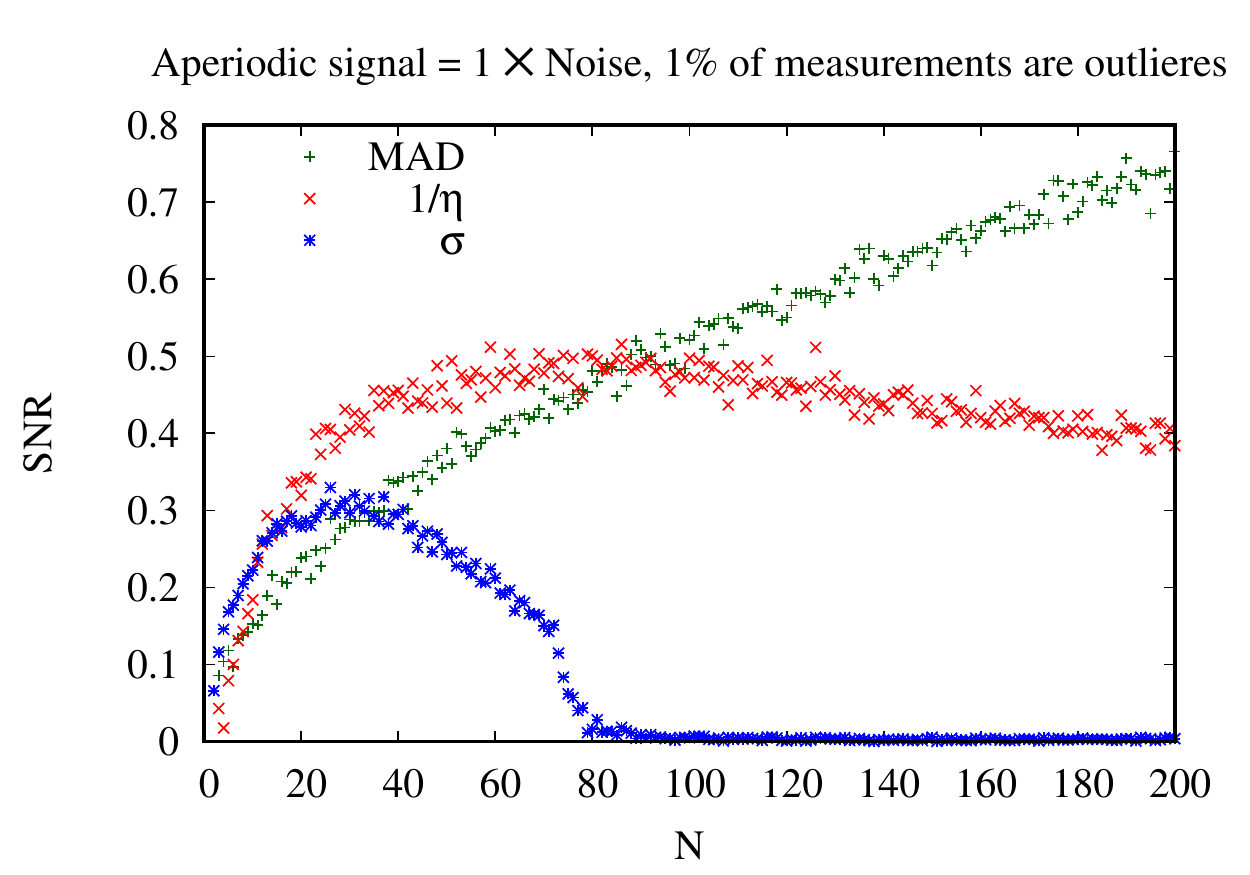}
\caption{The simulated median signal to noise ratio of variability detection as a function of the number of light curve points with no outlier measurements (top panel) and in the presence of outliers (bottom panel).}\label{simulations}
\end{figure}

We perform Monte-Carlo modeling to characterize the performance of various variability-detection statistics in the presence of photometric outliers. First, we model $i=1...10000$ light curves each containing N points randomly distributed in time.  At each point in the model light curve we assigned a brightness value drawn from the Gaussian distribution characterized by the variance $e^2$. In addition, 1\% of the points get a ``cosmic ray hit'' modeled by the additional increase in brightness by a value drawn from a uniform distribution between 0 and $100e$. We, then, compute the median value, $I^{\rm non-var}$ and the standard deviation $\sigma(I^{\rm non-var})$ scaled from the median absolute deviation of $I_{\rm non-var}$ values for each of the tested variability indices: 
$$\sigma(I^{\rm non-var}) = 1.4826 \times \median(|I^{\rm non-var}_i-\median(I^{\rm non-var}_i)|).$$
After that, we add to each light curve an aperiodic variation characterized by a power-law power spectral density with a slope of $-1$ and amplitude $e$ (equal to the noise level). We use these lightcurves to compute the median value of the variability index:
$$I^{\rm var} = \median(I^{\rm var}_i)$$ and the typical Signal-to-Noise ratio , SNR, of variability detection (among all the realizations of the noise and variability patterns):
$${\rm SNR} = (I^{\rm var} - I^{\rm non-var})/\sigma(I^{\rm non-var}).$$
The resulting values of SNR as a function of N are presented in Figure~\ref{simulations} for the three variability indices: the standard deviation $\sigma$, the median absolute deviation (MAD) that characterize the scatter of measurements in a light curve and the $1/\eta$ that quantifies the smoothness of a lightcurve. A detailed discussion of these variability indicators can be found in \cite{sokol2017b}.

Figure~\ref{simulations} highlights that in the absence of outlier measurements (i.e. non-periodic variability is being detected over a pure Gaussian noise) $\sigma$ and $1/\eta$ typically provide a higher SNR detection for a given number of light curve points than MAD. If outlier measurements are present in light curves, they dramatically affect the efficiency of $\sigma$ as a variability indicator rendering its useless as soon as each light curve has so many points that it is likely to contain at least one outlier (recall, that in our model the variability amplitude is lower than the amplitude of outliers). The ability of $1/\eta$ to identify smooth variability is also reduced considerably by outliers, while MAD maintains the SNR that is steadily increasing with N.

The simulations described above confirm that MAD may serve as a variability indicator resistant to individual outlier measurements. It is also apparent that $\sigma$ is on average a more sensitive variability indicator than MAD as long as N is sufficiently low that each individual lightcurve is unlikely to contain even one outlier. However, if outliers are present in the data set, the light curves that contain outliers will predominantly be selected with $\sigma$ as candidate variables. The use of MAD is still preferred to select a clean sample of variable objects, even at the cost of a slightly lower detection efficiency compared to $\sigma$.

\subsection{Variability threshold dependence on the number of data points in light curves}
\label{simulationsN}

\begin{figure}
\label{simulN}
\includegraphics[width=0.50\textwidth]{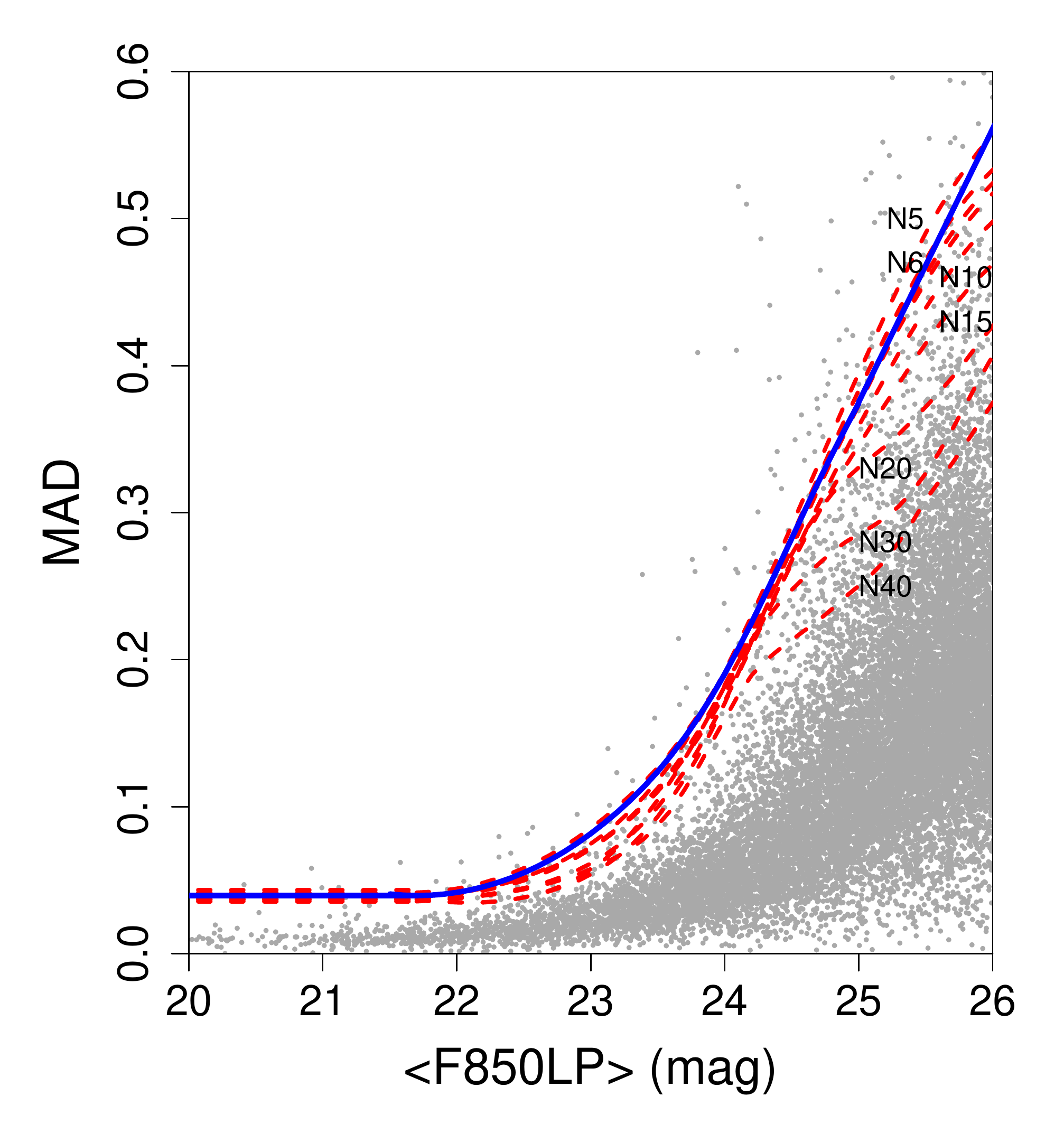}
\caption{MAD as a function of magnitude of all sources in our initial sample. The blue solid line represents the adopted variability threshold of 3.5$\sigma$, while the red dashed lines represent the thresholds for the median under-sampled sets for different number of data points in the light curves, N.}
\end{figure}

In order to test if the 3.5$\sigma$ threshold will be at a different level if we consider only light curves with a specific number of data points, we used simulations and checked if the thresholds derived from under-sampled data is lower than the adopted threshold in this work. If this is the case, then the sample of variable sources derived from applying the same threshold to all light curves (irrespective of the number of points in them) will not be affected.

We simulated ten sets of under-sampled data with the sets differing in the number of data points in the light curves (N=5, 6, 7, 8, 9, 10, 15, 20, 30 \& 40). The number of sources of the under-sampled data are 21022, 19346, 17428, 15891, 14207, 13128, 6790, 3677, 1769 \& 1350, respectively. For each set, we performed 1000 iterations and at each time, we randomly selected N data points for all the sources of our initial sample. We then calculated the MAD values and followed the procedure described in Section~\ref{vi} to find the 3.5$\sigma$ thresholds. We took the median values of the thresholds with N data points in the light curve and we compared the results with different N. In Figure~\ref{simulN}, we plot the MAD as a function of magnitude for all the sources of our sample and the 3.5$\sigma$ threshold adopted in this work (solid blue line). We over-plot the median values of 3.5$\sigma$ thresholds for different number of data points in the light curve, N (dashed black lines).

We find no extreme differences between the thresholds derived from different N. From the bright end of the magnitude distribution up to $\sim$24 mag, the thresholds follow the same trend with small scatter with each other. Above $\sim$24 mag, the scheme changes as the simulated thresholds are getting lower for increasing N. The adopted threshold in this work is above all the simulated ones but the set with N=5 above 24 mag that has slightly higher values. As long as the number of sources that have only five data points in their light curves is small, and the differences in the thresholds is not significant, we expect no false variability induced by the different number of data points in the light curves.

\section{Observational properties and light curves of Supernovae}\label{AppendixSNe}

\begin{figure}
\centering
\begin{tabular}{  l }
\includegraphics[trim=0cm 0cm 0.cm 0cm,width=0.45\textwidth]{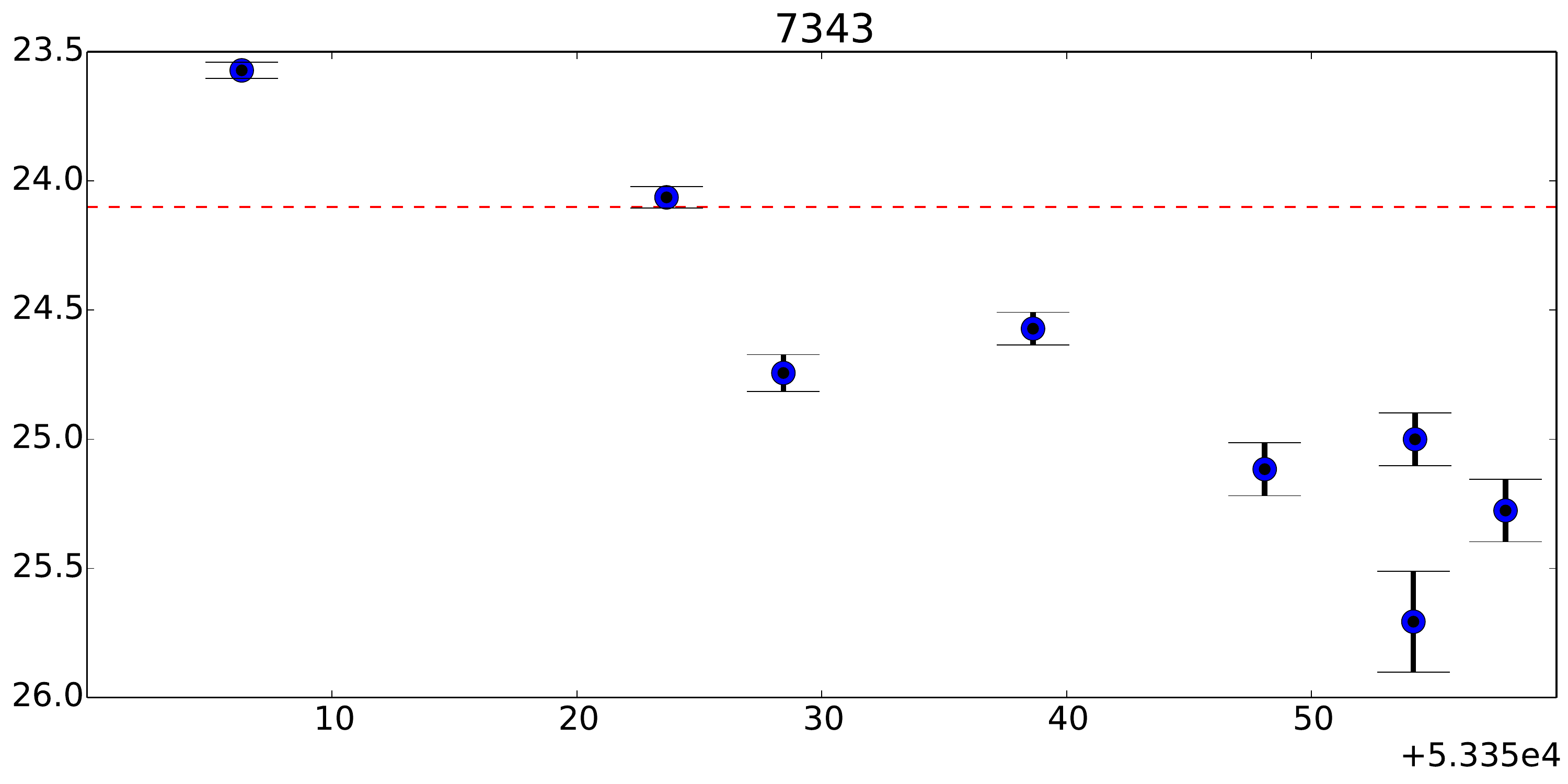}\\ 
\includegraphics[trim=1.45cm 0cm 0.cm 0cm,width=0.45\textwidth]{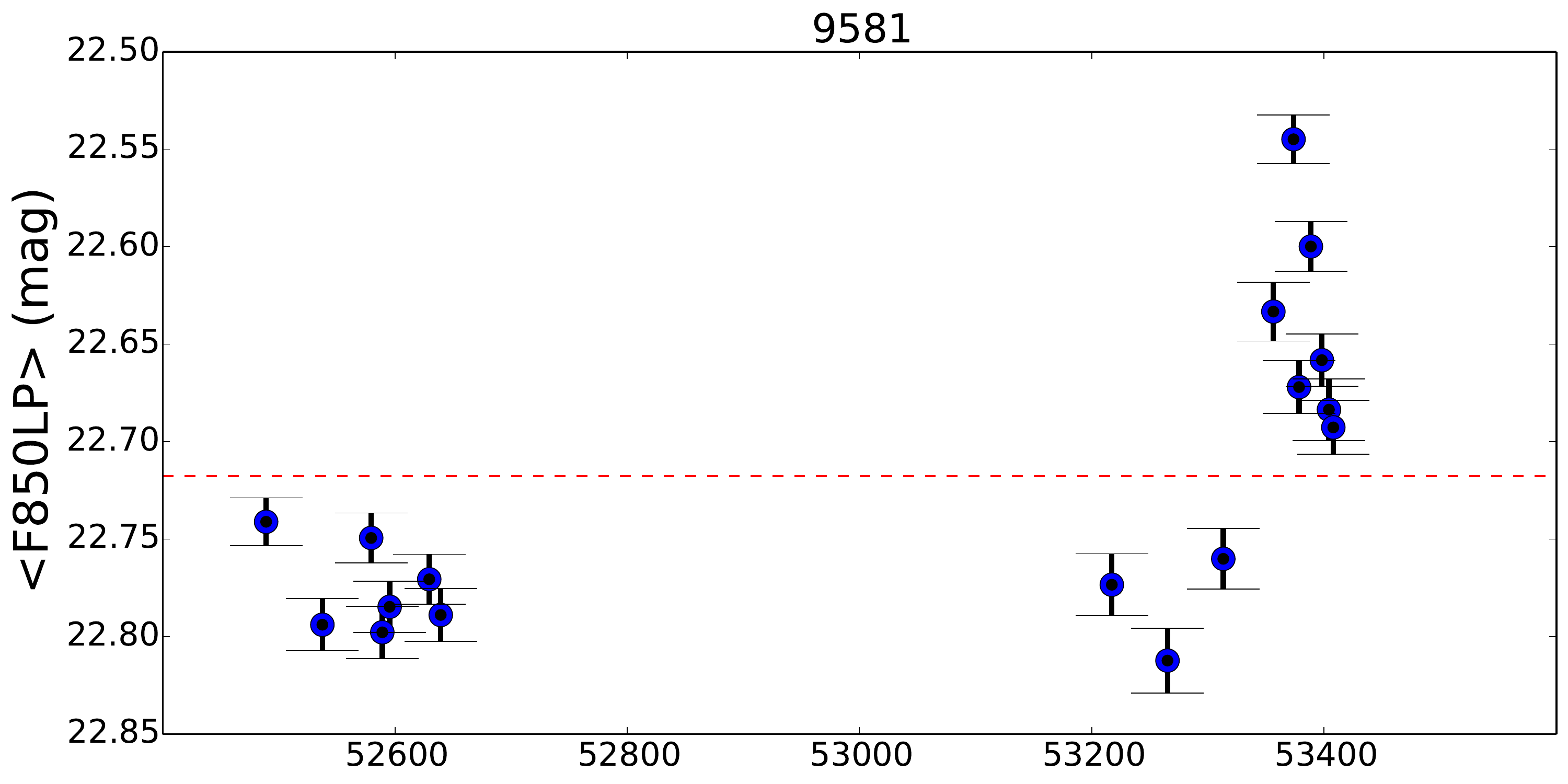}\\ 
\includegraphics[trim=0cm 0cm 0.cm 0cm,width=0.45\textwidth,height=0.255\textwidth]{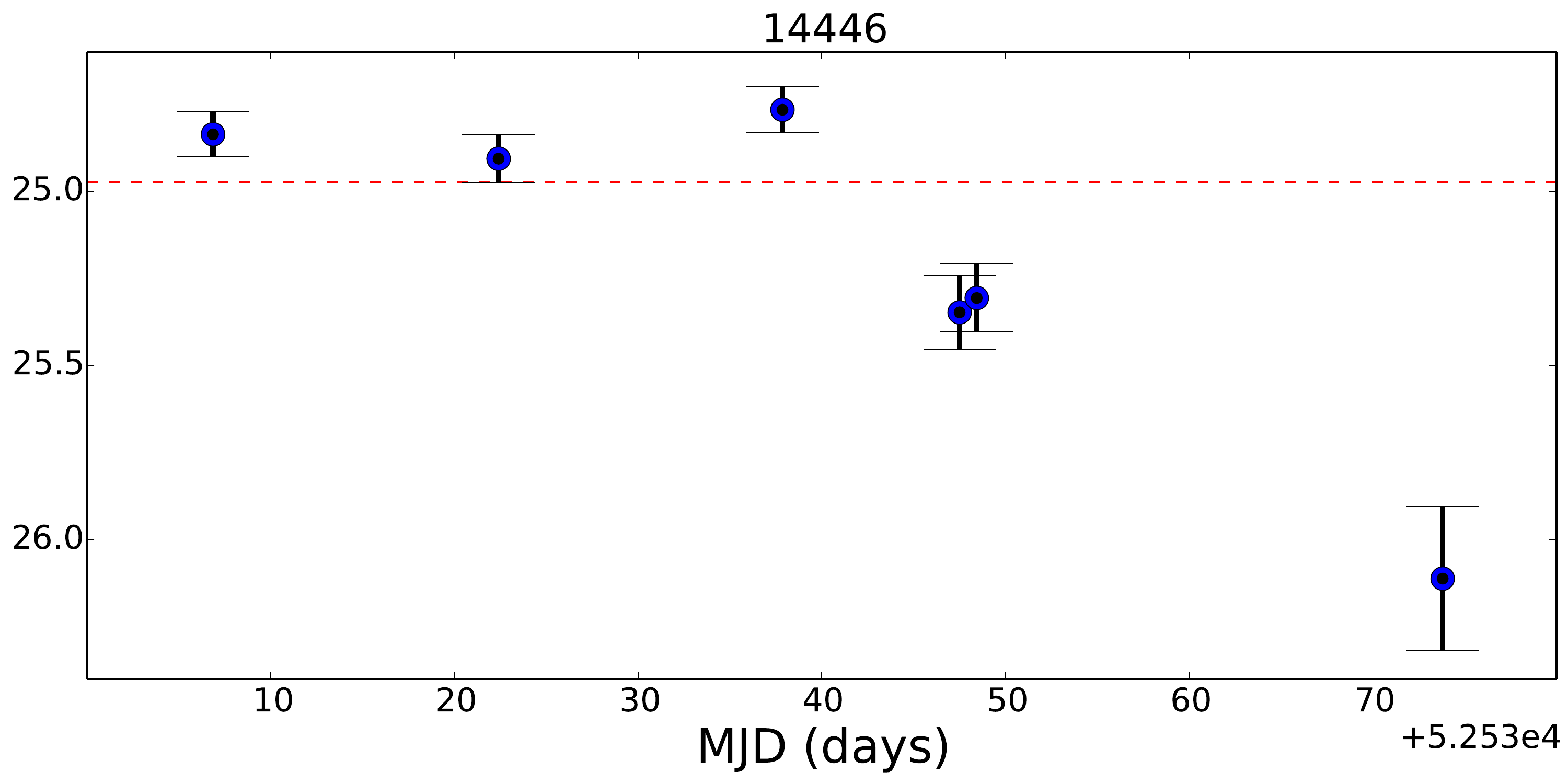}
\end{tabular}
\caption{Light curves of the confirmed SNe identified in this study. The dashed line indicates the median magnitude and the number on the top of each plot indicates the identifier of the source.}\label{lc2}
\end{figure}

Table ~\ref{SN} lists the observational properties, while Figure~\ref{lc2} presents our photometry of the identified SNe. All three SNe have been reported previously in the literature \citep{strolger2004,riess2007}.
\begin{table}
\caption{Catalogue of confirmed SNe identified in our survey.}
\centering
\label{SN}  
\begin{tabular}{ c c c c c c c c c c c}
\hline
ID  & RA & Dec & N\textsubscript{p} & T\textsubscript{bas} & <F850LP> & ${\rm MAD}$*\\
 & (J2000) & (J2000) &  & (years) & (mag) & ($\sigma$)
\\ \hline\hline
    7343 & 53.07570 & -27.73630 & 8  & 0.14  & 24.10 & 11.1\\
    9581 & 53.10564 & -27.75084 & 17 & 2.51  & 22.71 & 4.72\\
    14446 & 53.15638 & -27.77966 & 6  & 0.18  & 24.97  & 4.20\\ \hline
\end{tabular}
\end{table}

\bsp	
\label{lastpage}
\end{document}